\documentclass[12pt,twoside,a4paper]{report}
\usepackage[english]{babel}
\usepackage[T1]{fontenc}
\usepackage[utf8]{inputenc}
\usepackage[usenames,dvipsnames]{xcolor}

\definecolor{color1}{rgb}{0,0,0.7}
\definecolor{color2}{rgb}{0.85,0,0}

\PassOptionsToPackage{hyphens}{url}
\usepackage[breaklinks=true,colorlinks,citecolor=red,linkcolor=color2,urlcolor=color1]{hyperref}
\hypersetup{colorlinks = true}

\usepackage[onehalfspacing]{setspace}

\usepackage{graphicx}
\usepackage{pgfplots}
\usepackage[export]{adjustbox}
\usepackage{lastpage}
\usepackage{float}
\usepackage{afterpage}
\usepackage[firstpage=True]{background}
\usepackage{tikz}

\usepackage{amsmath}
\usepackage{amssymb}
\usepackage{tensor}
\usepackage{bm}
\usepackage{amsthm}
\usepackage{braket}
\usepackage{dsfont}
\usepackage{subfiles}
\usepackage{bbm}
\DeclareMathAlphabet{\mathbbold}{U}{bbold}{m}{n}
\allowdisplaybreaks

\usepackage[style=alphabetic,backend=biber,maxcitenames=3,maxbibnames=99,giveninits=true]{biblatex}
\renewcommand*{\bibfont}{\footnotesize}
\DeclareUnicodeCharacter{0301}{\'{e}}
\DeclareFieldFormat[article,book,inproceedings,misc,incollection]{title}{\mkbibemph{#1}}

\renewbibmacro*{in:}{}
\DeclareFieldFormat[article]{journaltitle}{#1}
\DeclareFieldFormat[article]{issue}{(#1)}
\DeclareFieldFormat[article]{number}{(#1)}
\DeclareFieldFormat[article]{volume}{\textbf{#1}}
\DeclareFieldFormat[inproceedings]{series}{\mkbibemph{#1}}
\DeclareFieldFormat{pages}{pages #1}
\DeclareFieldFormat{url}{\url{#1}}
\DeclareFieldFormat{date}{%
	\ifboolexpr{
		test {\iffieldundef{month}} 
	}
	{(\printfield{year})} 
	{\mkbibparens{\mkbibmonth{\thefield{month}} \addspace \thefield{year}}} 
}

\DeclareBibliographyDriver{article}{%
	\usebibmacro{author}%
	\setunit{, }%
	\usebibmacro{title}%
	\setunit{, }%
	\iffieldundef{doi}{%
		\href{\thefield{url}}{%
			\usebibmacro{journal}%
			\setunit{ }%
			\printfield{volume}%
			\setunit{}%
			\iffieldundef{issue}{%
				\printfield{number}%
			}{%
				\printfield{issue}%
			}%
			\setunit{ }%
			\usebibmacro{date}%
			\setunit{.}%
		}%
	}{%
		\href{http://doi.org/\thefield{doi}}{%
			\usebibmacro{journal}%
			\setunit{ }%
			\printfield{volume}%
			\setunit{}%
			\iffieldundef{issue}{%
				\printfield{number}%
			}{%
				\printfield{issue}%
			}%
			\setunit{ }%
			\usebibmacro{date}%
			\setunit{.}%
		}%
	}%
	\finentry
}

\DeclareBibliographyDriver{book}{%
	\usebibmacro{author}%
	\setunit{, }%
	\usebibmacro{title}%
	\setunit{, }%
	\printlist{publisher}%
	\setunit{, }%
	\printfield{year}%
	\setunit{.}%
	\finentry
}

\DeclareBibliographyDriver{incollection}{%
	\usebibmacro{author}%
	\setunit{, }%
	\usebibmacro{title}%
	\setunit{, in }%
	\printfield{booktitle}%
	\setunit{, }%
	\printlist{publisher}%
	\setunit{, }%
	\printfield{pages}%
	\setunit{, }%
	\printfield{year}%
	\setunit{.}%
	\finentry
}

\DeclareBibliographyDriver{inproceedings}{%
	\usebibmacro{author}%
	\setunit{, }%
	\usebibmacro{title}%
	\setunit{, in }%
	\printfield{booktitle}%
	\setunit{, }%
	\printfield{volume}%
	\setunit{ of }%
	\printfield{series}%
	\setunit{, }%
	\printfield{pages}%
	\setunit{, }%
	\printlist{publisher}%
	\setunit{ }%
	\usebibmacro{date}%
	\setunit{.}%
	\finentry
}

\DeclareBibliographyDriver{misc}{%
	\usebibmacro{author}%
	\setunit{, }%
	\usebibmacro{title}%
	\setunit{, }%
	\printfield{year}%
	\setunit{, }%
	\printfield{url}%
	\setunit{.}%
	\finentry
}

\usepackage{epigraph}
\usepackage{pdfpages}
\usepackage{enumerate}
\usepackage{subcaption}

\usepackage{pifont}
\newcommand{\cmark}{\text{\ding{51}}}
\newcommand{\xmark}{\text{\ding{55}}}

\newenvironment{changemargin}[2]{%
	\begin{list}{}{%
			\setlength{\topsep}{0pt}%
			\setlength{\leftmargin}{#1}%
			\setlength{\rightmargin}{#2}%
			\setlength{\listparindent}{\parindent}%
			\setlength{\itemindent}{\parindent}%
			\setlength{\parsep}{\parskip}%
		}%
		\item[]}{\end{list}}

\usepackage[english,bt]{ethidsc} 

\definecolor{identifiercolor}{rgb}{.4,.6,.56}
\definecolor{stringcolor}{gray}{0.5}
\definecolor{inactivecolor}{rgb}{0.15,0.15,0.5}
\usepackage{listings}
\def \Tr {{\rm Tr}}
\def \eps {\varepsilon}

\newcommand{\<}{\langle}
\renewcommand{\>}{\rangle}
\newcommand{\ev}[1]{\langle #1\rangle}
\newcommand{\tr}[1]{\Tr\!\left[#1\right]}

\graphicspath{{img/}{../img/}}

\newcommand{\eref}[1]{\textcolor{color2}{\hyperref[#1]{eq.$\,$(\ref{#1})}}}
\newcommand{\eqsref}[2]{\textcolor{color2}{\hyperref[#1]{eqs.$\,$(\ref{#1}-\ref{#2})}}}
\newcommand{\Eref}[1]{\textcolor{color2}{\hyperref[#1]{Eq.$\,$(\ref{#1})}}}
\newcommand{\fref}[1]{\textcolor{color2}{\hyperref[#1]{Fig.$\,$\ref{#1}}}}

\newcommand{\sfref}[2]{\textcolor{color2}{\hyperref[#1]{Fig.$\,$\ref{#1}(#2)}}}
\newcommand{\tref}[1]{\textcolor{color2}{\hyperref[#1]{Table~\ref{#1}}}}

\newcommand{\aref}[1]{\textcolor{color2}{\hyperref[#1]{App.$\,$\ref{#1}}}}
\newcommand{\sref}[1]{\textcolor{color2}{\hyperref[#1]{Sec.$\,$\ref{#1}}}}
\newcommand{\cref}[1]{\textcolor{color2}{\hyperref[#1]{Chapter$\,$\ref{#1}}}}

\theoremstyle{plain}
\theoremstyle{definition}
\newtheorem{defn}{Definition}

\DeclareMathOperator*{\argmax}{arg\,max}

\newcommand{\defref}[1]{\textcolor{color2}{\hyperref[#1]{Definition }\bfseries{\ref{#1}}}}

\makeatletter
\newcommand\footnoteref[1]{\protected@xdef\@thefnmark{\ref{#1}}\@footnotemark}
\makeatother

\usepackage{etoolbox}
\makeatletter
\newlength\epitextskip
\pretocmd{\@epitext}{\em}{}{}
\apptocmd{\@epitext}{\em}{}{}
\patchcmd{\epigraph}{\@epitext{#1}\\}{\@epitext{#1}\\[\epitextskip]}{}{}
\makeatother

\begin{document}
		\backgroundsetup{
			scale=1,
			opacity=0.3,
			angle=0,
			hshift=73pt,
			vshift=-2.18cm, 
			contents={\includegraphics[width=\paperwidth]{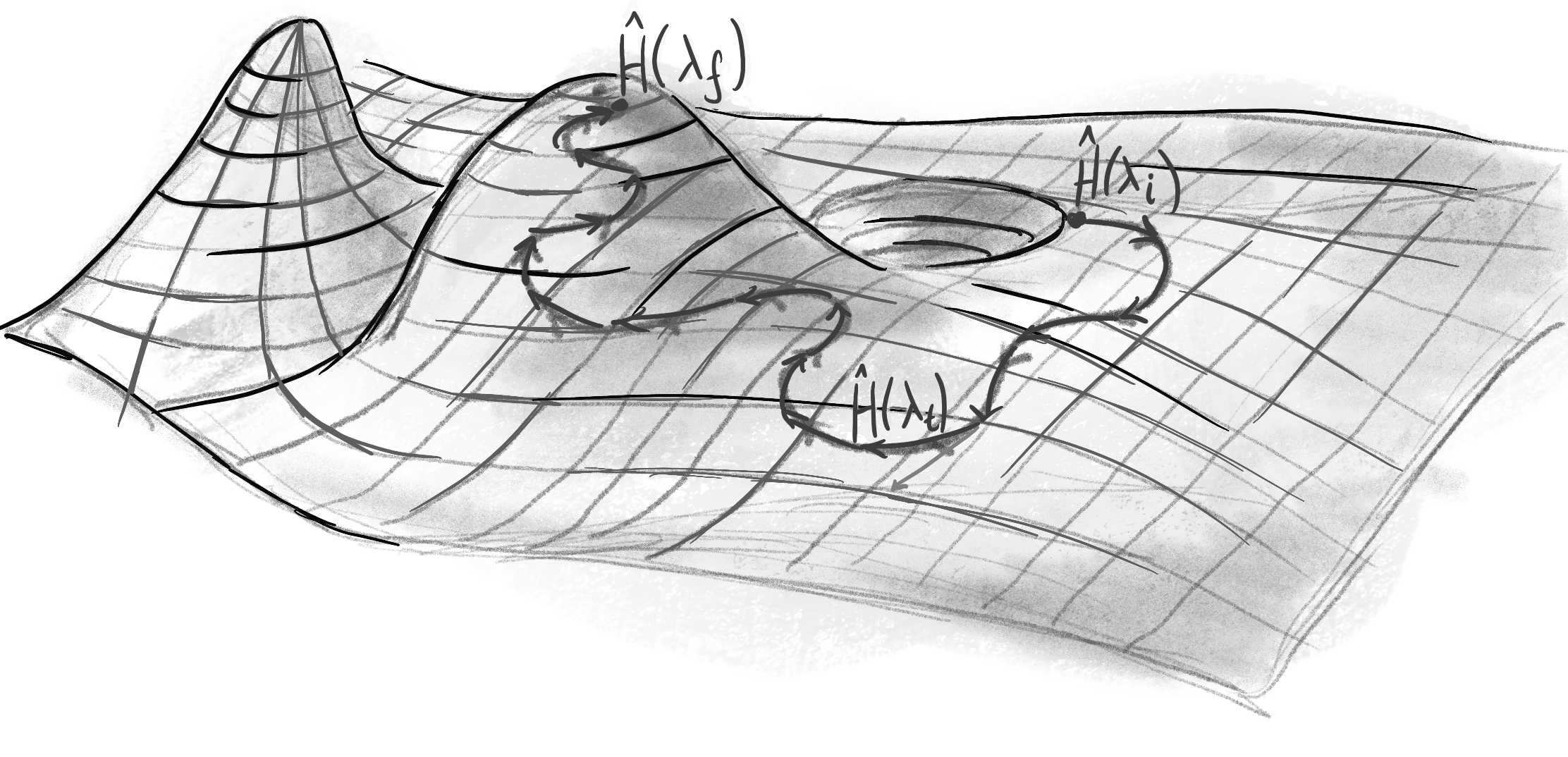}}
		}
		
		\begin{titlepage}
		\newcommand{\HRule}{\rule{\linewidth}{0.5mm}} 
		\center 		
		\begin{changemargin}{-1cm}{-1cm}
			\begin{center}
				\vspace*{-2cm}
				\begin{minipage}{0.6\textwidth}
					\begin{flushleft} 
						\vspace*{-0.68cm}\large UNIVERSIT\'E DE GEN\`EVE\\
						D\'epartement de Physique Appliqu\'ee
						
					\end{flushleft}
				\end{minipage}
				\hspace{1cm}
				\begin{minipage}{0.4\textwidth}
					\begin{flushright} \large
						FACULT\'E DES SCIENCES \\
						Professeur N. Brunner\\
						Docteur M. Perarnau-Llobet 
					\end{flushright}
				\end{minipage}\\[2cm]
				\vspace*{-1.8cm}\HRule 
				\vspace{1cm}
				
				{\huge Finite-Time Processes In Quantum\\ Thermodynamics: The Limits Of Irreversibility}
				
				\vspace*{2cm}
				\begin{minipage}{0.9\textwidth}
					\begin{center} \large
						\textsc{Th\`ese}\\
						pr\'esent\'ee \`a la Facult\'e des sciences de l'Universit\'e de Gen\`eve\\ pour obtenir le grade de
						Docteur \`es sciences, mention physique\\
						\vspace{2.9 cm}
						par\\\vspace{0.6cm}
						Alberto ROLANDI\\
						de Anacapri (Italie)
					\end{center}
				\end{minipage}\\[2cm]
				
				\vspace{20pt}
				\large{Th\`ese N$^\text{o}$ 5846}\\ 
				\vspace{2cm}
				\large{GEN\`EVE}\\
				\large{Centre d'impression Uni Mail}

				{\large \ 2024}\\[2cm] 
				
				\vfill 
			\end{center}
		\end{changemargin}
	\end{titlepage}

	\pagenumbering{roman} 		 
	
	\pagenumbering{gobble} 

\vspace*{3cm}
\thispagestyle{empty}
\setlength\epigraphwidth{.65\textwidth}
\epigraph{Physics is like sex: sure, it may give some practical results, but that's not why we do it.}{--- Richard Feynman}
\vspace*{3.5cm}
\epigraph{Everything we call ``real'' is made of things that cannot be regarded as real.}{--- Niels Bohr}
\setlength\epigraphwidth{.8\textwidth}

\cleardoublepage

\pagenumbering{roman} 
\begin{center}
    Soutenue le 27 septembre 2024 devant le jury compos\'e de:\\ 
    \vspace{2cm}
    Dr. Mart\'i Perarnau-Llobet\\
    Prof. Nicolas Brunner\\
    Prof. Antonio Ac\'in\\
    Prof. Marcus Huber \\
    Dr. Patryk Lipka-Bartosz
\end{center}
\cleardoublepage

 
\chapter*{Acknowledgements}
\addcontentsline{toc}{chapter}{Acknowledgements}
First and foremost, I would like to express my unending gratitude to my supervisor Martí Perarnau-Llobet. Thank you for these amazing and formative four years. I cannot overstate how amazing it was to be supervised by someone like you: your planning, guidance, and trust played an ever so important role in making me the person and researcher I am today, for which I can be nothing but thankful.\\

To a great extent, this thesis is the result of wonderful teamwork and collaborations: Nicolas, Paolo (A), Paolo (E), Harry, Pavel, Kush, Yikai, Vasso, Tiff, and Sjaak; thank you all, you have my deepest gratitude for working with me.\\

Then I wish to thank my friends and colleagues from Geneva: Ivan, Lorenzo, Gianmichele, Flavio, Ophelia, Rafa, Joey, Mingsong, Maria, Yiorgos, Roope, Victor, Michael, Théo, Patryk, Bora, Alexandre, Giovanni, Rebecka, Sadra, Shishir, Geraldine, Sophie, Towsif, Javier, Davide, and Dmitry;
it is difficult for me to express how important you have been in making me feel at home in this city.\\

I would also like to thank friends and researchers from around the world, from my time in Lausanne to my trips during my PhD, and including timeless friends. Whether it is for the support they gave me, the ``good vibes'', or the fruitful discussions: Aakash, Alessandro, Andres, Antonia, Boris, Christian, Edgars, Eloic, Emilio, Evi, Fabian, Florian, Gabriel, Gabriele, Gerry, Hadrien, Hassan, Hendrik, Henrik, Jake, Jeanne, Kai, Kevin, Laetitia, Leonardo, Ludovico, Maja, Mark, Martina, Marzio, Matheus, Mathieu, Matteo, Michael, Nauris, Nicola, Noshin, Pierre, Quentin, Ricard, Tommaso, Victor, and Willem.\\

Finally, I would like to thank my parents and step-parents. To all of you, thank you for believing in me, thank you for supporting me from the very start. To Ariella and Mike, thank you for teaching me how to be critical of the world and of myself -- and how to be a somewhat practical person. \emph{Per Mamma e Papà, grazie per avermi introdotto a tutto quello che so, a come ragiono e come vedo il mondo. Grazie per avermi voluto sempre bene e grazie per avermi sempre guidato in qualsiasi avventura della mia vita. Non bastano parole per esprimere quanto io vi sia riconoscente...}

\cleardoublepage

\chapter*{Abstract}
\addcontentsline{toc}{chapter}{Abstract}
The emergence of irreversibility in physical processes, despite the fundamentally reversible nature of quantum mechanics, remains an open question in physics. This thesis explores the intricate relationship between quantum mechanics and thermodynamics, with a particular focus on minimizing entropy production in finite-time processes. By employing tools from quantum information theory and geometric thermodynamics, we tackle the challenge of deriving irreversible thermodynamic behavior from the reversible microscopic framework of quantum mechanics.

We begin with a comprehensive review of the laws of thermodynamics, setting the stage for the subsequent analyses. We introduce novel developments in quantum thermodynamics through a generalized framework for geometric thermodynamics, which enables the derivation of finite-time corrections beyond the Markovian regime.

Building on this foundation, we extend Landauer's principle by incorporating a finite-time correction that highlights the necessity of strong coupling for optimal information erasure processes. This result underscores the emergence of Planckian time as a fundamental speed limit to thermalization. Additionally, we explore how collective effects can be harnessed to reduce energy dissipation in thermodynamic operations, revealing that classical correlations between systems can significantly mitigate dissipation, though this may pose new questions regarding the third law of thermodynamics. Finally, we optimize thermodynamic processes in mesoscopic systems, including quantum dot engines and information engines. 

These findings not only enhance our understanding of the fundamental limits of irreversibility but also open new avenues for research. Future works will focus on fully exploiting collective effects, aligning these with the third law of thermodynamics, and understanding the thermodynamic consistency of master equations.

\chapter*{R\'esum\'e}
\addcontentsline{toc}{chapter}{R\'esum\'e}
L'émergence de l'irréversibilité dans les processus physiques, malgré la nature fondamentalement réversible de la mécanique quantique, demeure une question ouverte en physique. Cette thèse explore la relation complexe entre la mécanique quantique et la thermodynamique, en se concentrant particulièrement sur la minimisation de la production d'entropie dans les processus thermodynamiques en temps fini. En utilisant des outils de la théorie de l'information quantique et de la thermodynamique géométrique, nous abordons le défi de dériver un comportement thermodynamique irréversible à partir du cadre microscopique réversible de la mécanique quantique.

Nous commençons par une revue complète des lois de la thermodynamique, établissant les bases pour les analyses suivantes. Nous introduisons des développements nouveaux en thermodynamique quantique à travers un cadre généralisé pour la thermodynamique géométrique, qui permet la dérivation de corrections en temps fini au-delà du régime Markovien.

En nous appuyant sur cette base, nous étendons le principe de Landauer en incorporant une correction en temps fini qui met en évidence la nécessité d'un couplage fort pour des processus d'effacement de l'information optimaux. De plus, nous explorons comment les effets collectifs peuvent être exploités pour réduire la dissipation d'énergie dans les opérations thermodynamiques, révélant que les corrélations classiques entre systèmes peuvent atténuer substantiellement la dissipation d'énergie. Enfin, nous optimisons les processus thermodynamiques dans les systèmes mésoscopiques, y compris les moteurs à points quantiques et les moteurs d'information.

Ces résultats non seulement enrichissent notre compréhension des limites fondamentales de l'irréversibilité, mais ouvrent également de nouvelles pistes de recherche. Les travaux futurs se concentreront sur l'exploitation complète des effets collectifs, leur alignement avec la troisième loi de la thermodynamique, et la compréhension de la cohérence thermodynamique des équations maîtresses.

\cleardoublepage

\section*{List of Relevant Publications}
\addcontentsline{toc}{chapter}{Relevant Publications}
This thesis is based on the following works (by order of appearance).
\begin{enumerate}
	
	\item \textit{Finite-time Landauer principle beyond weak coupling}\\
	\textbf{A. Rolandi}, M. Perarnau-Llobet.\\ \href{https://doi.org/10.22331/q-2023-11-03-1161}{Quantum 7, 1161 (2023).}
	
	\item \textit{Collective advantages in finite-time thermodynamics}\\
	\textbf{A. Rolandi}, P. Abiuso, M. Perarnau-Llobet.\\  \href{https://doi.org/10.1103/PhysRevLett.131.210401}{Phys. Rev. Lett. 131, 210401 (2023)}.
	
	\item \textit{Pareto-optimal cycles for power, efficiency and fluctuations of quantum heat engines using reinforcement learning}\\
	P. A. Erdman, \textbf{A. Rolandi}, P. Abiuso, M. Perarnau-Llobet, F. Noé.\\
	\href{https://doi.org/10.1103/PhysRevResearch.5.L022017}{Phys. Rev. Research 5, L022017 (2023).}
	
	\item \textit{Optimal finite-time Szilard quantum dot engine}\\
	\textbf{A. Rolandi}, K. Aggarwal, Y. Yang, N. Ares, M. T. Mitchison, M. Perarnau-Llobet.\\
	(In preparation)
	
	\item \textit{Optimal control of dissipation and work fluctuations for rapidly driven systems}\\
	\textbf{A. Rolandi}, M. Perarnau-Llobet, H. J. D. Miller.\\
	\href{https://doi.org/10.1088/1367-2630/ace2e3}{New J. Phys. 25, 073005 (2023).}
	
\end{enumerate}
The following are works in which I took part during my PhD, but are not covered in the thesis.
\begin{enumerate}
	\item \textit{Extensive Rényi entropies in matrix product states}\\
	\textbf{A. Rolandi}, H. Wilming.\\ \href{https://doi.org/10.48550/arXiv.2008.11764}{arXiv preprint (2020).}
	
	\item \textit{Proof of entanglement via dynamics exclusion in a quantum dots pair}\\
	\textbf{A. Rolandi}, V. Angelopoulou, T. Brydges, C. J. van Diepen, P. Sekatski.\\
	(In preparation)
\end{enumerate}

\cleardoublepage

\setcounter{tocdepth}{4}
\hypersetup{linkcolor=color1}
\tableofcontents
\hypersetup{linkcolor=color2}

\cleardoublepage

	
	
	\pagestyle{fancy}
	\pagenumbering{arabic}
	\setlength{\parindent}{20pt}

	\chapter{Introduction}\label{introduction}
\epigraph{The more success the quantum theory has, the sillier it looks.}{--- Albert Einstein}

One of the most puzzling open problems in physics is the emergence of irreversibility. The phenomena that we observe in the world around us are evidently irreversible. This irreversibility is even thought to be the origin of the arrow of time. However, at the microscopic scale the world is described by a reversible theory: quantum mechanics. It is difficult to overstate the success of quantum physics, which has led us to a better understanding of the most fundamental aspects of nature, enabling technological discoveries that revolutionized the world of computing, communication and medicine. Therefore, it is surprising that the irreversible processes that we observe at the macroscopic scale should emerge from the fundamentally reversible microscopic processes described by quantum mechanics -- or more precisely, quantum electrodynamics. It is worth noting that, however, the theory of quantum mechanics was born in an attempt to describe a thermodynamic -- and fundamentally irreversible -- process: black-body radiation. Indeed, the origins of quantum mechanics are deeply rooted in thermodynamics: the quantization of light arises by imposing the thermodynamic consistency of Maxwell's equations~\cite{Einstein1905}.

Given this context, it is natural to ask if one can derive irreversibility -- and more generally, the laws of thermodynamics -- by starting from a microscopic and reversible theory like quantum mechanics. Indeed, foundational insights on the topic date back to the early days of quantum mechanics with the works of Von Neumann~\cite{vonNeumann2010,vonNeuman1932}, which extend the works of Boltzmann in classical statistical mechanics to quantum mechanics. A key insight that they brought to our modern understanding of both quantum mechanics and thermodynamics is the role of \emph{information} in the two theories. Specifically, they illuminate how the lack of information on a system relates to entropy. Perhaps, the role of information for irreversibility is best exemplified by Landauer's principle~\cite{Landauer1961}, as it shows that the entropy of the universe increases when information is erased.

Since then, the question of defining entropy production and proving the irreversibility of thermodynamics starting from a reversible classical theory has been a central focus of the field of stochastic thermodynamics, which has developed modern mathematical tools to tackle the topic. Among its most impactful results are the fluctuation theorems~\cite{Jarzynski1997,Crooks1999}, which generalize the laws of thermodynamics to non-equilibrium processes. These prove the validity of the second law of thermodynamics for a wide variety of scenarios.

In more recent years, with the advent of quantum information and a variety of experimental breakthroughs that have demonstrated an unprecedented level of control over quantum systems -- such as trapped ions~\cite{Ronagel2016} and optomechanical systems~\cite{Rademacher2022} -- the focus has increasingly shifted towards quantum thermodynamics. This emerging field, at the intersection of quantum information, stochastic thermodynamics, and many-body physics, has led to a variety of significant advancements ranging from the topic of the equilibration and thermalization of quantum systems to thermodynamic uncertainty relations. For instance, it has been demonstrated that small subsystems of a large pure quantum state can effectively evolve towards a thermal state under certain conditions, highlighting the robustness of thermalization in quantum settings. Moreover, the special role of the observer/external agent in both quantum mechanics and thermodynamics naturally leads to information-theoretic interpretations in quantum thermodynamics. In particular, irreversibility in thermodynamic processes is now understood as the loss of information about the state of the system. 

The question of irreversibility is especially relevant for finite-time processes where the system is brought out of equilibrium and entropy production is thought to be always strictly positive. Indeed, the main goal of this thesis is to reach a better understanding on the gap between reversible processes with no entropy production and irreversible processes with finite entropy production. Therefore, it is in this context -- and armed with quantum information theoretic tools -- that we approach multiple aspects of the minimization of the entropy production of physical tasks.

\section*{Thesis Outline}
\addcontentsline{toc}{section}{Thesis Outline}


In this thesis we start by presenting a self-contained review on the laws of thermodynamics in \cref{chap:thermo_laws}. We then present the mathematical formalism that is used in the rest of the thesis. In particular, in \sref{sec:geom} we generalize the results of \cite{Scandi19} to obtain a derivation of the thermodynamic geometry technique that applies to all types of dynamics -- i.e. for both open and closed quantum systems; thus giving a formal setting for the work minimization of thermodynamic protocols in the slow-driving regime, which is used multiple times in the rest of the thesis.

In \cref{chap:landauer} we apply this extended formalism to obtain a finite-time correction to Landauer's bound on the dissipated energy required to erase information. While previous works obtaining a finite-time correction to the bound already existed, these were limited to the Markovian regime~\cite{Scandi19,VanVu2022,Zhen2021}. The main result of the chapter  \eref{MainResult} is valid beyond such approximations~\cite{Rolandi2023Quantum}. Indeed, it unveils the need for strong coupling for the optimal energy management of erasure processes. Furthermore, despite the result being derived in a specific setting, it reveals the general form of a universally valid finite-time correction to Landauer's bound with the emergence of \emph{Planckian time}.

Remarkably, the results of \cref{chap:collective} show how one can use the framework of geometric thermodynamics to unveil a novel type of collective effects that allow to drastically reduce the amount of energy that is dissipated when performing a task on a collection of systems simultaneously~\cite{Rolandi2023Collective}. Collective effects are a well known and important phenomenon in physics, ranging from phase transition to quantum entanglement~\cite{Campaioli2018,Vroylandt2017,Souza2022Collective,Niedenzu_2018,Chiaracane2020}. It is often the case that one can exploit these effects to improve the result or cost of a given task, leading to a notion of collective advantages: the outcome of a task is improved when performed globally on a collection of systems instead of each system individually. Indeed, our results show that if one wishes to perform multiple tasks on multiples systems, then it is possible to exploit classical correlations between these systems to drastically suppress the amount of energy that is dissipated into the environment. To put it simply, one manages to ``share the losses'' between the individual systems. With sufficient amount of control, this suppression is strong enough to obtain a \emph{vanishing dissipation per system} in the thermodynamic limit. This is an astounding result, as it seems to go counter to the third law of thermodynamics, but on a technical level there seems to be no contradiction as the remaining finite-time dissipation is always strictly positive.

Finally, in \cref{chap:opt} we focus on the thermodynamic optimization of mesoscopic systems. We show how to fully optimize two types of nano-scale thermal engines and fast operations on a system. The optimization of thermal engines and, more generally, operations at the nano-scale are questions that allows us to probe the fundamental limits of the trade-offs between relevant physical costs and desiderata~\cite{Holubec2014,Mehboudi2022,pietzonka2018,Abiuso2020,Miller2023}. In steady state heat engines, these trade-offs are captured by thermodynamic uncertainty relations~\cite{barato2015,pietzonka2018}.
First, we study how one can fully optimize all the relevant thermodynamic aspects of the simplest possible thermal engine: a periodically driven quantum dot engine~\cite{Erdman2023Pareto}. Indeed, with the use of analytical techniques and reinforcement learning we characterize the optimization trade-off of power, efficiency and power fluctuations of a quantum dot engine in all driving regimes. Second, we characterize and optimize the performance of an information engine at all driving speeds. Information engines stand in a class of their own, as they make use of measurements on the system instead of a cold reservoir. Furthermore, we confirm the feasibility of the optimal protocols we compute with an experimental implementation thanks to the collaboration of the group of prof. N. Ares. Last, but not least, we develop a general framework for the optimization of work and work fluctuations in rapidly driven systems~\cite{Rolandi2023Fast}. Our results show that optimal fast protocols, in both open and closed system dynamics, consist of two instantaneous jumps in the control parameters -- one at start and one at the end of the protocol. Interestingly, this is true whether we are aiming to optimize work or work fluctuations -- or any trade-off of the two quantities, but it is seldom the case that the optimal ``jump point'' is the same for both objectives. We showcase this phenomenon by applying the framework to compute the optimal fast protocols across phase transitions of many-body quantum systems.

In \cref{chap:conclusion} we conclude on the presented work and reflect on the future perspectives of research that were opened by this thesis.

	\cleardoublepage
	\chapter{The Laws of Thermodynamics}\label{chap:thermo_laws}
\epigraph{The law that entropy always increases holds, I think, the supreme position among the laws of Nature. If someone points out to you that your pet theory of the universe is in disagreement with Maxwell's equations – then so much the worse for Maxwell's equations. If it is found to be contradicted by observation – well, these experimentalists do bungle things sometimes. But if your theory is found to be against the second law of thermodynamics I can give you no hope; there is nothing for it but to collapse in deepest humiliation.}{--- Sir Arthur Eddington, \emph{The Nature of the Physical World}}

\section{Thermodynamic Operations and Work}\label{sec:thermo_op}
\begin{figure}[tb!]
	\centering
	\includegraphics[width=\textwidth]{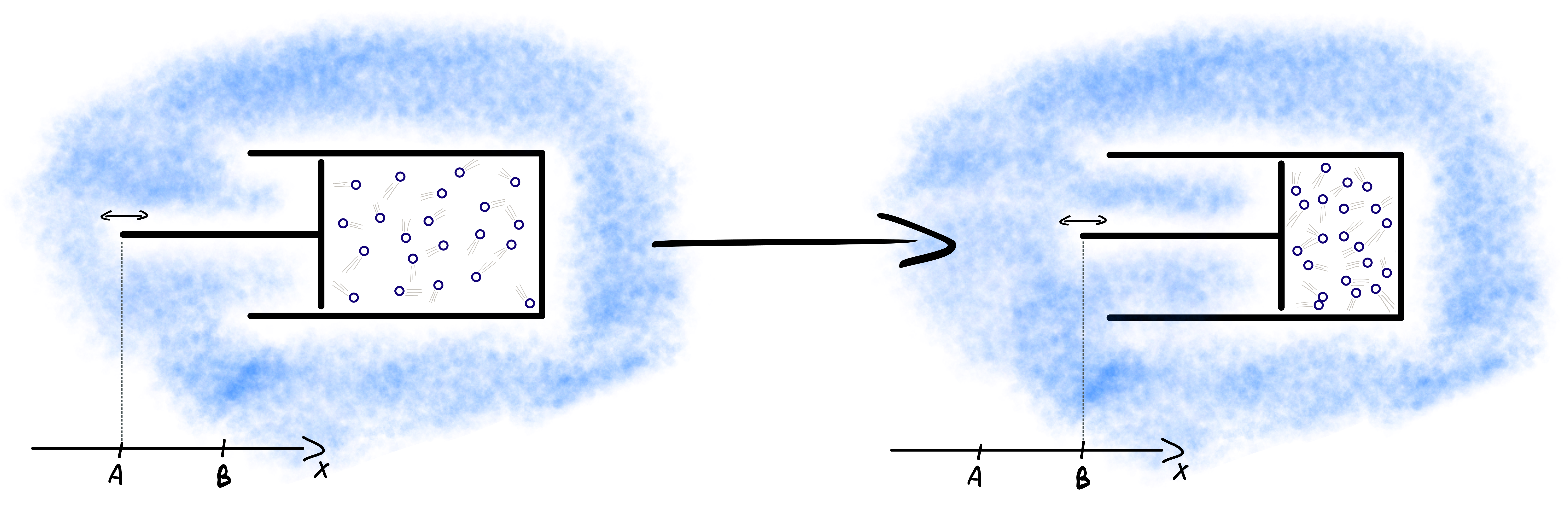}
	\caption{Example of a thermodynamic operation: compressing a piston. The studied system is a gas of $N$ particles inside a piston. The whole system is in an environment that is depicted in blue. On the left we have the gas in its initial state with the piston in position $A$ and on the right the gas after the piston has been compressed to position $B$, reducing the volume of the system.}
	\label{fig:thermo_op}
\end{figure}

One of the main objectives of Thermodynamics is to describe flows of energy in a physical process~\cite{Fermi56}. This question is so general that it can be applied to any field of physics, which makes thermodynamics a so-called ``universal'' theory. Before stating the laws of thermodynamics, we will define what is an ``operation'' in the thermodynamic sense. In a typical thermodynamic setting one is studying a system $S$ with a set of parameters that can be controlled by an external agent. For example, if we consider the gas in a piston (cf. \fref{fig:thermo_op}), an agent can control the volume of this gas by moving the piston rod. Another example -- in the world of Quantum Mechanics -- would be an electron in the magnetic field of an experiment.\\
We define a thermodynamic operation as the changing of one or more parameters of the system from an initial value to a final value -- which could be the same as the initial value. It is worth noting that any physical process could be framed as a thermodynamic operation. In \fref{fig:thermo_op} we give a typical textbook example of such a task: the compression of a gas in a piston. In mathematical terms, this corresponds to changing the Hamiltonian of the system $S$ from its initial value $\hat H_i$ to its final value $\hat H_f$. We will be calling \emph{protocol} the function $\hat H(t)$ that describes how this change is implemented.
This leads naturally to the following definition:
\begin{defn}[Thermodynamic Work]\label{def:work}
	The \emph{thermodynamic work} $W$ of an operation on a system $S$ is the amount of energy that the agent has to spend in order to perform the aforementioned operation on the system $S$.
\end{defn}
If we denote by $\hat\rho_{tot}(t)$ and $\hat H_{tot}(t)$ the state and Hamiltonian of the system and everything that it is interacting with -- except the agent -- we have a very natural definition for the expected value of work
\begin{equation}\label{eq:work_def}
	\<W\> = \Tr\!\left[\hat\rho_{tot}(\tau)\hat H_{tot}(\tau) - \hat\rho_{tot}(0)\hat H_{tot}(0)\right]~,
\end{equation}
where we denoted by $\tau$ the amount of time it took to perform the operation. The study of work and how to minimize it is a central question in thermodynamics. Typically, one can expect the work cost to increase for decreasing values of $\tau$ and to decrease for increasing values of $\tau$. In fact, one can show that the work cost is minimal when $\tau$ goes to infinity (cf. \sref{sec:third_law}). Since we will be treating stochastic systems throughout this thesis and will never explore single shot scenarios, we will only be speaking of the expected work cost $\< W\>$ instead of the work cost $W$. Therefore, to lighten the notation and text, we will drop the average symbol $\<\cdot\>$ for this quantity and will be writing ``work cost'' implying that the average has been taken.

\section{The First Law and the Definition of Heat}
As in the example shown in \fref{fig:thermo_op}, during a thermodynamic operation one can distinguish three parties that are involved: an agent, the system being described, and the environment.
If one considers the energy flow that happens between these three parties during a thermodynamic operation then, by conservation of energy, when the agent provides $\delta W$ of energy it is split between the system and the environment.\\

\noindent{\bf The First Law of Thermodynamics.} During a thermodynamic operation, the following holds by conservation of energy
\begin{equation}\label{eq:first_law}
	\delta W = dE + \delta Q~,
\end{equation}
where $\delta W$ is the amount of energy provided by the agent during an infinitesimal amount of time $dt$, $dE$ is the energy gained by the system $S$ and $\delta Q$ is the energy gained by the environment. \\

Typically one identifies the environment as ``whatever is not the system and is interacting with it''. This usually coincides with something that cannot be controlled by the agent -- or at least not as well as the system. Therefore the energy that is gained by the environment cannot be directly accessed anymore by the agent. Which leads us to another definition:
\begin{defn}[Heat]\label{def:heat}
	The \emph{heat}
	\begin{equation}
		Q := W - \Delta E 
	\end{equation}
	of an operation is the amount of energy that the environment has gained when the agent performed a thermodynamic operation on a system $S$ --  which gained $\Delta E$ of energy\footnote{This partition of the energy is omitting any amount of energy that is stored in the interaction between system and environment. In all the scenarios we will study, the interaction energy at the start and end of the process will be the same, thus allowing us to write the energy balance as in \defref{def:heat}. This is still a lively topic in the community, see \cite{Dann2023} for a summary.}.
\end{defn}
It is interesting to note that by defining a different cut between what is being studied and what is the environment, one gets a different definition of heat. Which might make it seem arbitrary, but this a fundamental feature of thermodynamics that ties it to information-theory. By distinguishing between what can and cannot be acted upon by the agent, we are making an information-theoretic statement. And  by applying conservation of energy to this statement we obtain the first law of thermodynamics. Therefore we can see that, at least partially, the first law of thermodynamics is an information-theoretic statement.\\

If we denote by $\hat\rho (t)$ the state of the system $S$ at a time $t$ and its Hamiltonian by $\hat H(t)$, then its expected energy is given by $E(t) = \Tr\!\left[\hat\rho(t) \hat H(t)\right]$. If we differentiate this expression in time we get
\begin{equation}
	\frac{dE}{dt} = \Tr\!\left[\hat\rho(t) \frac{d\hat H(t)}{dt}\right] + \Tr\!\left[\frac{d\hat\rho(t)}{dt} \hat H(t)\right]~.
\end{equation}
By using the first law \eref{eq:first_law} and the definition of work, we can identify that work corresponds to the term with the change in the Hamiltonian: $\delta W =\Tr\!\left[\hat\rho(t) \hat H'(t)\right]$, where we used $'$ to denote the time derivative to lighten the notation. One can recover this term by taking the time derivative of \eref{eq:work_def}, but it can be also understood as the term that the agent has control over.
 This leads to the two integral formulas for the work and heat of a process
\begin{align}
	\label{eq:work}
	W &= \int_{0}^\tau\!dt~\Tr\!\left[\hat\rho(t) \frac{d\hat H(t)}{dt}\right]~, \\
	Q &= -\int_{0}^\tau\!dt~\Tr\!\left[\frac{d\hat\rho(t)}{dt} \hat H(t)\right]~.
	\label{eq:heat}
\end{align}
It is worth noting that these integral forms are not fully rigorous -- despite being widely used -- as they do not give the proper result in the case one introduces discontinuities at the start and end of the protocol. In particular, if we imagine a protocol where at $t=0$ there is a sudden change in the Hamiltonian -- which we will call \emph{quench} going forward -- then the integral in \eref{eq:work} will give an erroneous $1/2$ factor to the term corresponding to the quench. However this can be easily ``fixed'' if we keep in mind that the true boundaries of the integrals in \eref{eq:work} and \eref{eq:heat} are $-\varepsilon$ and $\tau+\varepsilon$ in the limit that $\varepsilon$ tends to $0^+$.

\newpage
\section{Equilibrium and the Second Law}
\label{sec:second_law}
\epigraph{
	And on the pedestal these words appear:\\
	``My name is Ozymandias, King of Kings:\\
	Look on my works, ye Mighty, and despair!''\\
	No thing beside remains. Round the decay\\
	Of that colossal wreck, boundless and bare\\
	The lone and level sands stretch far away.}{--- Percy Shelley, \emph{Ozymandias}}
As opposed to Maxwell's equations, the second law of thermodynamics is still today a source of lively arguments and discussion. The core physical phenomenon it addresses is the presence of irreversible phenomena in the macroscopic world -- e.g. the mixing of liquids -- despite the time-symmetry of the underlying physical laws. On the fundamental side, the second law has deep ties to the emergence of the \emph{arrow of time}. While on a more practical aspect, together with the first law it is essential for the design of thermal engines, power plants, and the analysis of chemical reactions~\cite{Jarzynski2011}.
It was first stated by Clausius in 1865 as ``Heat can never pass from a colder to a warmer body without some other change, connected therewith, occurring at the same time'' \cite{Clausius1865}. It is important to note that this is a law of averages: it does not forbid the transfer of energy from a cold body to a warmer body in a single instance, but rather it states that the energy transfer is more likely in a direction than the other. The comparison of the probability of an energy transfer occurring in one direction or the other is a well studied subject in the context of \emph{fluctuation theorems}~\cite{Jarzynski2011}.

The formulation of Clausius has been changed into many forms since he stated it, but today the most common statement of the second law of thermodynamics is in terms of \emph{entropy production}.\\

\noindent{\bf The Second Law of Thermodynamics.} The entropy production of a thermodynamic operation is non-negative
\begin{equation}\label{eq:second_law}
	\Sigma \geq 0~.
\end{equation}

Before defining entropy production, we will introduce a few concepts that build up to it. Clearly, the first of these concepts should be entropy. Entropy is an information-theoretic quantity that captures the lack of knowledge that an observer has on the system -- in this context the thermodynamic agent is the observer. This abstract notion of ``lack of knowledge'' can be exemplified by Gibbs' definition of entropy: $S = \ln \Omega$, where $\Omega$ is the number of (equally likely) configurations of the system that are compatible with the knowledge of the observer. With this definition we have a natural operational interpretation of the entropy of a system: it is the minimal amount of information (in number of bits) that we need to measure about the system to pin down the specific configuration of the system. For quantum systems we will be using a more general definition of entropy:
\begin{defn}[von Neumann Entropy]\label{def:entropy}
	The \emph{von Neumann entropy} of a system described by the state $\hat\rho$ is given by
	\begin{equation}
		S(\hat\rho) := -\Tr\left[\hat\rho\ln\hat\rho\right]~.
	\end{equation}
\end{defn}
It is worth noting how the von Neumann entropy reduces to the Gibbs definition of entropy when we have a uniform mixture between $\Omega$ orthogonal states. Furthermore, this definition carries the same operational interpretation about ``lack of information''. In particular, it is minimal (equal to zero) if and only if the state is pure, and it is maximal if and only if the state is fully mixed.

The concept of entropy allows us to define a notion of ``usefulness'' of a state. To illustrate this, let us consider two scenarios of a gas with many particles. First suppose that we have no information on the state of the gas. This would imply that we describe the state of the gas with a completely mixed density matrix, which has maximal entropy. Because of our lack of information on the gas, there is no operation we can do to extract energy from it in a repeatable way. In the second scenario, suppose that thanks to some sophisticated measurement apparatus we acquired the information about the position and momentum of every particle in the gas. Therefore we can now describe the state of the gas with a pure state, which has zero entropy. Since we know the position and velocity of every particle in the gas, we can compute their future trajectories. And therefore, in principle, we could make a very contrived machine that extracts all of the kinetic energy from each gas particle and make use of all of the energy available. Since in both scenarios the energy of the system can be the same, but in one we can reliably extract all of it as opposed to the other, we naturally reach a notion of ``useful energy'', which is captured by the following definition.
\newpage
\begin{defn}[Non-Equilibrium Free Energy]\label{def:neq_free_energy}
	The \emph{non-equilibrium free energy} $F_{n.eq}(\hat\rho)$  of a system $\hat\rho$ with Hamiltonian $\hat H$ is defined as
	\begin{equation}\label{eq:neq_free_energy}
		F_{n.eq}(\hat\rho) := \Tr[\hat\rho \hat H] - k_B T S(\hat\rho)~,
	\end{equation}
	where $k_B$ is the Boltzmann constant and $T$ is the temperature\footnote{Here the temperature can be defined as $k_B^{-1}\partial E/\partial S$. However it is worth noting that often it coincides with the temperature of the environment with which the system is interacting. See \cite{Strasberg2021} for a more detailed discussion.}.
\end{defn}
With this definition, one can interpret the second law as a statement about how, over time, we have less and less free energy available in the system, until the entropy is maximized and the system is ``useless''. We then say that the system has reached \emph{equilibrium}. For a given Hamiltonian and a given temperature there is a unique equilibrium state:
\begin{defn}[Gibbs state]\label{def:gibbs}
	For a system at temperature $T$ with Hamiltonian $\hat H$, its equilibrium state is given by the \emph{canonical ensemble}, also known as \emph{Gibbs state}
	\begin{equation}
		\hat\pi := \frac{e^{-\beta \hat H}}{\Tr[e^{-\beta \hat H}]}~,
	\end{equation}
	where $\beta := 1/k_B T$ is the \emph{inverse temperature}, and the normalization defines the \emph{partition function} $Z = \Tr[e^{-\beta \hat H}]$.
\end{defn}
We can use this definition to rewrite the non-equilibrium free energy of a state $\hat\rho$ in terms of the Gibbs state of the Hamiltonian
\begin{equation}
	F_{n.eq}(\hat\rho) = F_{n.eq}(\hat\pi) + S(\hat\rho||\hat\pi)~,
\end{equation}
where $S(\hat\rho||\hat\pi) := \Tr[\hat\rho(\log\hat\rho-\log\hat\pi)] \geq 0$ is the relative entropy. Since $\Tr[\hat\rho(\log\hat\rho-\log\hat\pi)] = 0$ if and only if $\hat\rho = \hat\pi$, it is clear that Gibbs states minimize the free energy. And thus shows how Gibbs states correspond to the aforementioned notion of \emph{equilibrium state}, which we will also refer to as \emph{thermal state}. It is also worth noting that for a given expected energy and temperature, it also follows that Gibbs states maximize the entropy. In the jargon of resource theory, thermal states are also known as the only \emph{completely passive states}~\cite{PerarnauLlobet2015}. This notion of equilibrium leads to a state-independent notion of free energy: the \emph{equilibrium free energy} $F_{eq} := F_{n.eq}(\hat\pi) = -k_BT\ln Z$.\\

At this point, one might be (very) tempted to take the change in von Neumann entropy of the state describing the system and environment as entropy production. However, von Neumann entropy is invariant under unitary transformations. Therefore the von Neumann entropy of a closed system remains constant when it evolves by the rules of quantum mechanics. Which means that we need to take a different approach to define entropy production.\\
The notion of ``usefulness'' given by the free energy is actually the key here: in a thermodynamic operation all the energy that can be recovered if we were to immediately revert the operation corresponds to changes in free energy. Therefore we can split the work of a thermodynamic operation as follows
\begin{equation}\label{eq:diss_def}
	W = \Delta F_{n.eq} + W_{diss}~,
\end{equation}
where $\Delta F_{n.eq} = F_{n.eq}(\hat\rho_S(\tau)) - F_{n.eq}(\hat\rho_S(0))$ is the difference in free energy of the system: the reversible component of work, and $W_{diss}$ is the dissipated energy: the irreversible component of work -- and is defined as the difference between work and free energy change. We also defined the reduced state $\hat\rho_S := \Tr_E[\hat\rho_{SE}]$, that describes the state of the system $S$ when one does not have knowledge of the total state of the system and environment together $\hat\rho_{SE}$.
This leads us to a very natural definition of entropy production: $\Sigma = \beta W_{diss}$. With this definition one could already express the second law of thermodynamics as $W \geq \Delta F_{n.eq}$. Using \eref{eq:diss_def} with \defref{def:neq_free_energy} and \defref{def:heat} we obtain a very common formulation of entropy production and the second law
\begin{equation}\label{eq:entropy_prod_neq}
	\Sigma = \beta Q + \Delta S \geq 0~,
\end{equation}
where $\Delta S = S(\hat\rho_S(\tau)) - S(\hat\rho_S(0))$ is the change of entropy of the system, and $\beta$ is the inverse temperature of the environment. The term $\beta Q$ corresponds to a flow of entropy from the system towards the environment. It is worth noting how the second law allows for a reduction in entropy of the system, as long as it is compensated with a sufficient flow of entropy into the environment (cf. \cref{chap:landauer}). There are many situations in which it is not immediate to assign a unique temperature to the environment -- e.g. an engine. However, one can always divide the environment into multiple uncorrelated parts $E_1,...,E_n$ and assign a temperature $T_i$ to each of these parts~\cite{Brunner2012}. In this case, one looses a clear definition of non-equilibrium free energy, because depending on the configuration of the system and environments there is not necessarily an equilibrium state towards which the state naturally evolves -- e.g. non-equilibrium steady states~\cite{zulkowski2013optimal}. Since the entropy is additive between independent systems, each environment contributes $\beta_i Q_i$ to the entropy production. Which allows us to generalize \eref{eq:entropy_prod_neq} to
\begin{equation}\label{eq:entropy_prod}
	\Sigma = \sum_i \beta_i Q_i + \Delta S \geq 0~.
\end{equation}
One often assumes that at the start of the process the environment -- or each of its subdivisions -- is at thermal equilibrium and uncorrelated to the system, which allows us to rewrite $\Sigma$ in terms of information theoretic quantities
\begin{equation}\label{eq:entropy_prod_info}
	\Sigma = I_\tau(S:E) + S(\hat\rho_E(\tau)||\hat\pi_E)~,
\end{equation}
where $I_t(S:E) :=  S(\hat\rho_{SE}(t)||\hat\rho_S(t)\otimes\hat\rho_E(t))$ is the mutual information, which quantifies the amount of information that is stored in the correlations between system and environment, which is lost when one has lo longer access to either $E$ or $S$. And the relative entropy $S(\hat\rho_E||\hat\pi_E)$ quantifies how much the environment was ``pushed away'' from the equilibrium\footnote{Strictly speaking the relative entropy is not a distance because it does not respect the triangular inequality. However it is non-negative, and equal to zero if and only if the two entries are equal.}. It is interesting to note how \eref{eq:entropy_prod_info} shows clearly this notion of the entropy production matching the concept of ``lost information''. Furthermore, since the quantities in \eref{eq:entropy_prod_info} are non-negative, the second law is mathematically proven for thermal environments.\\

Going back to processes where the system interacts with a single bath at temperature $T$, very often one also assumes that also the system is initialized in a thermal state. For these situations, usually the system is also left to thermalize after the operation is over at time $\tau$. And therefore the ``useful energy'' that is in the system at time $\tau$ is also dissipated, which leads to the following definition of dissipation
\begin{equation}
	k_B T\Sigma =  W - \Delta F_{eq}~.
\end{equation}
It is worth noting that $W - \Delta F_{n.eq} = W - \Delta F_{eq} - k_B T S(\hat\rho_S(\tau) || \hat\pi_S(\tau)) \leq W - \Delta F_{eq}$, where the difference is exactly entropy associated to the leftover free energy of the system at the end of protocol.\\

Finally, we will briefly cover a different approach to defining dissipation, which is purely information-theoretic. The main idea stems from finding a generalization of \eref{eq:entropy_prod_info}. Starting from an initially uncorrelated system and environment, if one considers the amount of lost information during the thermodynamic process, then there are two contributions: the correlations generated between the system and environment, and the perturbation of the environment~\cite{Landi2021rev}. Which leads to the following definition
\begin{equation}
	\Sigma = I_\tau(S:E) + S(\hat\rho_E(\tau)||\hat\rho_E(0))~,
\end{equation}
which coincides with \eref{eq:entropy_prod_info} when the environment -- or its sub divisions -- is assumed to start in a thermal state. However, the downside of this approach is that it is not equivalent to \eref{eq:entropy_prod} as these expressions do not coincide whenever the assumption of a thermal environment is not satisfied.

\section{The Third Law}\label{sec:third_law}
The third law of thermodynamics has been formulated in many different forms over the years and the subject of intense discussions~\cite{Kox2006}. Unlike the second law, a fully general proof of the third law -- or as general as the second law -- is still lacking~\cite{Masanes2017,Wilming2017}. The first formulation, given by Nernst, is now known as the heat theorem. It states ``At zero temperature, a finite size system has an entropy S, which is independent of any external parameters $x$, i.e. $\lim_{T\rightarrow 0} S(T,x_1) - S(T,x_2) = 0$.''~\cite{Nernst1906}. However, in quantum mechanics it is clear that the zero-temperature entropy is equal to the logarithm of the ground state degeneracy. Therefore the validity of the heat theorem is dependent on the possibility of changing this degeneracy with the parameter $x$. Today, the third law is understood as Nernst's \emph{unattainability principle}~\cite{Nernst1906}.\\

\noindent{\bf The Third Law of Thermodynamics.} It is impossible to bring a system into its ground state without either \emph{infinite time}, \emph{infinite energy}, or \emph{infinite complexity}.\\

This version of the third law -- that includes complexity as a resource -- is based on the most recent advances of the field~\cite{Taranto2023}. However, one can find a mathematical formulation of it in \cite{Masanes2017} and \cite{Wilming2017}. In quantum mechanics the third law implies that it is impossible to prepare a pure state without one of the three aforementioned quantities going to infinity.\\

An interesting consequence of the third law is that the second law \eref{eq:second_law} can only be saturated when the work, heat, complexity, or time diverge. In particular, when the time of the operation $\tau$ goes to infinity, then the system is at all times in the instantaneous equilibrium state -- typically given by $\hat\pi(t)$. Therefore the work cost of the operation will be given exactly by the free energy change, which implies that the dissipation is exactly $\Sigma = 0$. In this case we say that the operation is \emph{reversible}. Conversely, for $\tau < \infty$, whenever we change the Hamiltonian the system will be pushed away from its equilibrium state. Which implies that there will be a non-zero dissipation $\Sigma > 0$, and we say that the operation is \emph{irreversible}.

\cleardoublepage
\chapter{Mathematical Tools}\label{chap:math}
\epigraph{Ludwig Boltzmann, who spent much of his life studying statistical mechanics, died in 1906 by his own hand. Paul Ehrenfest, carrying on the work, died similarly in 1933. Now it is our turn to study statistical mechanics.}{--- David Goodstein, \emph{States of Matter}}
\section{Evolution of Quantum Systems}\label{sec:evolution}
The evolution of a quantum system described by a state $\hat\rho$ is given by a the dynamical equation
\begin{equation}\label{eq:dyn_gen}
	\frac{d}{dt}\hat\rho(t) = \mathcal G_t[\hat\rho(t)]~,
\end{equation}
where $\mathcal G_t$ is the time-dependent generator of the dynamics. The solution of this equation is given by
\begin{equation}
	\hat\rho(t) = G(t,0)[\hat\rho(0)]
\end{equation}
where $G(t,t')[~\cdot~]$ is a \emph{completely positive trace preserving} map known as the \emph{propagator} of the dynamics from $t'$ to $t$. It is defined as
\begin{equation}\label{eq:prop_def}
	G(t,t') := \overleftarrow{\mathcal T}\!\exp\!\left[\int_{t'}^t\!ds~\mathcal G_s\right]~,
\end{equation}
where we introduced the \emph{time ordering operator} $ \overleftarrow{\mathcal T}$. The exponential operator is defined via its usual series; that, when expanded, yields the Dyson expansion
\begin{equation}\label{eq:prop_dyson}
	G(t,t') = \mathbbm 1 + \sum^{\infty}_{n=1}\int^t_{t'}\! dt_{n} \int^{t_{n}}_{t'}\! dt_{n-1}~...\int^{t_2}_{t'}\! dt_1~  \mathcal{G}_{t_{n}}\circ\mathcal{G}_{t_{n-1}}\circ...\circ\mathcal{G}_{t_1}~,
\end{equation}
where $\mathbbm 1$ denotes the identity super-operator, and $\circ$ denotes the map composition.\\

When we consider an isolated quantum system -- in the thermodynamic setting this is the case when we consider a system together with its environment -- then its evolution is described by unitary dynamics: $\mathcal G_t[~\cdot~] = \frac{-i}{\hbar}[\hat H(t),~\cdot~]$. Therefore \eref{eq:dyn_gen} reduces to
\begin{equation}
	\frac{d}{dt}\hat\rho(t) = \frac{-i}{\hbar}\left[\hat H(t),\hat\rho(t)\right]~.
\end{equation}
The solution of this equation leads to the unitary evolution of the system
\begin{equation}\label{eq:unitary_evolution}
	\hat\rho(t) = \hat U(t)\hat\rho(0) \hat U^\dagger(t)~,
\end{equation}
where the unitary operator is given by
\begin{equation}\label{eq:unitary}
	\hat U(t) = \overleftarrow{\mathcal T}\!\exp\!\left[-\frac{i}{\hbar}\int_{0}^t\!ds~\hat H(s)\right]~,
\end{equation}
which has a similar expansion as in \eref{eq:prop_dyson}.\\

Taking the typical thermodynamic scenario, we are often only interested in describing the dynamics of the system without the environment. In this case the evolution of the system alone is naturally given by
\begin{equation}\label{eq:exact_local_dyn}
	\hat\rho_S(t) = \Tr_E\left[\hat U(t)\hat\rho_{SE}(0) \hat U^\dagger(t)\right]~.
\end{equation}
This problem is can still be very challenging to solve analytically, in particular when the traced out part is large, which is typically the case for the environment. However, one can usually make two approximations that greatly simplify \eref{eq:exact_local_dyn}: first, that the environment is well described by a thermal state at the start of the evolution $\hat\rho_{SE}(0) = \hat\rho_S(0)\otimes\hat\pi_B$. Second, that the interaction between system and environment is weak and Markovian. It is worth noting that this second approximation is often made also in classical thermodynamics, where the interactions between system and environment are attributed to boundary effects. With these two approximations one can simplify the the time derivative of \eref{eq:exact_local_dyn} to the famed \emph{Lindblad master equation} (c.f. \cite{Manzano2020} for a detailed derivation) 
\begin{equation}\label{eq:Lindblad}
	\frac{d}{dt}\hat\rho_S(t) = \mathcal L[\hat\rho_S(t)]~,
\end{equation}
where $\mathcal L$ is the \emph{Lindbladian}. It is defined as
\begin{equation}
	\mathcal L[\hat\rho_S(t)] := \frac{-i}{\hbar}\left[\hat H_S(t),\hat\rho_S(t)\right] + \sum_i\gamma_i\left(\hat L_i\hat\rho_S(t)\hat L_i^\dagger - \frac{1}{2}\left\{\hat L_i^\dagger \hat L_i,\hat\rho_S(t) \right\}\right)~,
\end{equation}
where $\hat H_S(t)$ is the Hamiltonian of the system -- without the parts acting on the environment. The $\hat L_i$ are the \emph{jump operators}, they describe the dissipative aspects of the dynamics and how the environment acts on the system. Finally, the $\gamma_i \geq 0$ are the \emph{damping rates} associated to the jump operators, they encode the strength of the effect of the corresponding jump operator. It is difficult to overstate the success and importance of this equation: it plays a crucial role in the field of quantum optics, quantum biology, quantum information and condensed matter.

\section{Geometric Thermodynamics}
\label{sec:geom}
\subsection{General Derivation of the Thermodynamic Metric}
\begin{figure}[tb!]
	\centering
	\includegraphics[width=\textwidth]{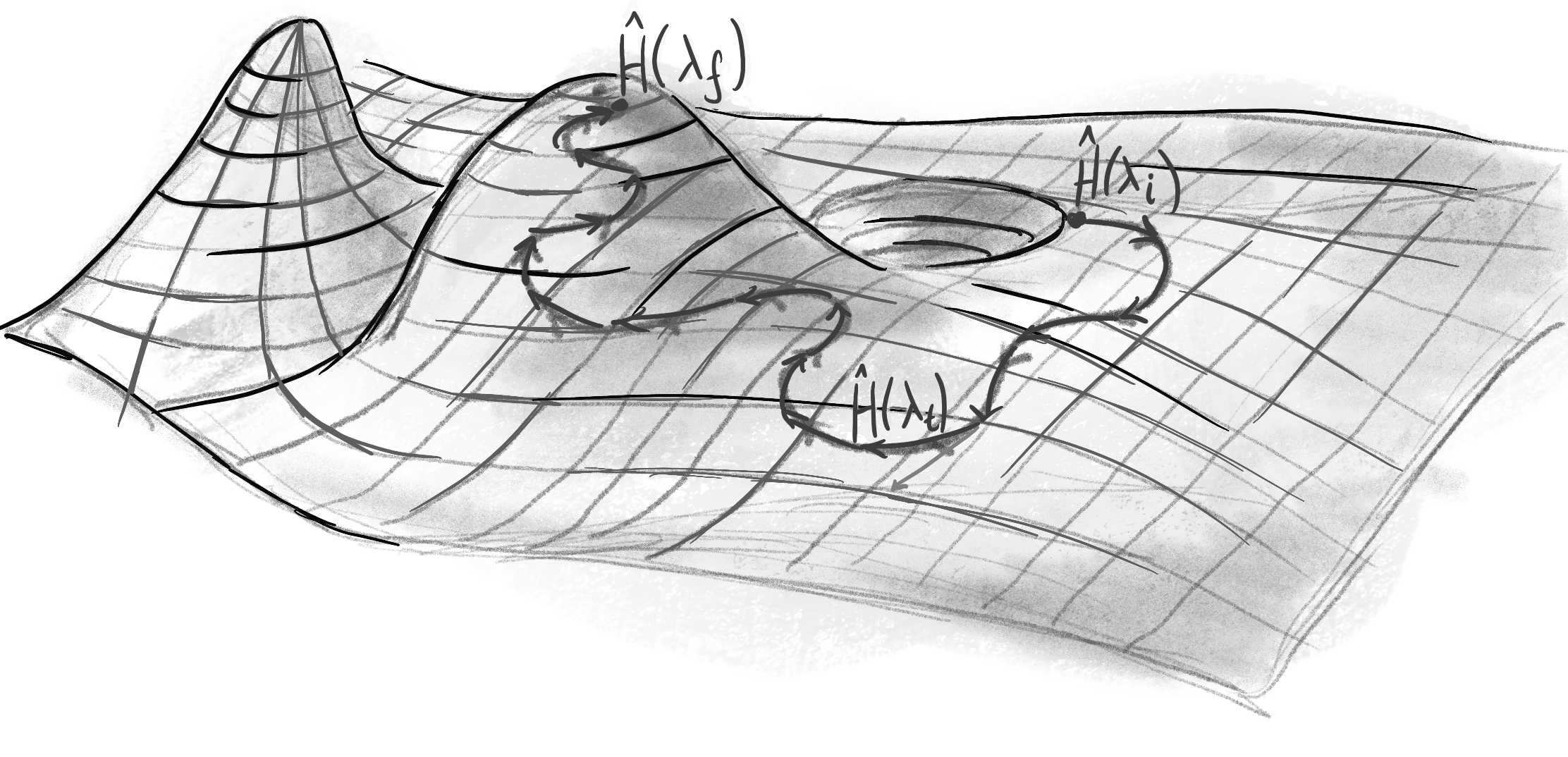}
	\caption{Curve of a protocol on the manifold of thermal states.~\cite{jake_drawing}}
	\label{fig:geometry}
\end{figure}

The framework of quantum thermodynamic geometry~\cite{Salamon1983,Scandi19,Abiuso2020Geometric} allows us to minimize the entropy production $\Sigma$ for protocols that are slow compared to their relaxation time-scale. We start by considering a system described by a state $\hat\rho(t)$ with an externally driven Hamiltonian $\hat H(t)$ that is undergoing the dynamics described by \eref{eq:dyn_gen} with the generator $\mathcal G_{\lambda(t)}$, where we made the dependence on the control parameters explicit. This description could be only about the system -- thus in the setting of the Lindblad master equation, or about the system and environment -- thus undergoing unitary dynamics. Since the Hamiltonian is a linear operator over the Hilbert space it can be decomposed in the following manner
\begin{equation}
	\hat H(t) = \hat H_{0}+\sum_{j=1}^n\lambda^j(t)\hat X_j~,
\end{equation}
where $\hat H_0$ contains the parts of the Hamiltonian that cannot be controlled, $\{\lambda^j\}_{j=1}^n$ are externally controllable parameters and $\{\hat X_j\}_{j=1}^n$ are the corresponding observables. These control parameters can be constrained, and we will denote by $M\subseteq \mathbb{R}^n$ the manifold of the allowed values for these parameters. A thermodynamic protocol corresponds to a curve within the space $M$, as is illustrated in \fref{fig:geometry}. In order to apply the framework of geometric thermodynamics, we need to satisfy three properties.\\

\noindent{\bf Requirement 0: A Single Thermal Environment.} We assume that the environment is initialized in a thermal state at a single well defined temperature $T$ and that it is infinitely large. So that, by the second law, there is a well defined thermal equilibrium for the state $\hat\rho$ if we were to freeze the control of the Hamiltonian. This is implies that $\hat\pi_t$ is a fixed point of the dynamics: $\mathcal G_{\lambda(t)}[\hat\pi_t] = 0$.

For unitary dynamics this is automatically the case since the thermal state commutes with its corresponding Hamiltonian. While for the Lindblad master equation we need to require that the whole environment is already at thermal equilibrium.

\noindent{\bf Requirement 1: Thermalization of the Work Observables.} In absence of driving, the expectation values of the observables $\{\hat X_j\}$ thermalize. More precisely, if at time $t$ we were to freeze the driving, then the propagator of the evolution becomes $G_{\lambda(t)}^{fz}(s,r) = e^{(s-r)\mathcal G_{\lambda(t)}}$ for $s\geq r\geq t$. We then say that the expectation value of $\hat X_j$ thermalizes if
\begin{equation}
	\lim_{s\rightarrow\infty}\Tr[\hat X_j e^{s\mathcal G_{\lambda(t)}}[\hat \rho(t)]] =  \Tr[\hat\pi(t)\hat X_j]~.
\end{equation}
This condition is typically satisfied by open quantum systems~\cite{Cresser2021} and by non-integrable systems that satisfy the \emph{eigenstate thermalization hypothesis}~\cite{Eisert2015}.

These two requirements are usually combined into one for the Lindblad master equation by simply stating that the state of the system converges to the thermal state in absence of driving. However, this is not the case for unitary dynamics, and thus we have to weaken this assumption into Requirements 0 and 1. Before stating the last requirement, let us consider the following decomposition of the state: $\hat \rho(t) = \hat\pi(t) + \Delta\hat\rho(t)$, where $\Delta\hat\rho(t)$ is defined as the difference between the thermal state and the state of the system. Inserting this decomposition into the dynamics \eref{eq:dyn_gen} we obtain
\begin{equation}\label{eq:expan_deriv_0}
	\left(\mathcal G_{\lambda(t)} - \frac{d}{dt}\right)\![\Delta\hat\rho(t)] = \frac{d}{dt}\hat\pi(t)~.
\end{equation}
Since $\mathcal G_{\lambda(t)}$ has a non-trivial kernel it is not invertible. However, we can solve this equation by introducing the concept of \emph{Drazin inverse}~\cite{Boullion1971}. In this case, it reduces to taking the inverse over the support of $\mathcal G_{\lambda(t)}$:
\begin{equation}
	\mathcal G_{\lambda(t)}^D = (\mathcal G_{\lambda(t)}|_{\text{supp}(\mathcal G_{\lambda(t)})})^{-1}\oplus \bm{0}_{\text{ker}(\mathcal G_{\lambda(t)})}~,
\end{equation}
where $\oplus$ denotes the direct sum and $\bm{0}_{\text{ker}(\mathcal G_{\lambda(t)})}$ denotes the zero operator over the kernel of $\mathcal G_{\lambda(t)}$. It is useful to note that $\mathcal G_{\lambda(t)}^D\mathcal G_{\lambda(t)} = \mathcal G_{\lambda(t)}\mathcal G_{\lambda(t)}^D$ and is equal to the projector over the support of $\mathcal G_{\lambda(t)}$: $\mathcal P_{\text{supp}(\mathcal G_{\lambda(t)})}$.
Therefore we can apply the Drazin inverse to both sides of \eref{eq:expan_deriv_0} to obtain
\begin{equation}
	\left(\mathcal P_{\text{supp}(\mathcal G_{\lambda(t)})} - \mathcal G_{\lambda(t)}^D\frac{d}{dt}\right)\![\Delta\hat\rho(t)] = \mathcal G_{\lambda(t)}^D\frac{d}{dt}[\hat\pi(t)]~.
\end{equation}
Since, by construction, the image and support of $\mathcal G_{\lambda(t)}^D$ are the support of $\mathcal G_{\lambda(t)}$, we can write
\begin{equation}
	\left(\mathbbm 1 - \mathcal G_{\lambda(t)}^D\frac{d}{dt}\right)\![\Delta\hat\rho_s(t)] = \mathcal G_{\lambda(t)}^D\frac{d}{dt}[\hat\pi(t)]~,
\end{equation}
where we decomposed $\Delta\hat\rho(t)$ into $\Delta\hat\rho_s(t)$ its projection onto the support of $\mathcal G_{\lambda(t)}$ and $\Delta\hat\rho_k(t)$ its projection onto the kernel of $\mathcal G_{\lambda(t)}$.
Finally, with $(\mathbbm 1 - X)^{-1} = \sum_{n=0}^\infty X^n$ we find
\begin{equation}\label{eq:state_expansion}
	\hat \rho(t) = \Delta\hat\rho_k(t) +  \sum_{n=0}^{\infty}\left(\mathcal G_{\lambda(t)}^D\frac{d}{dt}\right)^{\!\! n}\![\hat\pi(t)]~.
\end{equation}
Each term in the sum is proportional to $\|\dot\lambda(t)\|^n$, therefore if $\|\dot\lambda(t)\|$ is progressively going to $0$ the only surviving term in the sum is $\hat\pi(t)$. Nonetheless the state will not necessarily thermal because of the term $\Delta\hat\rho_k(t)$. If we are in the case of a Lindblad master equation where the only fixed point of the evolution is $\hat\pi(t)$, then $\delta\hat\rho(t) = 0$. However, in the case of a full unitary description $\Delta\hat\rho_k(t)$ cannot be vanishing because only the thermal state maps to the thermal state in a unitary evolution. In particular, if we were to freeze the evolution as described in Requirement 1, since $\Delta\hat\rho_k(t)$ is a fixed point of the frozen evolution, we have 
\begin{equation}
	\lim_{s\rightarrow\infty} e^{s\mathcal G_t}[\hat\rho(t)] = \Delta\hat\rho_k(t) + \hat\pi(t)~.
\end{equation}
Combining this last equation with Requirement 1, we get
\begin{equation}\label{eq:fixed_point_null}
	\Tr[\Delta\hat\rho_k(t)\hat X_j] = 0~.
\end{equation}
The limit where $\|\dot\lambda(t)\|$ goes to zero corresponds to the quasi-static limit with the protocol time $\tau$ going to infinity. But here we are interested in the case where $\tau$ is finite and $\|\dot\lambda(t)\|$ is sufficiently small to drop most of the terms in the sum of \eref{eq:state_expansion}. Which naturally leads us to the last requirement:

\noindent{\bf Requirement 2: Slow driving.} The system is driven in such a way that its evolution is slow compared to the relaxation timescales of evolution of the expectation values of the observables $\{\hat X_j\}$. More precisely, if we denote by $\tau^{eq}_j(t)$ the relaxation timescale of $\Tr[\hat X_j e^{s\mathcal G_{\lambda(t)}}[\hat \rho(t)]]$ as we let $s$ increase. Then by defining $\tau_{eq} := \max_j\sup_{0\leq t\leq \tau}\tau^{eq}_j(t)$, we are in the slow driving regime when $\tau_{eq}/\tau \ll 1$.

The expansion for a varying $\tau$ should be understood as done for a fixed curve $\{\lambda(t)\} \subset M$, so that changing the value of $\tau$ only changes the speed over which the path on the curve is covered. Which allows us to identify $\|\dot\lambda\| = \mathcal O(\tau_{eq}/\tau)$, we then get from \eref{eq:state_expansion}
\begin{equation}\label{eq:state_slow}
	\hat \rho(t) = \Delta\hat\rho_k(t) + \hat\pi(t) - \beta\mathcal G_{\lambda(t)}^D[\mathbb D_{\hat\pi(t)}[\hat H'(t)]] + \mathcal O(\tau_{eq}^2/\tau^2)~,
\end{equation}
where we used $\frac{d}{dt}\hat\pi(t) = -\beta\mathbb D_{\hat\pi(t)}[\hat H'(t)]$, with
\begin{equation}
	\mathbb D_{\hat\rho}[\hat A] := \int_0^1\!ds~\hat\rho^s\left(\hat A-\Tr[\hat\rho\hat A]\mathbbm 1\right)\hat\rho^{1-s}~.
\end{equation}
From \eref{eq:work}, \eref{eq:fixed_point_null} and \eref{eq:state_slow} we can find the following expression for work
\begin{equation}
	W = \Delta F_{eq} -\beta\int_0^\tau\! dt~\Tr\!\left[H'(t)\mathcal G_{\lambda(t)}^D[\mathbb D_{\hat\pi(t)}[\hat H'(t)]]\right] + \mathcal O(\tau_{eq}^2/\tau^2)~,
\end{equation}
where we obtained $\Delta F_{eq}$ from integrating $\Tr[\hat\pi(t)\hat H'(t)]$. By using $\hat H'(t) = \sum_j \dot\lambda^j(t) \hat X_j$, we can now easily identify the leading order for the entropy production in the slow driving regime
\begin{equation}\label{eq:entropy_prod_geom}
	\Sigma = \sum_{ij}\int_0^\tau\! dt~\dot\lambda^i(t)\dot\lambda^j(t)g_{ij}(\lambda(t))~,
\end{equation}
where we defined the symmetric tensor $g(\lambda)$ as
\begin{equation}\label{eq:metric}
	g_{ij}(\lambda) := -\frac{\beta^2}{2}\Tr\!\left[\hat X_i \mathcal G_{\lambda}^D[\mathbb D_{\hat\pi(\lambda)}[\hat X_j]] + \hat X_j \mathcal G_{\lambda}^D[\mathbb D_{\hat\pi(\lambda)}[\hat X_i]]\right]~.
\end{equation}
From the second law of thermodynamics we have $\Sigma\geq 0$. Therefore, since \eref{eq:entropy_prod_geom} is valid for any choice of $\lambda$ as long as $\tau$ is large enough, then we necessarily have that the tensor $g$ is positive definite -- and because of the third law we know that it is strictly positive definite. Since it also depends smoothly on $\lambda$ and is independent of the velocity $\dot\lambda$, we can notice that $g$ has all the properties of a metric over $M$. In fact, it is known in the literature as the \emph{thermodynamic metric} and defines the length of a curve $\gamma : [0,\tau]\mapsto M$
\begin{equation}\label{eq:thermo_length_curve}
	L[\gamma] := \int_\gamma\!dt\sqrt{\dot\gamma^i(t)\dot\gamma^j(t)g_{ij}(\gamma(t))}~,
\end{equation}
where we adopted the Einstein summation convention, which we will keep through the rest of this thesis.

\subsection{Minimally Dissipating Curves}
We will now discuss how to use the metric \eref{eq:metric} to find minimally dissipating protocols from an initial configuration $\hat H_i$ and initial state $\hat\rho(0) = \hat\pi_i$ to a final configuration $\hat H_f$.
The entropy production and length associated to a specific protocol $\lambda : [0,\tau]\mapsto M$ are related via a Cauchy-Schwarz inequality:
\begin{equation}
	\Sigma[\lambda] \geq \frac{1}{\tau}L[\lambda]^2~,
\end{equation}
where we made explicit the dependence of $\Sigma$ on $\lambda$. This inequality can always be saturated by making the integral of the entropy production $\dot\lambda^i(t)\dot\lambda^j(t)g_{ij}(\lambda)$ constant, which can always be achieved by modulating the speed along the curve -- and it does not affect its length. Therefore to minimize the dissipation $W_{diss} = k_B T\Sigma$ we need to find the curve in the space $\mathcal C_{\lambda_i,\lambda_f}(M)$ of smooth paths connecting $\hat H_i$ and $\hat H_f$ that minimizes its length. This corresponds to a geodesic path on $M$, which satisfies the geodesic equation
\begin{equation}
	\ddot\lambda^i + \Gamma^i_{jk}\dot\lambda^j\dot\lambda^k = 0~,
\end{equation}
where $\Gamma^i_{jk}$ are the Christoffel symbols
\begin{equation}
	\Gamma^i_{jk} = \frac{1}{2}g^{il}\!\left(\partial_j g_{kl} + \partial_k g_{lj} - \partial_l g_{jk}\right)~,
\end{equation}
where $g^{il}$ denotes the elements of the inverse of the metric and $\partial_\alpha := \frac{\partial}{\partial\lambda^\alpha}$. We can therefore define $\mathcal L(\hat H_i,\hat H_f)$ as the length of the geodesic curve connecting $\lambda_i$ and $\lambda_f$. Automatically $\mathcal L$ satisfies all the properties of a distance function over the space $M$, furthermore it gives us an expression for the minimal dissipation $W_{diss}^* := \min_{\lambda\in \mathcal C_{\lambda_i,\lambda_f}(M)}W_{diss}[\lambda]$ and the minimal entropy production $\Sigma^* = \beta W_{diss}^*$
\begin{equation}\label{eq:min_diss}
	W_{diss}^* = k_BT\Sigma^* = \frac{k_BT}{\tau} \mathcal L^2 + \mathcal O(\tau_{eq}^2/\tau^2)~.
\end{equation}
It is interesting to note that the length of a curve in $M$ is independent of $\tau$, and therefore all the dependency of the dissipation on the protocol time is in the prefactor $1/\tau$.\\

It is well known that optimal finite-time protocols feature jumps~\cite{Schmiedl2007opt}. However, these jumps disappear near the reversible limit (cf. \cite{Esposito2010} and \fref{fig:WC_comparison}) and their contribution to the dissipated heat becomes either negligible or is disadvantageous. Let us first consider jumps at the start of the protocol $\Delta \hat H_i = \hat H(0^+) - \hat H_i$, the work cost of this jump is $\Tr[\hat\pi_i\Delta \hat H_i]$, which yields a contribution to the dissipation that is independent of $\tau$. Therefore, however small is the work cost of the rest of the protocol, by \eref{eq:min_diss} there exists $\tau^*$ such that for all $\tau \geq \tau^*$ the geodesic will dissipate less than the protocol with a jump at the start. A similar argument can be applied to the end of the protocol. What this shows is that the jumps need to at most be of magnitude $\mathcal O(\tau^{eq}/\tau)$, which implies that they are vanishing in the slow driving limit, and most of the time we will be ignoring their contribution.

	\cleardoublepage
	\chapter{Finite-Time Landauer Principle}\label{chap:landauer}
\epigraph{Ahimè, non mai due volte configura\\ il tempo in egual modo i grani! E scampo \\ n'è: che, se accada, insieme alla natura \\ la nostra fiaba brucerà in un lampo.}{--- Eugenio Montale, \emph{Vento e bandiere, Ossi di seppia}}

\noindent Any logical irreversible operation will incur a thermodynamic cost in the form of heat dissipated into the environment. On a fundamental level this is because logically irreversible operations lead to a loss of information, and in the words of Rolf Landauer ``information is physical''. Therefore a loss of information results in a loss of energy. Landauer's principle quantifies this relation between information processing and thermodynamics with the bound $Q\geq k_B T \ln 2$ for the erasure of a single bit of information~\cite{Landauer1961}. Here $Q$ is the dissipated heat, $k_B$ is the Boltzmann constant and $T$ is the absolute temperature at which the process is taking place. In recent years, this principle has been intensively studied within the fields of stochastic and quantum thermodynamics~\cite{Sagawa2009,Esposito2011}, and  has been approached in several experimental platforms~\cite{Brut2012,ciampini2021experimental,Scandi2022}. 

The third law of thermodynamics (cf. \sref{sec:third_law}) implies that Landauer's bound \emph{cannot} be saturated 
with finite resources, namely time and energy~\cite{Nernst1906,Masanes2017}. In finite time, using tools from optimal transport theory~\cite{vanvu2023} and thermodynamic geometry~\cite{Abiuso2020Geometric,VanVu2021}, optimal erasure protocols have been  derived both for classical systems described by over-damped Langevin dynamics~\cite{Zulkowski2014} and  open quantum systems described by  Lindblad master equations~\cite{VanVu2022,Zhen2022}.  
Such optimal protocols naturally lead to a finite-time correction to Landauer's bound in different physical set-ups, which has given rise to the term \emph{finite-time Landauer principle}~\cite{Proesmans2020}. 
For a slowly driven (quantum) two-level system  weakly coupled to a thermal bath, the finite-time bound takes the simple form~\cite{Scandi19}
\begin{equation}
	Q \geq k_B T \left(\ln 2 + \frac{ \pi^2 }{4 \Gamma \tau} \right)+ \mathcal{O}\left(\frac{1}{\Gamma^2 \tau^2}\right),
	\label{eq:finitetimeboundweak}
\end{equation}
where  $\tau$ is the total time of the process and  $\Gamma$ is the thermalization rate.
The finite-time correction is positive, in agreement with the second law of thermodynamics, and when $\Gamma \tau \rightarrow \infty$ we recover the standard bound.
We also note that the optimal protocol saturating the finite bound \eref{eq:finitetimeboundweak} has been recently implemented in a semiconductor quantum dot~\cite{Scandi2022}.  
More general versions of \eref{eq:finitetimeboundweak} have also been recently developed for Markovian systems driven at any speed~\cite{Zhen2021,VanVu2022}.

Despite this remarkable progress, previous works on the \emph{finite-time Landauer principle} have focused in Markovian systems which, for quantum systems, can be guaranteed by a sufficiently  weak interaction between system and bath. In the presence of strong coupling, we expect both new opportunities arising due to faster relaxation rates and non-Markovian dynamics~\cite{Rivas2020}, as well as  challenges due to the presence of new sources of irreversibility~\cite{PerarnauLlobet2018}. In this chapter we will derive the first order to a tight finite-time correction of Landauer's principle for a single fermion that can interact \emph{strongly} with a reservoir, as described by the  resonant-level model~\cite{schaller2014open}. The main result of this chapter can be summarized as follows.

Given a two-level system that can be  strongly coupled to a thermal bath, we find that the finite-time version of Landauer's principle can be expressed as 
\begin{align}
	Q \geq k_B T \left(  \ln 2 + a \frac{\tau_{\rm Pl}}{\tau}  \right) + \mathcal{O}\left(\frac{1}{\Gamma^2 \tau^2}\right) 
	\label{MainResult}
\end{align}
where $a\approx 2.57946$,   $\tau_{\rm Pl} = \hbar/k_B T$ is the so-called Planckian time~\cite{Hartnoll22}, and $\Gamma$ is the average thermalization rate (see details below). This expression generalizes \eref{eq:finitetimeboundweak} to strong system-bath couplings, with the transition between the two being characterized in \fref{fig:main}. The finite-time correction in \eref{MainResult} is of quantum-mechanical nature and independent of the coupling strength, hence prevailing even for   arbitrarily strong system-bath coupling  (roughly speaking, $\Gamma \rightarrow \infty$ in \eref{eq:finitetimeboundweak}).\\

The appearance of the Planckian time $\tau_{\rm Pl} = \hbar/k_B T$ in \eref{MainResult} is particularly interesting. This timescale encodes two fundamental constants of nature: Boltzmann's constant $k_B$ and Planck's constant $\hbar$. It arises in several contexts  in many-body physics, including quantum transport and quantum chaos; see Ref.~\cite{Hartnoll22} for a review. In analogy with the ``Planck time'' in quantum gravity, it is 
associated with  the shortest timescale of thermalization~\cite{Hartnoll22,Sachdev2011}; that is, the shortest time needed to redistribute energy between particles and reach thermal equilibrium. This gives an insightful context to our main result \eref{MainResult}: a fundamental finite-time quantum correction must appear to Landauer's bound due to a minimal time required for thermalization. This also suggests that the form of \eref{MainResult} has a broader range of applicability, with the dimensionless value $a$ depending on the specific many-body thermalizing dynamics considered. In order to obtain \eref{MainResult}, we exploit the framework of thermodynamic geometry explained in \sref{sec:geom}.
\section{Framework}
We consider a driven system $S$ that can be put in contact with a thermal bath $B$, so that the total time-dependent Hamiltonian reads:
\begin{equation}
	\hat H(t)= \hat H_S(t)+\hat H_{\rm int}(t)+\hat H_B.
	\label{eq:GeneralH(t)}
\end{equation}
Here, $\hat H_S(t)$, $\hat H_{\rm int}(t)$ are the externally controllable Hamiltonian of $S$ and the $SB$ coupling, whereas $\hat H_B$ is the Hamiltonian of $B$. 
The state $\hat \rho(t)$ of $SB$ evolves according to \eref{eq:unitary_evolution}. 
From \eref{eq:work}, the work cost induced by driving $\hat H(t)$, with $t\in [0,\tau]$, reads: 
\begin{equation}
	W = \int_0^\tau\!dt~\Tr[\hat\rho(t) \hat H'(t)] 
	= \Tr[\hat H(\tau)\hat\rho(\tau) - \hat H(0)\hat\rho(0) ]  
\end{equation}
Focusing on protocols where  $\hat H_{\rm int}(0)=\hat H_{\rm int}(\tau)=0$, we can naturally identify from the first law of thermodynamics \eref{eq:first_law} $W=Q+\Delta E$, with the change in energy of the system
\begin{equation}
	\Delta E=\Tr[\hat H_S(\tau)\hat\rho_S(\tau) - \hat H_S(0)\hat\rho_S(0)]
\end{equation}
with $\hat\rho_S(t) := \Tr_B[\hat\rho(t)]$, and the energy absorbed by the bath
\begin{equation}
	Q=\Tr[\hat H_B(\hat\rho_B(\tau)-\hat\rho_B(0))]
\end{equation}  
with $\hat\rho_B(t) := \Tr_S[\hat\rho(t)]$.

Assuming that the initial state of $SB$ is a thermal state: $\hat\rho(0)= \hat{\pi}(0) := \frac{e^{-\beta\hat H(0)}}{Z(0)}$ with $Z(t) := {\rm Tr} [e^{-\beta\hat H(t)}]$, we can be re-express the work cost as:
\begin{equation}
	W=\Delta F_{eq} + k_B T \Sigma ~,
	\label{eq:WSigma}
\end{equation}
where $\Delta F_{eq}=k_B T \ln [Z(0)/Z(\tau)] $ is the change of equilibrium free energy of SB, and the entropy production $\Sigma$ can be expressed as $\Sigma=S(\hat\rho(\tau)||\hat\pi(\tau))$ (cf. \sref{sec:second_law} for more detail). The entropy production  $\Sigma \geq 0$  accounts for the irreversible energetic contribution in finite-time processes, and depends on the particular driving path $\hat H(t)$ linking $\hat H(0)$ to $\hat H(\tau)$. Minimizing $\Sigma$ over all finite-time processes leads to thermodynamic protocols that minimize the work $W$. Furthermore, in an erasure process, $\Delta E = 0$ (see details below) therefore these protocols also minimize the dissipated heat $Q$. 

\section{The Resonant Level Model}
We now focus on finite-time driving processes of a single fermionic mode coupled to a fermionic bath, which can e.g. describe a single-electron quantum dot. The total Hamiltonian reads:
\begin{equation}
	\hat H(t)
	= \varepsilon(t) \hat a^\dagger \hat a + \sum_{k=1}^n \omega_k \hat b_k^\dagger \hat b_k+ \kappa(t) \sum_{k=1}^n \lambda_k \hat a^\dagger \hat b_k + \lambda_k^* \hat b_k^\dagger \hat a. 
\end{equation}
where $\hat a^\dagger$ is the creation operator of the two-level system and $\hat b_k^\dagger$ is the creation operator of a bath mode with frequency $\omega_k$. These operators follow the canonical anti-commutation relations: $\{\hat a^\dagger , \hat a\} = \mathbbm{1}$, $\{\hat b_j^\dagger , \hat b_k\} = \delta_{jk}\mathbbm{1}$, $\{\hat b_j , \hat b_k\} = \{\hat a , \hat b_k\} = \{\hat a^\dagger , \hat b_k\} = \{\hat a , \hat a\} = 0$. $\kappa(t)$ modulates the interaction Hamiltonian and  $\lambda_k$ are the interaction weights which define the spectral density function of the bath $\mathfrak J(\omega) = 2\pi\sum_k|\lambda_k|^2\delta(\omega-\omega_k)$. Finally, the energy $\eps(t)$ is the difference between the energy gap of the two-level system and the chemical potential of the bath\footnote{The chemical $\nu$ potential of the bath is incorporated by subtracting $\nu \hat a^\dagger\hat a$ to the system's Hamiltonian. Since here $\hat H_S = \eps a^\dagger\hat a$ (with $\eps$ the energy of the system), we can simply redefine $\eps$ to be the difference between the system's energy and the chemical potential and set $\nu = 0 $ without loss of generality.}.
We are assuming optimal control over the functions $\eps(t)$ and $g(t)$ so that we can fully optimize the protocol and reach the fundamental limit for this system. While this level of control is, in principle, ambitious experimentally in regards to the coupling, it has been achieved in quantum dots \cite{Rochette2019} where the tunneling rate (i.e. interaction strength) can be modified by several orders of magnitude.
We take the continuum limit and assume that the spectral density of the bath is a Lorentzian
\begin{equation}
	\mathfrak J(\omega) = \frac{\Lambda^2}{\Lambda^2 + \omega^2}~,
\end{equation}
where $\Lambda>0$ is a parameter characterizing its width. Exact and explicit solutions for the resonant-level model are known in the wide-band limit $\Lambda \rightarrow \infty$~\cite{schaller2014open}. This limit is commonly used to describe quantum systems in contact with fermionic macroscopic baths, e.g. in quantum dots or single-molecule junctions (see \cite{Covito2018} for an example). In essence, it neglects the structure of the density of states in the bath and, as a consequence, a  main limitation is that it fails to describe the short-time dynamics~\cite{Covito2018}. Nevertheless,  this problem does not affect this study since we are interested in large times. We should further note that the energy of the system-bath interaction is proportional to $\Lambda$, and therefore is divergent in this limit. We will therefore take $\Lambda$ to be finite but much larger than any other energy scale of the system. For our analysis to be valid we simply require  dynamics much slower than $\Lambda^{-1}$ \cite{schaller2014open}.

The dynamics are solved via a quantum Langevin approach \cite{Rolandi2023Quantum,Bettmann2024Meso}. For an uncorrelated initial state $\hat\rho(0) = \hat\rho_S(0)\otimes\hat\pi_B$, we compute the expectation values that are relevant to the work cost: the probability of occupation of the excited level of the system $p(t) = \<\hat a^\dagger \hat a \>$ and the system-bath interaction energy $v(t) = \sum_k \lambda_k\<\hat a^\dagger \hat b_k\> + h.c.$.
\begin{align}
	\label{eq:p_exact}
	p(t) =& \left|G(t,0)\right|^2 p(0) + \frac{1}{2\pi}\int_{-\infty}^{\infty}\!\!d\omega~ f_\beta(\omega)\left|\int_0^t ds~ \kappa(s) G(t,s) e^{i\frac{\omega}{\hbar}(t-s)}\right|^2  \!\!,\\
	\label{eq:v_exact}
	v(t) =& \frac{1}{\pi}\Im\! \int_{-\infty}^\infty\!\! d\omega~f_\beta(\omega) \int_0^t ds~\kappa(s)G(t,s)e^{i\frac{\omega}{\hbar}(t-s)},
\end{align}
where $f_\beta(\omega) = (1+e^{\beta\omega})^{-1}$ is the Fermi-Dirac distribution. We defined the propagator
\begin{equation}
	G(t,s) = \exp\!\left[-\frac{1}{\hbar}\int_s^t dr~ \mu(r) + i\eps(r)\right]~,
\end{equation}
where we introduced $\mu(t):=\frac{1}{2}\kappa(t)^2$ in order to have more concise equation in the rest of the chapter. We therefore obtain the expected thermodynamic work cost to be
\begin{align}
	W = \int_0^\tau dt~ \dot\eps(t)p(t) + \dot \mu(t)v(t)/\kappa(t).
\end{align}
The framework of quantum thermodynamic geometry allows us to minimize the dissipated work $W_{diss} =  W - \Delta F_{eq}$, and therefore the dissipated heat $Q$ for erasure, for protocols that are slow compared to their relaxation time-scale. In order to apply the framework presented in \sref{sec:geom}, we have to ensure that the requirements are satisfied. Since we are in unitary dynamics we already satisfy requirement 0. The proof that requirement 1 is satisfied is quite convoluted, which is why we left it for the end of the chapter in \sref{sec:proof_therm}. While requirement 2 can be easily imposed by choose $\tau$ appropriately. In particular, using that the thermalization rate of the system is $\Gamma := \frac{2}{\hbar\tau}\int_0^\tau dt\, \mu(t)$ we can perform the slow driving expansion of \eref{eq:p_exact} and \eref{eq:v_exact} in orders of $1/(\tau\Gamma)$.
We then obtain an expansion for $W$ analogous to \eref{eq:WSigma} where the entropy production $\Sigma$ is described by \eref{eq:entropy_prod_geom} with $\lambda(t) = (\eps(t),\mu(t))$ and the thermodynamic metric
\begin{equation}\label{eq:metric_L}
	g(\lambda(t)) = \frac{\beta\hbar}{\pi}\int_{-\infty}^\infty d\omega~f_\beta(\omega) m_\omega(\eps(t)-\omega,\mu(t))~,
\end{equation}
where
\begin{equation}\label{eq:metric_omega}
	m_\omega(\eps,\mu) := \frac{1}{\left(\mu^2 + \eps^2\right)^3}
	\begin{pmatrix}
		4\eps\mu^2 &
		-\mu(\mu^2-3\eps^2) \\
		\mu(\mu^2-3\eps^2) &
		2\eps(\eps^2-\mu^2)
	\end{pmatrix}~.
\end{equation}
This metric gives a geometrical description of slow thermodynamic protocols performed on the system. By solving the geodesic equations, we can find the geodesic length $\mathcal L$ and hence the minimal entropy production. 

\section{Special Limits of the Metric}

Before attempting to solve the geodesic equations for the  case of  erasure, we now study the high and low temperature limits, as well as the limit of weak coupling, to gain further analytical insights on the form of optimal protocols and the associated entropy production.

\subsection{High Temperature Limit ($\beta \eps, \beta \mu \ll 1$)}

Since the terms of \eref{eq:metric_L} quickly decay at high frequencies, we can perform the high temperature expansion  $f_\beta(\omega) = \frac{1}{2} - \frac{1}{4}\beta\omega + \mathcal O(\beta^3\omega^3)$ directly in the metric. At leading order, we find:
\begin{equation}\label{eq:metric_HT}
	g_{HT} = \frac{\beta^2\hbar}{8\mu}\mathbbm{1}~.
\end{equation}
This enables an analytical solution of the geodesic equations.
Given the boundary conditions $\{\eps(0) = \mu(0) = \mu(\tau)=0,\,\eps(\tau)=\eps_*>0 \}$, which  will  later  match  those of an erasure protocol\footnote{\label{fn:init_cond}Usually, the initial condition for erasure would be $\eps(0) = \nu$ for $\nu$ the chemical potential of the bath and $\eps$ the energy of the two-level system (so that the corresponding thermal state is the fully mixed state). But since here we defined $\eps$ to be the difference to the chemical potential we take $\eps(0) = 0$ without loss of generality.}, we find the following geodesic path
\begin{align}
	\label{eq:geo_HT_eps}
	\eps(t) &= \eps_*\left(t/\tau - \frac{\sin(2\pi t/\tau)}{2\pi}\right)~, \\
	\label{eq:geo_HT_mu}
	\mu(t) &= \frac{\eps_*}{\pi}\sin(\pi t/\tau)^2~.
\end{align}
In the regime $\beta \eps(t) \ll 1$, we observe that minimising entropy production requires a maximal coupling strength~$\eps(\tau)/\pi$. The  entropy production of the geodesic protocol is 
\begin{equation}\label{eq:work_HT}
	k_B T\Sigma^* = \frac{\pi\hbar\beta\eps_*}{2\tau} + \mathcal O(\beta^3 \eps_*^3)~,
\end{equation}
which linearly scales with the final energy $\beta\eps_*$.

\subsection{Zero Temperature Limit ($\beta \eps$ or  $\beta \mu \rightarrow \infty$)}
In the limit of $T=0$\footnote{The zero temperature limit is achieved whenever the energy gaps of the system are too large for thermal fluctuations to occur between the energy levels. Bringing either $\beta\eps$ or $\beta\mu$ to infinity achieves this effect. It is the opposite in the infinite temperature limit, where the thermal fluctuations need to overcome any energy gap, therefore in that limit, both $\beta\eps$ and $\beta\mu$ need to be brought to zero.} we have $f_\beta(\omega)\rightarrow f_\infty(\omega) = \Theta(-\omega)$, where $\Theta$ is the Heaviside step function. Therefore the metric becomes
\begin{equation}\label{eq:metric_T0}
	g_{T=0} = \frac{\beta\hbar}{\pi}\frac{1}{\left(\mu^2 + \eps^2\right)^2}
	\begin{pmatrix}
		\mu^2 &
		-\eps\mu \\
		-\eps\mu &
		\eps^2
	\end{pmatrix}~,
\end{equation}
which coincides with the metric of an angle distance in the $(\eps,\mu)$ space -- hence the metric is singular.
If we re-parameterize $(\eps,\mu)$ as $(r \cos{\phi},r \sin{\phi})$ we find $k_B T\Sigma = \frac{1}{\pi}\int_0^\tau dt \hspace{1mm} \dot\phi(t)^2$. Therefore any protocol that keeps $\dot{\phi}(t)$ constant is a geodesic, leading to the minimal entropy production: 
\begin{equation}\label{eq:work_LT}
	k_B T\Sigma^*\bigg|_{T=0} = \frac{\hbar (\Delta \phi)^2}{\pi\tau},
\end{equation}
with $\phi=\arctan (\mu/\eps)$.
Note that there are multiple (infinitely many) geodesics for any pair of boundary points.
This fact prevents us from continuing the expansion to further orders in temperature. Nevertheless, this limit provides analytical insights on optimal protocols with $\beta \eps$ or  $\beta \mu \gg 1$. In particular, we note that  there is no need for a diverging coupling even when $\eps(\tau) \rightarrow \infty$ as, once $\mu$ has become large, \eref{eq:work_LT} shows  that it is optimal to reduce the coupling while increasing the energy. Furthermore, \eref{eq:work_LT} shows that at zero temperature, while the reversible cost of the operation goes to zero, the dissipation $W_{diss}^* = k_B T\Sigma^*$ remains strictly positive. This result is complementary to the findings of Ref.~\cite{Timpanaro2020} which demonstrate a finite-size correction to Landauer's bound that does not disappear in the zero-temperature regime.

\subsection{Weak Coupling Limit}
Lastly, we take take the weak coupling limit to compare to previous erasure results that are obtained via Lindbladian dynamics, which is a common assumption in previous works on optimal thermodynamic control in the quantum regime.
In this limit the coupling is taken to be small and constant, therefore the metric becomes a scalar:
\begin{equation}\label{eq:metric_wc}
	g_{weak}(\eps) = \frac{\beta^2\hbar}{\Gamma}f_\beta(\eps)(1-f_\beta(\eps))~.
\end{equation}
Which matches the metric one obtains from the rate equation
\begin{equation}\label{eq:rate_eq}
	\frac{dp(t)}{dt} = -\Gamma\left(p(t) - \frac{1}{1+e^{\beta\eps(t)}} \right)~.
\end{equation}
In this regime, protocols that minimize dissipated heat at arbitrary speed were found by~\cite{Esposito2010}. Therefore we will compare the results one obtains in slow driving and the results of~\cite{Zhen2021,VanVu2022} to the exact minimization of~\cite{Esposito2010}.\\ 

\begin{figure}[t]
	\centering
	\includegraphics[width=0.9\textwidth]{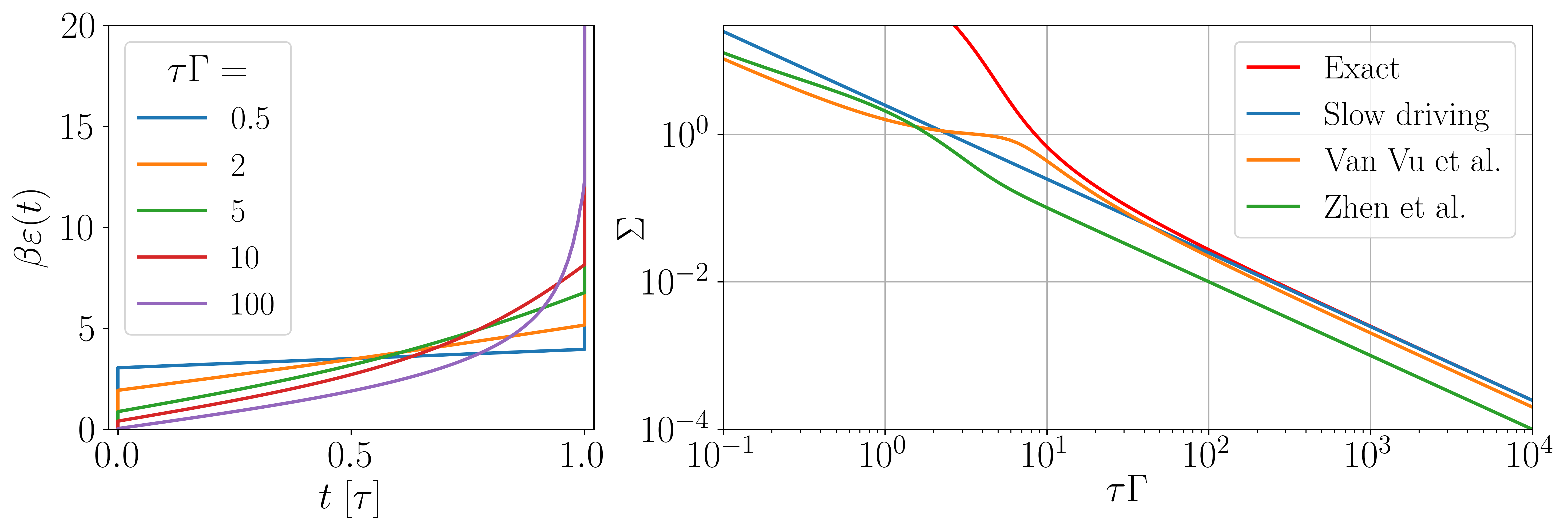}
	\vspace{-10pt}
	\caption{\label{fig:WC_comparison}(left) Examples of optimal protocols computed from~\cite{Esposito2010} for boundary conditions  $\eps(0) = 0$ and $\beta\eps(\tau) = 20$ at different values of $\tau$. (right) Comparison of the entropy production of the optimal protocols from~\cite{Esposito2010} with boundary conditions $\eps(0) = 0$ and $\beta\eps(\tau) = 100$ to the lower bounds given by \eref{eq:finitetimeboundweak}, Van Vu et al.~\cite{VanVu2022} and Zhen et al.~\cite{Zhen2021}.}
\end{figure}
We are interested in erasure processes, where $\eps(t)$ is driven from\footnoteref{fn:init_cond} $\eps(0)=0$ to $\eps(\tau)=\eps_*$ with $\eps_* \gg k_B T$ in a time $\tau$. Optimal finite-time protocols are those which minimize the work cost $W = \int_0^\tau dt~ \dot\eps(t)p(t)$, and hence the heat dissipated to the environment $Q= W - \Delta E$. 
The results of~\cite{Esposito2010} provide an exact solution to this problem, which is shown in \fref{fig:WC_comparison}. As is well-known in finite time stochastic thermodynamics~\cite{Schmiedl2007opt}, jumps appear in the optimal solution. However, as we approach the quasi-static limit where $\tau \Gamma \gg 1$, the jumps progressively disappear (cf. \fref{fig:WC_comparison}). In \sref{sec:geom} we prove why the jumps should also disappear in the long times limit at strong coupling.
The optimal driving solution in this limit has the simple analytical form 
\begin{equation} 
	\varepsilon(t) = 2k_B T \ln\tan\!\left[\frac{\pi}{4}(t/\tau+1)\right]~,
\end{equation}
leading to the work cost 
\begin{equation}\label{eq:work_wc}
	W = k_B T\left( \ln 2 + \frac{\pi^2}{4\tau \Gamma} \right)~,
\end{equation} 
from where we can directly recover \eref{eq:finitetimeboundweak} through the first law of thermodynamics (note that $\Delta E = 0$). 
In \fref{fig:WC_comparison} we notice that the exact solution of~\cite{Esposito2010} agrees well with this analytical form in the slow driving limit. For completeness, we also show recent results of~\cite{Zhen2021,VanVu2022}. These results apply more generally to any Markovian master equation -- here we apply them to the particular case of \eref{eq:rate_eq}, and one can see that they provide a bound to the exact numerical -- and approximate analytical -- solutions. 

\section{Optimized Erasure}
We now focus on erasure outside any approximation, where we will optimize the driving over both the energy and coupling. In what follows, we focus on minimizing $\Sigma$ in an erasure process, which imposes specific boundary conditions to the geodesic equations. We assume that we have no prior knowledge of the system, therefore its initial state is $\rho_S(0) = \mathbbm{1}/2$. This translates in taking $\eps(0) = 0$ so that it coincides with the thermal state of $\hat H(0)$. For the qubit to be erased we want its final state to be $\rho_S(\tau) \approx \ket{0}\!\!\bra{0}$ (i.e. $p(\tau)\approx0$). Since the driving is done slowly, $p(t)$ is always close to its thermal expectation value. Therefore by choosing $\beta\eps(\tau) \rightarrow \infty$ we ensure $p(\tau) \approx 0$\footnote{Strictly speaking, in order to ensure consistency with the slow driving limit,   $\beta\eps(\tau)$ has to remain finite (so that the speed $\dot{\lambda}$ remain finite). However, the final population $p(\tau)$ is exponentially small with  $\beta \eps(\tau)$, leading to exponentially small corrections.  Our  results are valid up to such corrections, and for sufficiently large~$\tau$ to ensure the validity of the approximation. }. For the coupling, the boundary conditions are $\mu(0) = \mu(\tau) = 0$, because we want to think of this as an ``erasure machine'' that the qubit is ``brought to'' at the start and ``retrieved from'' at the end. Given this family of protocols, we recognize from \eref{eq:WSigma} that $W = Q = k_B T (\ln 2 + \Sigma)$.
\emph{After} the qubit has been decoupled (i.e. at $t>\tau$), we bring the Hamiltonian of the system back to its starting value ($\eps = 0$) to close the cycle. Since  $p(\tau)\approx 0$, this step requires no work, and it can be done arbitrarily quickly.

The geodesic equations we obtain for this process are not solvable analytically. The integral of \eref{eq:metric_L} can be solved to give us an expression of the metric in terms of polygamma functions but it does not simplify the geodesic equations into an analytically solvable form. We therefore turn to numerical tools to obtain the optimal protocol and compute the dissipated work. However, in our case, we want to impose the aforementioned boundary conditions; this is known as a Boundary Value Problem (BVP), which is famously hard to solve numerically~\cite{Fox87}. Though we can use the fact that the high temperature limit is accurate at the start of an erasure protocol, therefore the initial conditions of the optimal protocol for erasure will match the initial conditions of~\eref{eq:geo_HT_eps} and \eref{eq:geo_HT_mu}. This allows us to turn the BVP into an Initial Value Problem which is much simpler to solve.

\begin{figure}[t]
	\centering
	\includegraphics[width=\textwidth]{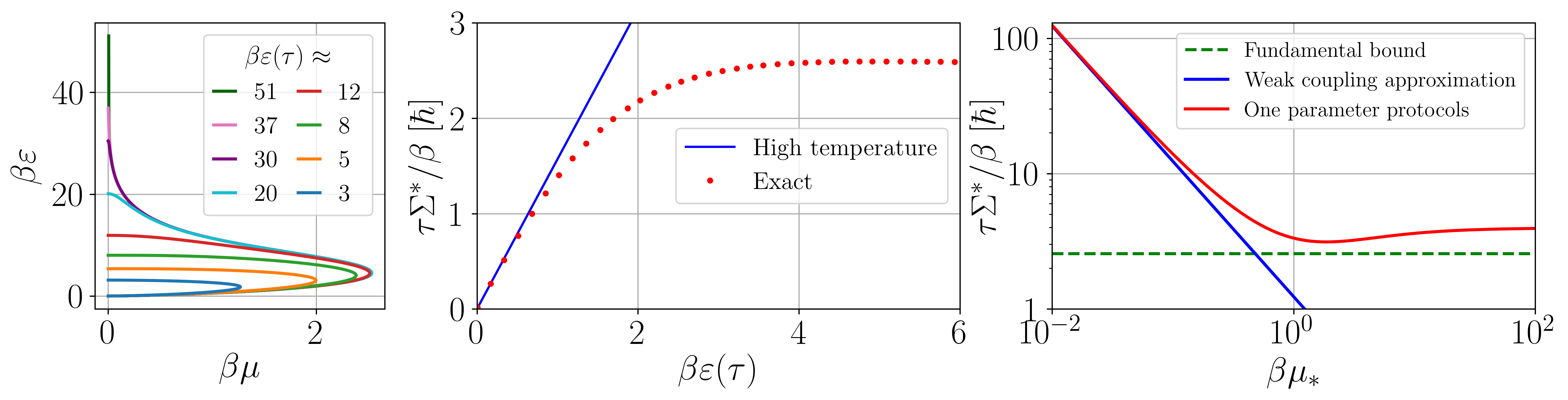}
	\vspace{-15pt}
	\caption{(left) A series of optimal protocols depicted in the $(\mu,\eps)$ space. They all start with zero energy and coupling and end with finite energy and zero coupling. In the limit of large $\beta\eps(\tau)$ they can be considered as erasure protocols. (middle) The entropy production of the optimal erasure protocol as a function of the final energy, compared to the high temperature regime cost \eref{eq:work_HT}. (right)  
		Comparison of the entropy production for a geodesic protocol in which one parameter is varied at a time  (with $\mu$ being increased until $\mu^*$)  and the weak coupling approximation \eref{eq:finitetimeboundweak}; the minimal possible entropy production, $\tau\Sigma^*/\beta = 2.57946 \pm 1\cdot 10^{-5}\;[\hbar]$, obtained when both parameters are changed simultaneously is also shown.
	}
	\vspace{-17pt}
	\label{fig:main}
\end{figure}

In \fref{fig:main} we show  optimal erasure protocols in the $(\mu,\eps)$ space for different final values of $\beta\eps$. We can notice that the predictions of the high and low temperature limit are verified: at the start of the protocols the coupling is increased, but once we reach a certain value there is no more need to increase it, regardless of the final value of $\beta\eps$ we try to reach. Interestingly, the maximal value reached by $\beta\mu$ is larger than $1$.
This shows that reaching the strong coupling regime is needed to achieve optimal erasure, which is one of the main insights of our work.
In the same figure we also show the value of $\tau \Sigma^*/\beta$ for the optimal protocol as a function of the final energy. We can see that for small values of $\beta\eps(\tau)$ the high temperature limit \eref{eq:work_HT} gives an accurate description of the work cost, but as we reach higher values it saturates to $\tau\Sigma^*/\beta = 2.57946 \pm 1\cdot 10^{-5}\;[\hbar]$. This provides a  finite-time correction to Landauer's principle in this setup:
\begin{equation}\label{eq:finitetimebound_gen}
	Q \geq k_B T \left(\ln 2 + a\frac{\tau_{\rm Pl}}{\tau}\right) + \mathcal{O}\left(\frac{1}{\Gamma^2 \tau^2}\right).
\end{equation}
with $a\approx 2.57946$ and $\tau_{\rm Pl} = \beta\hbar$. This is one of the main results of this work  and can be seen as a generalization of \eref{eq:finitetimeboundweak}. As opposed to the results of~\cite{Zhen2021} and~\cite{VanVu2022}, \eref{eq:finitetimebound_gen} is only valid for large protocol times; yet, it has the advantage of taking into account strong coupling effects (including any possible variation of the coupling strength), having a much simpler form for the correction (which is independent of any chosen relaxation timescale), and we provide an explicit protocol to  achieve it.  By turning around \eref{eq:finitetimebound_gen} one can highlight a quantum speed limit for erasure of a qubit, furthermore this speed limit is of the order of the Planckian time $\tau_{\rm Pl} = \hbar/k_B T$ which is conjectured to be the fastest relaxation timescale for thermalization~\cite{Hartnoll22}. In particular, one can see that \eref{eq:finitetimebound_gen} bounds the speed of erasure by the order of $\tau_{\rm Pl}$ regardless of how large is the coupling strength used in the protocol.

Interestingly, we now argue that the form of the correction \eref{eq:finitetimebound_gen} is in fact general of any Landauer erasure protocol with control on S and the SB coupling.  Indeed, first note that -- in natural units -- the minimal dissipation $\Sigma^*$ is dimensionless and  can only depend on $\beta$ and the boundary conditions as we optimize over $\mu$ and $\eps$. In an erasure process, the boundary conditions read:  $\eps(0)=0$, $\eps(\tau) \rightarrow \infty$, and $\mu_i(0)=\mu_i(\tau)=0$ where $i$ runs over all the possible control parameters on SB.  But this implies that $\Sigma^*$ is independent of them and hence of $\beta$. Therefore $W_{diss}^*$ will take the form of a constant, independent of any parameter of the system and bath, divided by $\tau$. This is a crucial difference from~\eref{eq:finitetimeboundweak}. 

This simple argument based on dimensional analysis thus shows that  \eref{eq:finitetimebound_gen} is rather general, with the value of $a$ depending on the specific implementation (e.g. the ohmicity of the bath). It is important to highlight that the bound \eref{eq:finitetimebound_gen}  implies that, even when having access to arbitrary strong SB interactions  (naively taking $\Gamma \rightarrow \infty$ in \eref{eq:finitetimeboundweak}), infinite time is still required for perfect erasure due to the quantum-mechanical correction derived here. 

Finally, we analyze a scenario where the coupling is kept constant while $\eps(t)$ is driven, which is  motivated both by experimental set-ups and for a comparison with the weakly interacting case. 
Therefore, we restrict to \emph{one-parameter protocols} consisting of the three following steps: 1.~while keeping $\eps$ at $0$ we turn on the coupling to some value $\mu_*$; 2.~while keeping the coupling fixed we bring $\eps$ from $0$ to some value $\eps_* \gg k_B T$; 3.~while keeping $\eps$ constant we turn off the coupling. 
Each step contributes positively to the entropy production, and their minimization leads to the following dissipation
\begin{equation}
	\Sigma^*_{\rm one~param} = \frac{1}{\tau}\!\left(\int_0^{\mu_*}\!d\mu~[g_{\mu\mu}(0,\mu)]^{1/2} + \int_0^{\infty}\!d\eps~[g_{\eps\eps}(\eps,\mu_*)]^{1/2}\right)^2~.
\end{equation}
In \fref{fig:main}, we show $\Sigma^*$ for different values of  $\mu_*$, ranging from the weak  to the super-strong coupling regime. It can be appreciated how~\eref{eq:finitetimeboundweak} breaks down, and also how such one-parameter protocols become close to the fundamental limit~\eref{eq:finitetimebound_gen} for $\beta \mu_*>1$.

\section{Conclusions and Outlook}

Deriving finite-time corrections to the seminal Landauer bound is a challenging endeavor in stochastic and quantum thermodynamics. Previous works have focused on Markovian systems only, which in the quantum regime is obtained through the weak coupling limit~($\beta \kappa^2 \rightarrow 0$). However, should a general finite-time correction exist, it will require the presence of strong coupling at some point during the process as the dissipation generated in finite time is proportional to $\kappa^{-2}$ when $\kappa$ is small\footnote{This can be seen by expanding the finite time dissipation around $\kappa^2 = 0$ for long times: $k_B T\Sigma \propto 1/\kappa^2\tau$, which follows by noticing that the relaxation time-scale is of the order of $\kappa^{-2}$.}. Motivated by this observation, we have developed new insights into the form of optimal protocols for erasure beyond the weak coupling limit.

We have focused on a bit encoded in the occupation of a single fermionic mode, which can be strongly coupled to a reservoir. We have derived analytically the thermodynamic metric, which governs the dissipation rate in the slow driving regime, and showed that it takes a simple form in the high and low temperature limits. From the general form of the metric we obtained the optimal erasure protocol, which requires increasing the coupling strength to $\kappa^2 \sim k_B T$, which corresponds to a relaxation timescale of the order of the Planckian time $\tau_{\rm Pl}$. The corresponding dissipation yields a finite-time correction to Landauer's bound for this setup, which is substantially lower than similar results in the weak coupling regime. Furthermore, by using the obtained bound as a quantum speed limit, this result adds further evidence to the conjecture \cite{Hartnoll22} that $\tau_{\rm Pl}$ is fastest relaxation timescale many-body systems can achieve.

While our results were derived in a fermionic model, there are some general insights that follow from our work. First there is a fundamental quantum correction that prevails, see \eref{eq:finitetimebound_gen}, which can  be compared with \eref{eq:finitetimeboundweak} derived in the weak coupling regime. While the specific value of $a$ in \eref{eq:finitetimebound_gen} will depend on the specific setup, it will never approach $0$ (even for diverging system-bath coupling) due to the inherent cost of changing the interaction strength. Furthermore, to obtain these results we adapted the framework of thermodynamic geometry to system-bath unitary dynamics in which the coupling can be arbitrarily large or small. Finally, as was argued before, our results make evident the need for strong coupling in a general finite-time correction to Landauer's principle.

This work opens exciting directions for the future. On the one hand, the level of experimental control required to implement such protocols is in principle possible in quantum dots, where   the energy-level $\eps(t)$ and coupling $\kappa(t)$ can be independently controlled, even by several orders of magnitude~\cite{Rochette2019}.  On the other hand, it would be interesting to characterize the dependence of $a$ in the nature of the bath and the SB coupling -- e.g. its spectral density,  more generally  to derive similar quantum-mechanical finite-time corrections that are independent of the specific implementation, and to gain further insights in the connection between Landauer erasure and the Planckian time.

\newpage
\section{Proof of Thermalization}\label{sec:proof_therm}
In this section we will prove that, in absence of driving, $p(t)$ and $v(t)$ thermalize. We do so in two steps, we first simplify the expressions of \eref{eq:p_exact} and \eref{eq:v_exact} for $\eps(t) = \eps$ and $\kappa(t) = \kappa$ and compute the infinite time limit. Then we compute the thermal expectation value of the corresponding observables and prove that the obtained expressions are the same.
\subsection{Infinite Time Limit in Absence of Driving}
By assuming that the driving parameters are kept constant the propagator becomes 
\begin{equation}\label{frozenH_prop}
	G(t,s) = e^{-(t-s)(\frac{1}{2}\kappa^2 + i\eps)}~.
\end{equation}
This allows us to compute the time integrals in \eref{eq:p_exact} and \eref{eq:v_exact}:
\begin{align}
	p(t) &= p(0)e^{-\kappa^2 t} + \frac{\kappa^2}{2\pi} \int_{-\infty}^{\infty}d\omega~ f_\beta(\omega) \frac{1-2e^{-\kappa^2t/2}\cos([\omega-\eps]t) + e^{-\kappa^2 t}}{\kappa^4/4 + (\omega-\eps)^2}~,\\
	v(t) &= \frac{\kappa}{\pi} \int_{-\infty}^{\infty}d\omega~ f_\beta(\omega) \frac{(\omega-\eps)\left[ 1 - e^{-\kappa^2 t/2}\cos([\omega-\eps]t) \right]- \frac{1}{2}\kappa^2 e^{-\kappa^2 t/2}\sin([\omega-\eps]t)  }{\kappa^4/4 + (\omega-\eps)^2}~.
\end{align}
By taking the limit $t\rightarrow\infty$, we find
\begin{align}
	\label{p_thm_0}
	\lim_{t\rightarrow\infty} p(t) &=\int_{-\infty}^{\infty} \frac{d\omega}{\pi} f_\beta(\omega) \frac{\kappa^2/2}{\kappa^4/4 + (\omega-\eps)^2}~,\\
	\label{v_thm_0}
	\lim_{t\rightarrow\infty} v(t) &= \kappa \int_{-\infty}^{\infty} \frac{d\omega}{\pi} f_\beta(\omega) \frac{(\omega-\eps) }{\kappa^4/4 + (\omega-\eps)^2}~.
\end{align}
Here we can notice that if we take the Laplace transform of the propagator we obtain
\begin{equation}
	\tilde G(z) := \int_0^\infty dt~ G(t,0)e^{-zt} = \frac{1}{z+i\eps+\kappa^2/2}~,
\end{equation}
which allows us to rewrite \eref{p_thm_0} and \eref{v_thm_0} as
\begin{align}
	\label{p_thm}
	\lim_{t\rightarrow\infty} p(t) &=\int_{-\infty}^{\infty} \frac{d\omega}{\pi} f_\beta(\omega) \Re\!\left[\tilde G(-i\omega)\right]~,\\
	\label{v_thm}
	\lim_{t\rightarrow\infty} v(t) &= \kappa\int_{-\infty}^{\infty} \frac{d\omega}{\pi} f_\beta(\omega) \Im\!\left[\tilde G(-i\omega)\right]~.
\end{align}

\subsection{Thermal Expectation Value}
We  now compute the expectation value of $\hat a^\dagger \hat a$ and $\hat V = \sum_{k=1}^n \lambda_k \hat a^\dagger \hat b_k + \lambda_k^* \hat b_k^\dagger \hat a$ when the state is a Gibbs state.
Therefore we want to find $p_{th} := \Tr[\hat\pi \hat a^\dagger \hat a]$ and $v_{th} := \Tr[\hat\pi \hat V]$. Using the fact that the total Hamiltonian is quadratic, we can diagonalize it to rewrite it in the following way
\begin{equation}
	\hat H  = \sum_k \eps_k \hat c_k^\dagger\hat c_k~,
\end{equation}
where $\eps_k$ are eigen-energies and $\hat c_k$ are fermionic ladder operators that follow the CAR: $\{\hat c_j^\dagger,\hat c_k\} = \delta_{jk}\mathbbm{1}$, $\{\hat c_j,\hat c_k\} = 0$. They are related to the original ones by
\begin{equation}
	\hat a = \sum_k \braket{0|\hat a|k}\hat c_k~,\qquad
	\hat b_j = \sum_k \braket{0|\hat b_j|k}\hat c_k~,
\end{equation} 
where $\ket{k} = \hat c_k^\dagger \ket{0} $ are $1$-particle eigenstates of the Hamiltonian with eigenvalue $\eps_k$. Inserting this relation in the expression for the thermal expectation of the probability of occupation we find
\begin{equation}
	p_{th} = \frac{1}{Z}\sum_{jk} \braket{j|\hat a^\dagger|0}\!\braket{0|\hat a|k}\Tr\!\left[e^{-\beta \hat H}\hat c_j^\dagger \hat c_k\right]
	= \sum_{k} \left|\braket{k|\hat a^\dagger|0}\right|^2 f_\beta(\eps_k)~.
\end{equation} 
By its definition, we can write the propagator as follows
\begin{equation}
	G(t,0) = \braket{0|\hat a_H(t) \hat a^\dagger|0}
	= \sum_k e^{-i \eps_k t}\braket{0|\hat U^\dagger(t)\hat a|k}\!\braket{k|\hat a^\dagger|0}
	= \sum_k e^{-i \eps_k t}\left|\braket{k|\hat a^\dagger|0}\right|^2 ~,
\end{equation}
where we used the fact that the vacuum state does not evolve $\hat U(t)\ket{0} = \ket{0}$ and $\hat U(t) = e^{-it\sum_k \eps_k \hat c^\dagger_k \hat c_k}$. By now defining $\varphi(\omega) := \sum_k \left|\braket{k|\hat a^\dagger|0}\right|^2 \delta(\omega-\eps_k)$, we can identify
\begin{equation}
	\label{eq:G_ft_phi}
	G(t,0) = \int_{-\infty}^\infty d\omega~ \varphi(\omega) e^{-i\omega t} ~,\quad p_{th} = \int_{-\infty}^\infty d\omega~ f_\beta(\omega) \varphi(\omega)~.
\end{equation}
Considering \eref{p_thm} it is clear that if $\varphi(\omega) = \frac{1}{\pi}\Re\!\left[\tilde G(-i\omega)\right]$ then we have proven $p_{th} = \lim_{t\rightarrow\infty}p(t)$. Therefore we compute the Laplace transform of $G(t,0)$ using \eref{eq:G_ft_phi}
\begin{equation}
	\tilde G(-i\omega) = \int_0^\infty dt~ G(t,0)e^{i\omega t} = \pi \varphi(\omega) + i P.\int_{-\infty}^\infty d\omega'~\frac{\varphi(\omega')}{\omega-\omega'}~,
\end{equation}
where $P.$ denotes the Cauchy principal value. Since $\varphi(\omega)$ is by definition a real function we can see that $P.\int_{-\infty}^\infty d\omega'~\frac{\varphi(\omega')}{\omega-\omega'}$ is a real number. Therefore we can conclude $\varphi(\omega) = \frac{1}{\pi}\Re\!\left[\tilde G(-i\omega)\right]$. Which concludes the proof of the thermalization of $p(t)$.

To prove the thermalization of $v(t)$ we proceed in a similar fashion. We start by computing $v_{th}$
\begin{equation}
	v_{th} = \sum_{jk}f_\beta(\eps_k)\left(\lambda_j\braket{k|\hat a^\dagger|0}\! \braket{0|\hat b_j|k}+ \lambda_j^*\braket{k|\hat b_j^\dagger|0}\! \braket{0|\hat a|k}\right)~.
\end{equation}
To proceed we have to define the following cross-propagators
\begin{align}
	\lambda_j K_j(t) &:= \braket{0|\hat a_H(t) \hat b^\dagger_j |0} = \sum_k e^{-i\eps_k t}\braket{k|\hat b_j^\dagger|0}\! \braket{0|\hat a|k} = \int_{-\infty}^\infty d\omega ~\psi_j(\omega)e^{-i\omega t}~,\\
	\lambda_j^* H_j(t) &:= \braket{0|\hat b_{j,H}(t) \hat a^\dagger |0} = \sum_k e^{-i\eps_k t}\braket{k|\hat a^\dagger|0}\! \braket{0|\hat b_j|k} = \int_{-\infty}^\infty d\omega ~\psi_j^*(\omega)e^{-i\omega t}~,
\end{align}
where we defined $\psi_j(\omega) = \sum_k \braket{k|\hat b_j^\dagger|0}\! \braket{0|\hat a|k} \delta(\omega-\eps_k)$. By further defining $\psi_0(\omega): = \sum_k \lambda_k^*\psi_k(\omega)$ and $\psi(\omega) = \psi_0(\omega)+\psi_0^*(\omega)$ we can see that
\begin{align}
	K(t) &:= \sum_j |\lambda_j|^2\left(K_j(t) + H_j(t) \right) = \int_{-\infty}^\infty d\omega ~\psi (\omega)e^{-i\omega t}~, \\
	\label{v_thm_proof}
	v_{th} &= \int_{-\infty}^\infty d\omega~ f_\beta(\omega) \psi(\omega)~.
\end{align}
Therefore, similarly to the case of $G(t,0)$, we have
\begin{equation*}
	\tilde K(-i\omega) = \pi \psi(\omega) + i P.\int_{-\infty}^\infty d\omega' \frac{\psi(\omega')}{\omega-\omega'}~,
\end{equation*}
and in particular $\pi \psi(\omega) = \Re\left[\tilde K(-i\omega)\right]$ (since $\psi(\omega)$ is real by definition). Hence, by \eref{v_thm} and \eref{v_thm_proof}, the last step to prove that $v(t)$ thermalizes is to check that $\Re\left[\tilde K(-i\omega)\right] = \kappa\Im\left[\tilde G(-i\omega)\right]$. To do so we start by computing the components of $K(t)$: from the solution of the evolution of $\hat a_H$ and $\hat b_{j,H}$ (cf. \cite{Rolandi2023Quantum} for more detail) we can find
\begin{equation}
		K_j(t) = H_j(t) = -i\kappa\frac{e^{-i \omega_j t} - G(t,0)}{\frac{1}{2}\kappa^2 +i (\eps-\omega_j)}~.
\end{equation}
Since the time time dependence is contained in the exponentials, it is straightforward to compute the Laplace transform
\begin{equation}
	\tilde K_j(z) = \tilde H_j(z) = \frac{- i \kappa}{z+i \omega_j}\tilde G(z)~.
\end{equation}
Therefore we find
\begin{equation}
	\tilde K(-i\omega) = \frac{\kappa}{\pi} \tilde G(-i\omega) P.\int_{-\infty}^\infty  d\omega' \frac{1}{\omega-\omega'} = -i\kappa \tilde G(-i\omega) ~,
\end{equation}
which allows us to conclude $\psi(\omega) = \kappa\Im\!\left[\tilde G(-i\omega)\right]$. This concludes the proof of the thermalization of $v(t)$.

	\cleardoublepage
	\chapter{Collective Advantages}\label{chap:collective}
\begin{wrapfigure}{R}{0.4\textwidth}
	\centering
	\vspace{-15pt}
	\includegraphics[width=0.4\textwidth]{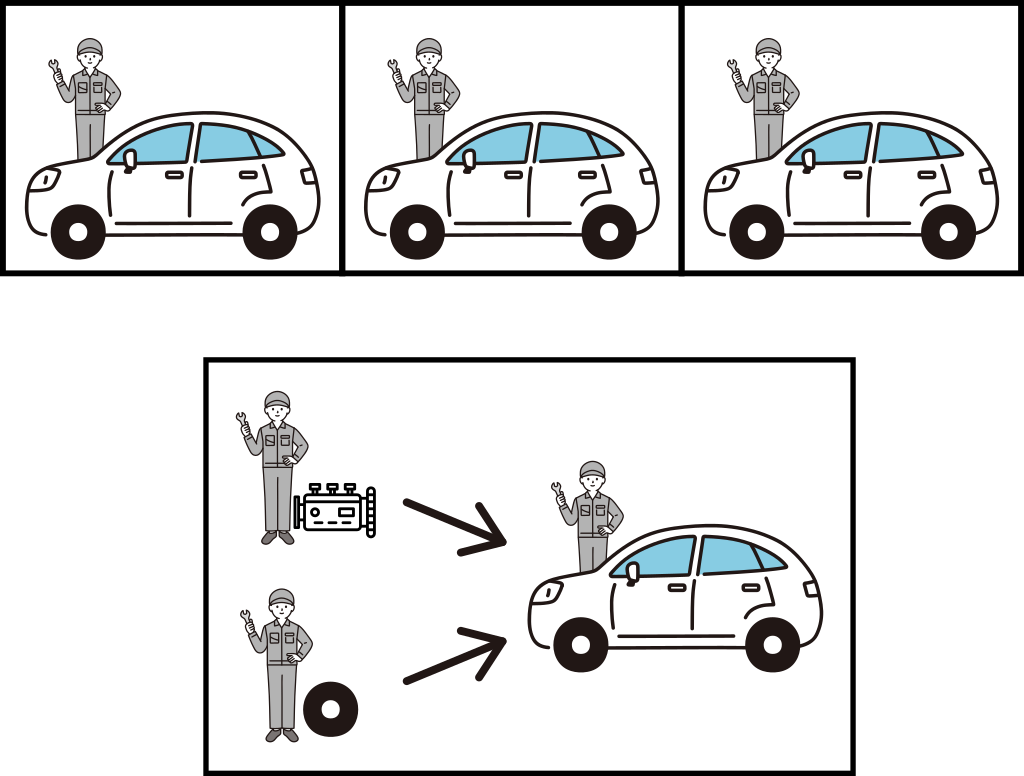}
	\vspace{-12pt}
	\caption{Analogue of collective advantages in the macroscopic world: the streamlining of production. Streamlined production lines (bottom) are much more efficient than a production line where each worker is creating the product independently (top).~\cite{maria_drawing}}
	\vspace{-18pt}
	\label{fig:collective_analogue}
\end{wrapfigure}

\epigraph{¡Hasta la victoria siempre!}{--- Ernesto Che Guevara}
Collective effects play a central role in physics, ranging from phase transitions to quantum entanglement. Often, they can be exploited for a useful task, such as ultra-precise measurements~\cite{Giovannetti2006}, leading to the notion of a collective advantage\footnote{The outcome of a task is improved when performed globally on a collection of systems compared than when realized on each system individually. There are multiple real life analogues of collective advantages in real life -- unrelated to physics -- such as collective action and bargaining or streamlined production (cf. \fref{fig:collective_analogue}).}.
In the growing fields of stochastic and quantum thermodynamics, such advantages have received notable attention: relevant examples are found in quantum batteries, where entangling operations have been proven to enable faster charging~\cite{Campaioli2018}; 
in many-body thermal engines, whose performance can be enhanced via phase transitions~\cite{Vroylandt2017}, many-body interactions~\cite{Souza2022Collective}, or superradiance~\cite{Niedenzu_2018}; and in quantum transport~\cite{Chiaracane2020}.

\noindent In this chapter, we uncover a new collective advantage in a crucial task in non-equilibrium thermodynamics: the minimization of dissipation in finite time. In general, the thermodynamic work $W$ required to transform a system, in contact with an environment, in a finite time $\tau$ can be split into two contributions (cf. \sref{sec:second_law})
\begin{equation}
	W= \Delta F_{eq} + W_{diss}
	\label{eq:Wdiss}
\end{equation}
a reversible contribution $\Delta F_{eq}$, the free energy change, and an irreversible positive contribution $W_{diss}$, the dissipated work. Whereas $\Delta F_{eq}$ is extensive with the size $N$ of the system, we will show here that $W_{diss}$ can grow sub-linearly in $N$. 
This is proven in the regime of slow-but-finite-time processes and becomes possible by exploiting many-body interactions suitably created along the process.

The advantage is dramatic: in principle, collective processes enable an $N$-fold reduction of $W_{diss}$ when compared to local processes (see \fref{fig}). While we will show that reaching this limit requires highly non-local or long-range interactions, a sub-linear growth of $W_{diss} $ can be achieved with two-body interactions and realistic control. To obtain these results, we rely on the framework of thermodynamic geometry (cf. \sref{sec:geom}). Our results show that geodesic protocols generically explore highly interacting Hamiltonians, even if interactions are absent at the beginning and end of the process. As an application, we focus on finite-time information erasure (cf. \cref{chap:landauer}) of $N$ qubits. We show that collective processing can substantially reduce dissipation in this relevant task, leading to a faster convergence to Landauer's bound.

Overall, these results uncover a genuine collective advantage in stochastic and quantum thermodynamics, which is not linked to standard collective phenomena such as quantum entanglement, phase transitions, or collective system-baths couplings (e.g. superradiance).

\section{Framework}
Following the formalism introduced in \sref{sec:geom}, let us consider a system in a $d$-dimensional Hilbert space $\mathbb C^d$ with an externally driven Hamiltonian~$\hat h(t)$. It can be parameterized as $\hat h(t)=\sum_{j=1}^n \lambda^j(t) \hat x_j$, $\{ \lambda^j \}$ are externally controllable parameters, and $\{\hat x_j \}$ are the corresponding observables. Additionally, the system is in contact with an external thermal bath at inverse temperature~$\beta$, so that the system is undergoing Lindblad dynamics \eref{eq:Lindblad}.

We focus on the task of driving $\hat h(t)$ from an initial configuration $\hat h(0) =\hat h_i$ to a final one $\hat h(\tau)= \hat h_f$ in a time~$t\in [0,\tau]$. External energy is needed to realize this transformation, quantified by the (average) thermodynamic work: 
\begin{equation}
	W= \int_0^\tau\!\!dt~ \Tr\Big[\frac{d\hat h(t)}{dt} \hat\rho(t)\Big]~,
\end{equation}
where $\hat\rho(t)$ is the state of the system. This expression can be split as in \eref{eq:Wdiss}, where $\Delta F_{eq}= \beta^{-1} \ln Z(0)/ Z(\tau)$. The minimal dissipated work $W_{diss}$ in a finite time $\tau$ can then be found by optimizing the driving protocol $\lambda : [0,\tau]\mapsto M\subset\mathbb R^n$. To address the non-trivial optimization we make some assumptions. 

First, we assume that the driving $\frac{d}{dt}\hat h(t)$ is slow compared to the relaxation rate. Then $W_{diss}$ can be expressed as a quadratic form at leading order in $\tau^{-1}$ (cf. \sref{sec:geom}): 
\begin{align}
	W_{diss}= k_B T \int_0^\tau dt~\dot{\lambda}^i(t)\dot{\lambda}^j(t) g_{ij}(\lambda(t)) +  \mathcal{O}\!\left(\tau^{-2}\right) ~,
	\label{eq:GeneralExpansionDiss}
\end{align}
where $g_{ij}(\lambda)$ is the so-called thermodynamic metric. By solving the geodesic equation that derive from the metric we can find the protocols that of minimal length $\mathcal L$ by the metric and minimal dissipation $W_{diss}^* = k_BT\mathcal L^2/\tau$.

As a second simplification, we assume there is a single relaxation timescale $\tau_{\rm eq}$\footnote{We assume that all driven observables decay exponentially to equilibrium with the same timescale \cite{Abiuso2020Geometric}.}, so that the metric becomes~\cite{Scandi19}: 
\begin{equation}
	\label{eq:metric_simple}
	g_{ij}  = -\tau_{\rm eq}\beta\frac{\partial^2 F_{eq}(\lambda) }{\partial \lambda^i \partial \lambda^j }~.  
\end{equation}
Note that $g_{ij}$ then becomes the standard thermodynamic metric for macroscopic systems~\cite{Salamon1985,Crooks2007}, which can also describes step-processes~\cite{Scandi2020}. In what follows, without loss of generality, we set $\tau_{\rm eq}=1$.

As a last simplification, we will assume that the initial and final Hamiltonian commute $[\hat h_i, \hat h_f] = 0$. This allows us to conclude that at all times $[\frac{d}{dt}\hat h(t), \hat h(t)] = 0$, as changes in the eigen-basis can only increase dissipation in the linear response regime~\cite{Abiuso2020Geometric}.

Let us now consider a scenario in which we perform the driving on $N$ copies of the system. We denote by $\hat H(t) = \hat H_0(t) + \hat H_{\rm int}(t)$ the total Hamiltonian for all the copies, where $\hat H_0(t) = \sum_{j=1}^N \hat h^{(j)}(t)$ and $\hat H_{\rm int}(t)$ contains the interaction between the copies. We parameterize $\hat H(t)$ similarly to $\hat h(t)$: $\hat H(t) = \sum_{i=1}^n \gamma^i(t)\hat X_i$, where the sum can have up to $n=d^N$ terms. The problem at hand imposes the following boundary conditions on the protocol: $\hat H_{\rm int} (0) = \hat H_{\rm int} (\tau) = 0$, $\hat h^{(j)}(0) = \hat h_i$, and $\hat h^{(j)}(\tau) = \hat h_f$ $\forall j$. Furthermore, by the same reasoning as in the case for a single copy, we have that $[\frac{d}{dt}\hat H(t), \hat H(t)] = 0$ for the geodesic protocol.

\section{Fundamental Limit of Collective Advantages}
Let us first note that $\Delta F_{eq}$ is extensive with $N$ which directly follows from the boundary conditions. Instead, $W_{diss}$ depends on the process and can exhibit a non-trivial behavior whenever $ \hat H_{\rm int} (t) \neq 0$. Indeed, we find that, in general, geodesic paths explore highly interacting Hamiltonians if the constraints allow for it.

To reach the fundamental limit of $W_{diss}^{*}$ we can assume full control on $\hat H(t)$, so that the $n = d^N$ different eigen-energies $\{\gamma^i\}$ can be externally controlled at will -- the corresponding $\{\hat X_i\}$ are chosen to be the corresponding eigen-projectors. In this case, the distance function corresponding to the thermodynamic metric \eref{eq:metric_simple} is known to be the quantum Hellinger angle: $\mathcal{L}(\gamma,\gamma') = 2 \arccos\Tr\big[\sqrt{\hat\pi(\gamma)} \sqrt{\hat\pi(\gamma')}\,\big]$ where $\hat\pi(\gamma) = e^{-\beta\hat H(\gamma)}/Z(\gamma)$ is the thermal state (cf. \cite{Rolandi2023Collective}). Therefore the optimal dissipation of a protocol is 
\begin{equation}
	\beta W_{diss}^{*}= \frac{1}{\tau}\left( 2 \arccos\Tr\!\left[\sqrt{\hat\pi(0)} \sqrt{\hat\pi(\tau)}\right]\right)^2 ~.
	\label{fundamental_bound}
\end{equation}
Since trivially $\arccos (x) \leq \pi/2$ for $x>0$, the minimal dissipation of a $N$-body system is bounded by a constant $W_{diss}^{*} \leq \frac{1}{\tau} \pi^2$ independent of $N$. This is somehow astonishing, as we expect the dissipation generated when driving a many-body system to increase extensively with its size. The corresponding protocol that achieves this limit is given by
\begin{equation}\label{FC_driving}
	\beta\hat H(t) \!=\! -2\log\!\left[\sin\!\!\left[\frac{ (\tau\!-\!t)\mathcal L^*}{2\tau}\right]\!\!\sqrt{\hat \pi(0)} + \sin\!\!\left[\frac{t \mathcal L^*}{2\tau}\right]\!\!\sqrt{\hat \pi(\tau)} \right]\!,
\end{equation}
where $\mathcal L^* = \mathcal L(\hat\pi(0),\hat\pi(\tau))$.
Crucially, this protocol generally requires all possible interacting terms available in the Hamiltonian space, including highly non-local $N$-body interactions. One can easily prove this, by contradiction, for generic protocols using the non linearity of the logarithm. This is illustrated in what follows for the paradigmatic task of erasing $N$ bits of information.

\section{Collective Erasure}
Let us consider $N$ qubits, each with local Hamiltonian $\hat h(t)=\eps(t) \hat\sigma_z$. We want to drive $\eps(t)$ from $\eps(0) = 0$ to $\eps(\tau) = E$ with $E\gg k_B T$, so that the state of each qubit evolves from a fully mixed state $\hat\pi(0)=\frac{1}{2}\mathbbm{1}$ to an (almost) pure state $\hat\pi(\tau) \approx \ket{0}\!\!\bra{0}$ due to the action of the external bath. We have $\Delta F_{eq} = N k_B T \ln 2$, corresponding to Landauer's bound.

Consider first the independent scenario, so that during the whole protocol $\hat H_{\rm int}(t) = 0$. For each qubit, the dissipation generated via an optimal driving can be computed from~\eref{fundamental_bound} with the aforementioned boundary conditions, yielding $\beta W^{*}_{diss} = \pi^2/4\tau $. The total dissipation of $N$ qubits then reads:
\begin{equation}
	W_{diss}^{\rm *,  local} = \frac{\pi^2}{4 \tau}N k_B T~, 
	\label{localbound}
\end{equation}
which  grows linearly with $N$.
The corresponding optimal driving reads $\beta\eps(t)=\ln\tan\!\left[\pi(t+\tau)/4\tau\right]$, which has been implemented experimentally in a single-energy driven dot~\cite{Scandi2022}. 

\begin{figure}[t]
	\centering
	\includegraphics[width=\textwidth]{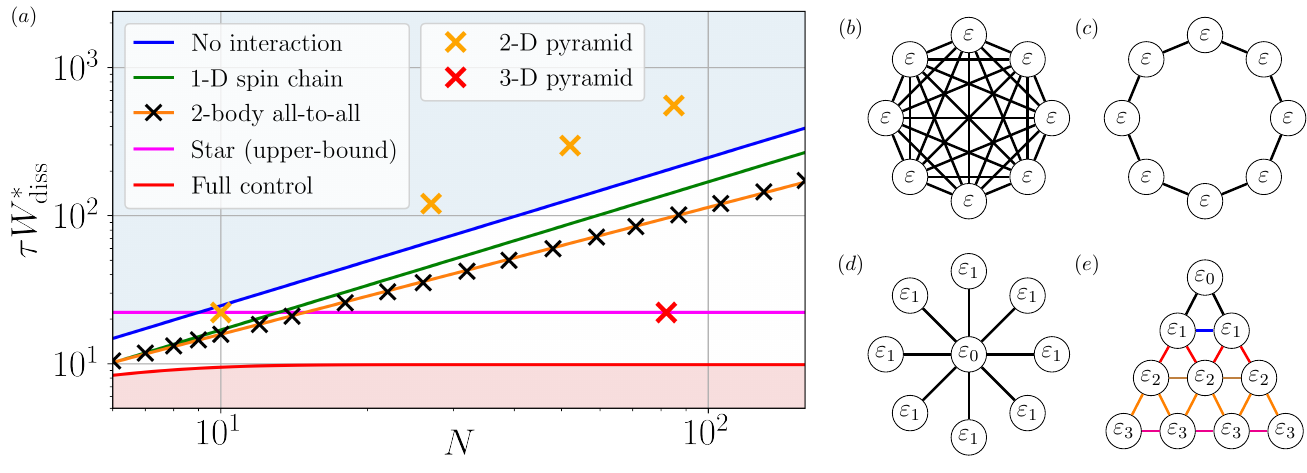}
	\vspace{-10pt}
	\caption{(a) Minimal dissipation for the erasure of $N$ spins for different control designs analyzed in this chapter. These are compared with the dissipations that are achievable with no interactions (\eref{localbound}, blue-shaded area), and with the dissipations that are not achievable regardless of the protocol (\eref{boundNqubits}, red-shaded area). We find $\tau W_{diss}^{\rm *, chain}\approx 1.69 N$,  $\tau W_{diss}^{\rm *, all} \approx 2.20 N^{0.857}$, while $\tau W^{\rm *, Star}_{diss} \leq 9\pi^2/4$. Single points are provided for 2-D and 3-D Pyramid models with few layers and an aperture of 8. (b-e) Depiction of the geometries of the interactions in \eref{eq:2B_hamiltonian} (equal colors/labels correspond to equal values of the local fields). (b) all-to-all model with $N=8$, (c) 1-D spin chain with $N=8$, (d) the Star model with $N=9$, (e) 2-D Pyramid model with $4$ layers and an aperture of $1$.}
	\vspace{-10pt}
	\label{fig}
\end{figure}

If we now allow for full control of the Hamiltonian, we can again use~\eref{fundamental_bound} to compute the minimal dissipation, but this time we use the global states $\hat\pi(0)= \frac{1}{2^N} \mathbbm{1}$ and $\hat\pi(\tau) \approx \ket{0}\!\!\bra{0}^{\otimes N}$ instead of the local ones. This leads to: 
\begin{equation}
	W_{diss}^{\rm *, global}  = 
	\frac{k_BT}{\tau} \left(2\arccos\!\left[\frac{1}{2^{N/2}}\right] \right)^2 
	= \frac{\pi^2}{\tau}k_B T + \mathcal{O}\!\left( 
	e^{-N/2} \right)~. 
	\label{boundNqubits}
\end{equation}
Therefore, an $N$-fold advantage can potentially be achieved by global processes, as illustrated in \fref{fig}. 

Let us now discuss the implications of this result for the reachability of Landauer's bound. From \eref{eq:Wdiss} we have $\Delta F_{eq} = N k_B T \ln 2$ whereas $W_{diss}$ can reach \eref{boundNqubits} at leading order in $\tau^{-1}$ (recall that our results are based on the slow driving assumption where the expansion~\eref{eq:GeneralExpansionDiss} is well justified). Hence, the work cost of erasure for each qubit can be written as:
\begin{equation}
	W_{\rm qubit}^* = k_B T\!\left(\ln 2 + \frac{\pi^2}{\tau N}\right) +\mathcal{O}\!\left(\tau^{-2}\right).
	\label{eq:ReachingLandauer}
\end{equation} 
Hence, in the thermodynamic limit $N\rightarrow \infty$, we can approach Landauer's bound with an error that scales as $\tau^{-2}$ instead of the standard $\tau^{-1}$ (cf. \cref{chap:landauer}).
We note that a link between complexity, as in higher level $k$-body interactions and faster information erasure has been suggested in Ref.~\cite{Taranto2023}. The optimal driving achieving the limit \eref{boundNqubits} can be computed from \eref{FC_driving}: 
\begin{equation}
	\label{eq:Nbody_prot}
	\beta \hat H (t)= \gamma(t)\sum_{j=1}^N (-1)^{j+1}\!\! \hspace{-4mm} \sum_{i_1<i_2<...<i_j}^N \hspace{-4mm} \hat x^{(i_1)} \hat x^{(i_2)} ... \hat x^{(i_j)}~,
\end{equation}
where $\hat x = \hat\sigma_+\hat\sigma_-$ and the control function can be written as 
\begin{equation}
\gamma(t) = 2\log\!\left[1+ 2^{N/2}\sin\!\left(\frac{\pi t}{2\tau}\right)\sin\!^{-1}\!\left(\frac{\pi(\tau-t)}{2\tau}\right)\right]~.
\end{equation}
It follows that highly non-local $N$-body interactions are required to saturate the bound~\eref{boundNqubits}. More specifically, one needs to activate every possible (classical) interaction present in the system. This makes reaching the fundamental bound \eref{boundNqubits} highly challenging in practice, and opens the question as to whether collective advantages beyond the local bound \eref{localbound} can be achieved via more realistic driven many-body systems featuring (local) few-body interactions. We address this relevant question in what follows. 

\section{Driven Many-Body Systems}
In this section, we constrain the total system to only feature at most 2-body interactions, to seek collective advantages in a more realistic model. Specifically, we consider a spin system with Hamiltonian of the form
\begin{equation}\label{eq:2B_hamiltonian}
	\hat H(t) = \sum_{i=1}^N\eps_i (t)\hat\sigma_z^{(i)} + \frac{1}{2}\sum_{i,j=1}^N J_{ij}(t) \hat\sigma_z^{(i)}\hat\sigma_z^{(j)}    
\end{equation}
We thus examine different degrees of control, reflected in the topologies represented in \fref{fig}: (i) an all-to-all spin model, (ii) a 1-D spin chain with nearest neighbor interaction (with periodic boundary conditions) (iii) a Star-shaped design, which we generalize to (iv) a multi-layer Pyramid scheme.
In practice, the energies $\eps_i(t)$ could be tuned via an external magnetic field whereas the interaction strength $J_{ij}(t)$ could be controlled by changing the distance between the spins, which are interacting via dipole-dipole coupling. Current quantum annealers have the capacity of tuning generic Hamiltonians of the form~\eref{eq:2B_hamiltonian}\footnote{See for example \href{https://www.dwavesys.com/}{D-Wave Systems}.}.

The all-to-all model corresponds to taking uniform magnetic fields and  spin interactions, i.e. $\eps_i(t)\equiv\eps(t)$ and $J_{ij}(t)\equiv J(t)$ in \eref{eq:2B_hamiltonian}. We can compute the partition function as follows 
\begin{equation}
	Z_{\rm all} = \sum_{k=0}^N \binom{N}{k} e^{-\beta E_k}~,
\end{equation}
where $E_k = \eps(2k-N) + \frac{1}{2}J(2k-N)^2$. The standard 1-D Ising model corresponds to uniform local terms $\eps$, and $J_{i,i+1}\equiv J$ for nearest neighbors and $0$ elsewhere. The partition function can be found by making use of the transfer matrix method:
\begin{equation}
	Z_{\rm chain} = z_+^N + z_-^N\;,
\end{equation}
where $z_\pm =e^{-\beta J/2} \cosh{\beta \eps}\pm\sqrt{e^{-\beta J} \sinh{\beta \eps} +e^{\beta J} }$.
Thirdly, we consider a Star topology corresponding to a central spin $\hat\sigma_z^{(1)}$ with local magnetic field $\eps_0(t)$ and uniform elsewhere $\eps_i(t)\equiv\eps_1(t)\; \forall\, i>1$, and uniform ``radial'' interaction $J_{1j}(t) = J_{j1}(t)\equiv J(t)$, and $0$ elsewhere. The partition function is easily computed as
\begin{equation}
	Z_{\rm Star} =e^{-\beta\eps}(2\cosh{\beta\lambda_+})^{N-1}+e^{\beta\eps}(2\cosh{\beta\lambda_-})^{N-1}\;,
	\label{eq:ZStar}
\end{equation}
where $\lambda_{\pm} =\eps_1\pm J$.

For the models above,  given the partition function, we compute the metric according to \eref{eq:metric_simple}, from which we can obtain the geodesic equations. Their solution provides us with the minimal length for given boundary conditions, from which we find the minimal dissipation. We implemented this procedure numerically for the task of approximate erasure, we take $\eps(0) = J(0) = J(\tau) = 0$ and $\eps(\tau) = 5k_BT$ which corresponds to an erasure process with an error of $4.5\cdot 10^{-5}$.

In \fref{fig} we present the resulting minimal dissipation for the different many-body models. The results are contrasted with the optimal non-interacting protocol \eref{localbound} and the fundamental bound obtained with full-control \eref{boundNqubits} (i.e. arbitrarily complex interactions). 

\begin{figure}[t]
	\centering
	\includegraphics[width=0.9\textwidth]{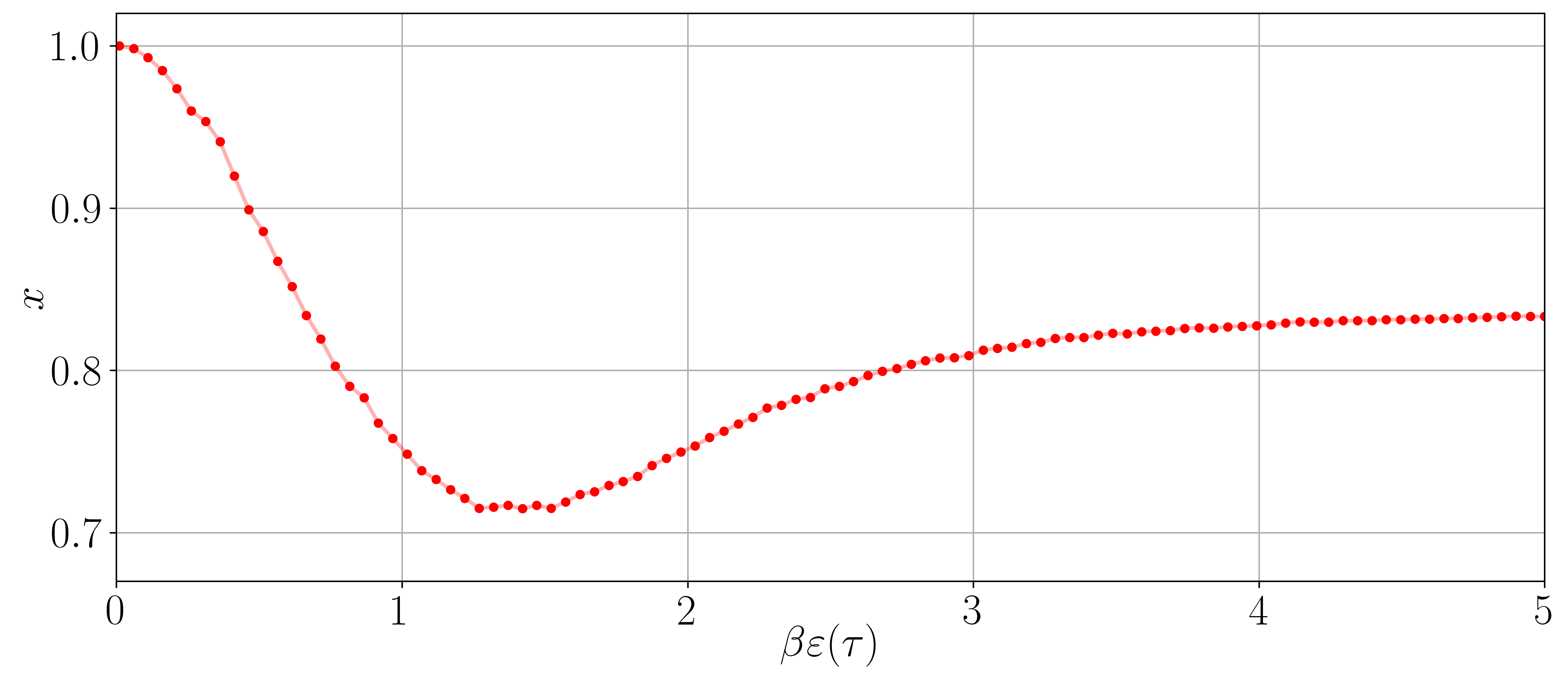}
	\vspace{-5pt}
	\caption{Dependence of the exponent $x$ of the dissipation $W_{diss}^{\rm *, all} = \alpha N^x$ on the boundary condition $\beta\eps(\tau)$ (with $\eps(0) = 0$). The fit is achieved for numerical data up to $N=150$.}
	\label{fig:growth}
\end{figure}
First, we observe that the nearest neighbor model displays a linear increase of the dissipation with $N$, but with a better pre-factor than the non-interacting case ($W_{diss}^{\rm *, chain}/W_{diss}^{\rm *, local} \approx 0.686$). On the other hand, the all-to-all model displays a sub-linear dependence on $N$: $W_{diss}^{\rm *, all} = \alpha N^x$ with $x\approx 6/7$. Furthermore, the exponent $x$ displays a non-trivial dependence on the specific boundary conditions, which can been seen in \fref{fig:growth}. Finally, quite remarkably, the Star model can achieve \emph{a finite value of the dissipation, independent of~$N$}. This feature is enabled by a 3-step protocol that suppresses specific terms in the otherwise-extensive $\log{Z_{\rm Star}}$. Interestingly, the Star model was found to be optimal in the context of two-body probes used for thermometry~\cite{Abiuso2024}.

The sub-linearity of the all-to-all's and Star-model's dissipation is remarkable as it allows for the same effect as in \eref{eq:ReachingLandauer}: it is possible to reach Landauer's bound in finite-time with an error that scales as $\tau^{-2}$ instead of $\tau^{-1}$ as one approaches the thermodynamic limit. However, both these models use long-range interactions between arbitrarily far spins as $N$ grows, and their scaling properties might thus be seen as inconsequential. However, we can generalize the Star model to a multi-layer structure, i.e. a Pyramid model (cf. \fref{fig}). By generalizing the Star protocol, it can be shown that (cf. Supp.~Mat.~of \cite{Rolandi2023Collective}) such model can achieve $W_{diss}^{\rm Pyr}\propto \ell^2$, where $\ell$ is the number of layers of the pyramid. Given that $N\propto \ell^D$ for pyramids in $D$ spatial dimensions, $W_{diss}^{\rm Pyr}\propto N^{2/D}$ follows asymptotically.

\section{Conclusions and Discussion}
\begin{table}[t]
	\centering
	\begin{tabular}{c|c|c|c|c}
		Model & 1D chain & All-to-All & \ Star \ & Pyramid\\
		\hline \hline
		Asymptotic $W_{diss}$ & $\mathcal{O}(N)$  & $\mathcal{O}(N^x)$  & $\mathcal{O}(1)$ & $\mathcal{O}(N^{2/D})$ \\  \hline 
		Short-range   & $\cmark$ & $\xmark$  & $\xmark$ & $\cmark$ \\ \hline 
	\end{tabular}
	\caption{All models studied in this work are based on two-body interactions (cf. \eref{eq:2B_hamiltonian}). The All-to-all and Star model feature long-range interactions that enable a sub-linear scaling of $W_{diss}$ , i.e. a \emph{collective advantage}. The Pyramid models can achieve such advantage in $D=3$ spatial dimensions using short-range interactions only.}
	\label{tab:models}
\end{table} 
In this chapter, we considered the task of minimizing dissipated work, $W_{diss}$, for an $N$-body system. We showed that, in contrast to~$\Delta F_{eq}$, $W_{diss}$ can grow sublinearly with $N$ by suitably creating interactions between the $N$ systems along the process. This leads to a finite-time reduction of dissipation induced by collective processes and has a clear potential for improving various thermodynamic tasks ranging from quantum/stochastic engines~\cite{Seifert2012,Myers2022} to the estimation of equilibrium free energy via non-equilibrium work measurements~\cite{Blaber2020Skewed}; or, as is shown here, for the erasure of information in finite time. There are several observations to be made about these collective processes. 

First, the derived collective processes are a genuine effect of finite-time thermodynamic protocols, which cannot be directly linked to other well-known collective phenomena such as entanglement, phase transitions, or superradiance. Indeed, (i) they do not require the presence of quantum correlations or coherence, but rather arise due to the interplay between interactions and dissipation to an external thermal environment; and (ii) they are process dependent -- i.e. depend on the whole driving protocol~$\hat H(t)$ -- unlike phase transitions which take place in a particular point in the parameter space.

Second, the results of this chapter suggest an interesting interplay between the complexity of the interactions and the associated reduction in dissipation, which is to be confronted with the results of Ref.~\cite{Taranto2023}. In particular, we argued that reaching the maximal advantage requires highly non-local $N$-body interactions. 
Despite this, we showed that similar reductions (in scaling)  can  be achieved with only two-body long-range interactions via the Star model. A sub-linear growth of $W_{diss}$ was found in the all-to-all model and, crucially, in the Pyramid model that only features short-range strong interactions. See \tref{tab:models} for a compact summary.

Third, being derived in the linear response regime, the dissipated work is directly related to the work fluctuations $\sigma_W^2$ via the work fluctuation-dissipation relation $\frac{\beta}{2} \sigma_W^2 = W_{diss}$. This  implies that the collective gains also lead to a reduction of work fluctuations, a desired property  in stochastic thermodynamics. 

Finally, it is important to stress that our results have been derived in the slow driving regime, i.e., for the leading order contribution of $W_{diss}$ in $\tau^{-1}$. For a finite (large) time $\tau$, the next order contributions of $\mathcal{O}(\tau^{-2})$ can become relevant when increasing $N$. An exciting future endeavor is to generalize such collective advantages for arbitrary non-equilibrium protocols. For this, it might be useful to exploit recent results on minimal dissipation and the Wasserstein distance~\cite{dechant2019thermodynamic} as well as new tools such as reinforcement learning~\cite{Erdman2023Pareto} or fast-driving expansions~\cite{Rolandi2023Fast} for finding optimal protocols.

Another future challenge is to understand how the collective advantages are modified beyond the simple model of thermalization used in \eref{eq:metric_simple} and by adding constraints on the strength of the couplings in \eref{eq:2B_hamiltonian}. In particular, whether such advantages can still be found for more realistic thermalization models where the relaxation timescale(s) is modified in the presence of interactions, which can lead to a critical slowdown of relaxation.
	\cleardoublepage
	\chapter{Thermodynamic Optimization}\label{chap:opt}
\epigraph{Give me a lever long enough and a fulcrum on which to place it, and I shall move the world.}{--- Archimedes}


The study of how to optimize the energetic cost of a physical task goes back to the very origin of Thermodynamics with Carnot optimizing the efficiency of thermal steam engines~\cite{Carnot1824}. This is a question that has a dual scope, on the one hand it can applied for very practical purposes such as optimizing car engines~\cite{Fermi56} or minimizing the overheating of the chips in super-computers~\cite{McMahon23}, while on the other hand it can applied to probe the most fundamental aspects of the laws of thermodynamics -- e.g. finding the fundamental limit between a finite-resource process and a reversible process~\cite{Taranto2023}.

Generally, optimizing a thermodynamic operation that brings the Hamiltonian of a system from $\hat H_i$ to $\hat H_f$ consists in finding the optimal time variation $\hat H(t)$ that minimizes the energy dissipated into the environment. This optimization can be done with varying degrees of constraints to achieve more applied or more fundamental lower bounds on dissipation. These bounds can then also be used as speed limits \cite{Das21} for performing the described operations, which is of particular interest for classical and quantum computers. However, it is always a very challenging problem as it requires functional optimization and boundary value problems to be solved \cite{Fox87} over solutions of the time-dependent Schrodinger (or Lindblad) equation. There are a variety of methods to tackle this problem -- analytical as well as numerical -- these range from optimal transport \cite{Aurell2011} and optimal control theory \cite{Esposito2010,cavina2018,erdman2019_njp} to geometric methods (cf. \sref{sec:geom} and \cite{Rolandi2023Quantum}). Because of the fundamental difficulty of this problem, many open questions remain that range from direct applications to more general and fundamental ones.

It is important to note that the minimal amount of work $W$ can be negative, thus resulting in a gain of energy for the agent. Therefore, in the proper setting, one can find thermodynamic protocols such that work is extracted from an environment that is \emph{out of equilibrium} (cf. \sref{sec:second_law}). Typically an out-of-equilibrium-environment is modeled as a collection of thermal baths at different temperatures -- often two of those. A \emph{thermal engine} is a device that implements periodically a work extracting protocol, so that it can convert a flow of heat between two (or more) thermal baths into ``useful energy'' (e.g. kinetic energy).\\
A relatively simple example to showcase how an engine can exploit a difference in temperature between two thermals baths to extract work is a periodically driven quantum dot engine. Since the quantum dot has only two energy levels, we can choose the Hamiltonian so that the lower energy state has zero energy and the higher energy state has energy $\eps$. Let us now consider the following process: we first let the dot thermalize with the colder bath, then we pay some energy to increase the energy gap $\eps$. At this point, we let the dot thermalize with the warmer bath and then gain some energy by taking $\eps$ back to its initial value. The main thing to
note, to see why one could gain energy in such a process,
is that one only pays or gains energy when the excited
state is occupied, as can be see from the integral form of work \eref{eq:work}. Since the probability of being in the excited state is lower when the dot is thermalized with the colder bath compared to when it's thermalized with the warmer bath, it is more likely to gain energy from this process than it is to pay energy. Therefore, on average, energy flows from the warmer bath to the cold bath and the agent.\\
The performance of an engine can be quantified in different ways. One very natural metric to do so is the average amount of energy extracted per unit time: power, which is defined as 
\begin{equation}\label{eq:power}
	\<P\> := -\frac{W}{\tau}~,
\end{equation}
where here we denote by $W$ the work cost of a cycle and by $\tau$ the length of the cycle. However, when we consider that some of the energy from the warm bath ends up in the cold bath during a cycle, it can also be natural to consider how efficient the engine is
\begin{equation}\label{eq:efficiency}
	\eta := \frac{W}{Q_h}~,
\end{equation}
where $Q_h$ denotes the energy \emph{gained} by the hot bath -- the sign choice of $Q_h$ might be making this definition less intuitive, but it was made so to be consistent with the first law as it is formulated in \eref{eq:first_law}. The optimization of these quantities is a topic which has been a widely studied in the literature~\cite{geva1992,feldmann2000,Ye2022}. However it is seldom the case that they can be optimized simultaneously. Finally, the systems we will be dealing with are stochastic in nature. As opposed to engines at the macroscopic scale, one cannot expect to extract the same amount of work at each cycle, therefore we will be quantifying the constancy -- or more precisely, the lack thereof -- with the power fluctuations $\<\Delta P\> := \sigma^2_W/\tau$, where $\sigma^2_W$ is the variance of work~\cite{Miller2019}
\begin{equation}\label{eq:var_work}
	\sigma_W^2:=2\, \Re\int^\tau_0 \! dt \int^t_0\! dt'~ \tr{\hat H'(t)G(t,t')\!\left[\Delta_{\hat \rho(t')}\hat H'(t') \hat\rho(t')\right]}~,
\end{equation}
for $\Delta_{\hat\rho}\hat A:=\hat A-\Tr[\hat A\hat \rho]$ and $G$ the propagator of the evolution (cf. \sref{sec:evolution}).

In this chapter, we will be looking at three examples of thermodynamic optimization: for a quantum dot engine \cite{Erdman2023Pareto}, for an information engine (a.k.a. Szilard engine), and finally for fast thermodynamic operations \cite{Rolandi2023Fast}.

\section{Quantum Dot Heat Engine}
Stochastic heat engines are devices that convert a heat flow into work at the nano-scale~\cite{giazotto2006}. We can distinguish between two classes of such engines: steady-state heat engines (SSHE) perform work against external thermodynamic forces (e.g. a chemical potential difference) after reaching a non-equilibrium steady state~\cite{Benenti2017}, while periodically driven heat engines (PDHE) perform work against external driving fields through time-dependent cycles.
Earlier works have started optimizing the power fluctuation of these engines~\cite{Holubec2014,Mehboudi2022}. However, a framework to fully optimize the performance of microscopic heat engines that accounts power, efficiency and power fluctuations is currently lacking.
In fact, an ideal engine operates at high power, high efficiency, and low power fluctuations; however, such quantities usually cannot be optimized simultaneously, but one must seek trade-offs. In SSHEs, a rigorous manifestation of this trade-off is given by thermodynamic uncertainty relations~\cite{barato2015,pietzonka2018}. For ``classical'' stochastic SSHE (i.e. in the absence of quantum coherence) operating between two thermal baths at inverse
temperatures $\beta_\text{C}$ (cold) and $\beta_\text{H}$ (hot), they read~\cite{pietzonka2018}:
\begin{equation}
	\xi \equiv \frac{2}{\beta_\text{C}} \frac{\< P\>}{\< \Delta P\> }\frac{\eta}{\eta_\text{c}-\eta} \leq 1,
	\label{eq:tur_ineq}
\end{equation}
where $\eta_\text{c}\equiv 1-\beta_\text{H}/\beta_\text{C}$ is the Carnot efficiency.
Such thermodynamic uncertainty relations imply, for example, that high efficiency can only be attained at the expense of low power or high power fluctuations. 
The thermodynamic uncertainty relation inequality \eref{eq:tur_ineq} can be violated with quantum coherence~\cite{Agarwalla2018} and in PDHEs~\cite{barato2016}.
This  has motivated various generalized thermodynamic uncertainty relations~\cite{Koyuk2018}, in particular for time-symmetric driving~\cite{proesmans2017} and  slowly driven stochastic engines~\cite{miller2021c}. Despite their importance, thermodynamic uncertainty relations provide an incomplete picture of the trade-off: while high values of $\xi$ may appear more favorable, this does not give us any information on the individual objectives.

In this section, we present a framework to optimize any trade-off between power, efficiency, and power fluctuations in arbitrary PDHE described by Lindblad dynamics~\cite{lindblad1976,breuer2002}. This allows us to find Pareto-optimal cycles, i.e. those cycles where no objective can be further improved without sacrificing another one.
We then show how reinforcement learning (RL) can be used to fully optimize a quantum dot (QD) engine~\cite{sothmann2015}. 
We characterize the Pareto front, i.e. the set of values $\{\< P\>,\< \Delta P\>, \eta\}$ corresponding to Pareto-optimal cycles, and  evaluate the thermodynamic uncertainty relation ratio $\xi$ on such optimal cycles. Furthermore,
we derive analytical results for the Pareto front and $\xi$ in the fast and slow driving regimes, i.e. when the period of the cycle is respectively much shorter or much longer than the thermalization timescale of the system.

\subsection{Multi-Objective Optimization of Heat Engines}
\begin{figure}[t]
	\centering
	\includegraphics[width=\textwidth]{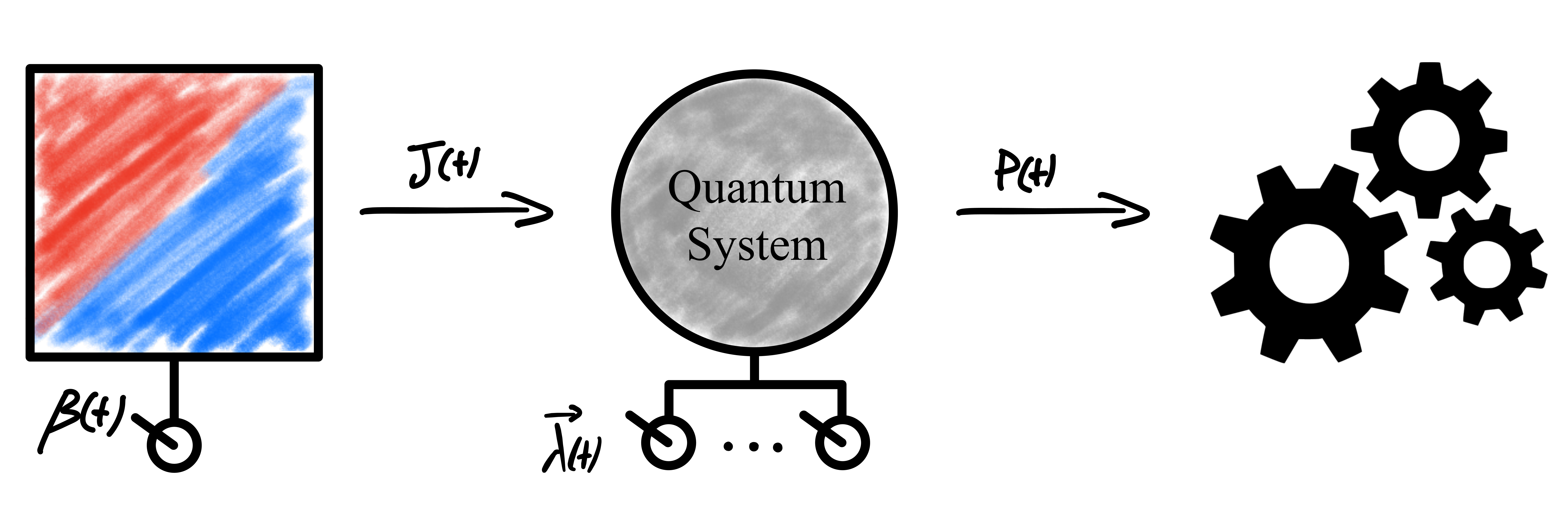}
	\caption{A quantum system (gray circle) is coupled to a thermal bath (left box) characterized by a controllable inverse temperature $\beta(t)$. The coupling produces a heat flux $J(t)$. Control parameters $\lambda(t)$ allow us to control the state of the system and the power $P(t)$ extracted from the system.}
	\label{fig:setup}
\end{figure}
In this case we will study an engine featuring a single thermal bath with tunable inverse temperature $\beta(t)$ between two extremal values $\beta_\text{H}$ and $\beta_\text{C}$, which can be regarded as having a large selection of baths to choose from -- thus allowing for further optimization. An illustration is presented in \fref{fig:setup}. The coupling between system and bath produces a heat flux $J(t)$ from the bath to the quantum system, while the system itself is controlled by the parameters $\vec{\lambda}(t)$ that allow exchanging work with the system. Therefore, in this framework, a thermodynamic cycle is described by periodic functions $\beta(t)$ and $\vec{\lambda}(t)$. This framework includes standard PDHEs, in which the system is sequentially put in contact with two baths (by abruptly changing the values of $\beta(t)$) and cases where $\beta(t)$ varies smoothly in time.
We assume that the dynamics of the system are described by a Markovian master equation (cf. \sref{sec:evolution}), and we consider PDHEs in the asymptotic limit cycle -- i.e. in the limit of infinite repetitions of the cycle. Because of this assumption, we can express that the time average $\ev{O}$ of an arbitrary quantity $O(t)$ as
\begin{equation}
	\ev{O} = \frac{1}{\tau}\int_0^\tau\!dt~O(t)~,
\end{equation}
where $\tau$ is the period of the cycle. We can therefore compute the average power $\ev{P}$, power fluctuations $\ev{\Delta P}$ and entropy production $\ev{\Sigma}$ by averaging
\begin{align}\label{eq:p_dp}
	P(t) &= -\tr{\hat\rho(t) \hat H'(t)}~,\\
	\Delta P(t) &= \tr{\hat s(t) \hat H'(t)},
	 \\
	\Sigma(t) &=  - \beta(t)\tr{\hat \rho'(t) \hat H(t)} = -J(t)\beta(t)~. 
	\label{eq:sigma}
\end{align}
Here, $\hat{\rho}(t)$ and $\hat H(t)$ are respectively the reduced density matrix and the local Hamiltonian of the quantum system, while $\hat s(t)$ is an auxiliary Hermitian operator satisfying $\Tr[\hat s(t)]=0$ that we introduce to compute the power fluctuations.
These satisfy
\begin{equation}
	\begin{aligned}
		\frac{d\hat\rho(t)}{dt} &= \mathcal{L}_{\vec{\lambda}(t),\beta(t)}[\hat \rho(t)], \\ 
		\frac{d\hat s(t)}{dt} &= \mathcal{L}_{\vec{\lambda}(t),\beta(t)}[\hat s(t)] + \{ \hat\rho(t), \hat H'(t)\} - 2\Tr[\hat\rho(t)\hat H'(t)]\hat \rho(t),
		\label{eq:ode}
	\end{aligned}
\end{equation}
where $\mathcal{L}_{\vec{\lambda}(t),\beta(t)}$ is the Lindbladian describing the evolution of the system and
$\{\cdot,\cdot\}$ is the anti-commutator.
In the asymptotic limit cycle, it can be shown that both $\hat\rho(t)$ and $\hat s(t)$ are periodic with the same period as the control, and can thus be determined by solving \eref{eq:ode} with periodic boundary conditions. While the expressions of $\ev{P}$ and $\ev{\Sigma}$ are standard in the limit cycle \cite{Alicki1979}, the expression for $\ev{\Delta P}$ in terms of $\hat s(t)$ is less common. But we use it here as it is crucial to efficiently minimize the power fluctuations of the non-equilibrium engine using RL. 

To identify Pareto-optimal cycles
we introduce the dimensionless figure of merit
\begin{equation}
	\label{eq:F_def}
	\ev{F} = a \frac{\ev{P}}{P_\text{max}} - b\frac{\ev{\Delta P}}{\Delta P(P_\text{max})} - c \frac{\ev{\Sigma}}{\Sigma(P_\text{max})},
\end{equation}
where $a,b,c\geq 0$ are three scalar weights, satisfying $a+b+c=1$, that determine how much we are interested in each of the three objectives, and  $P_\text{max}$, $\Delta P (P_\text{max})$ and $\Sigma(P_\text{max})$ are respectively the average power, fluctuations and entropy production of the cycle that maximizes the power.
Notice that, given the relation between entropy production and efficiency, cycles that are Pareto-optimal for $\{\ev{P},\ev{\Delta P},\eta\}$, are also Pareto-optimal for $\{\ev{P},\ev{\Delta P},\ev{\Sigma}\}$. 
The positive sign in front of $\ev{{P}}$ in \eref{eq:F_def} ensures that we are maximizing the power, while the negative sign in front of $\ev{\Delta {P}}$ and $\ev{\Sigma}$ ensures that we are minimizing power fluctuations and the entropy production. Pareto-optimal trade-offs are then found maximizing $\ev{F}$ for various choices of $a$, $b$ and $c$.

\subsection{Optimal Quantum Dot Heat Engine}
In the following, we compute Pareto-optimal cycles in a minimal heat engine consisting of a two-level system coupled to a Fermionic bath with flat density of states. This represents a model of a single-level QD 
\cite{geva1992}.
The Hamiltonian reads
\begin{equation}
	\hat H(t) = \lambda(t) \frac{\eps}{2} \hat\sigma_z,
\end{equation}
where $\lambda(t)$ is our single control parameter, $\eps$ is a fixed energy scale and $\hat\sigma_z$ is the z-Pauli matrix. Denoting with $\ket{1}$ the excited state of $\hat H(t)$, and defining $p(t):= \braket{1|\hat\rho(t)|1}$ as the probability of being in the excited state, the Lindblad equation \eref{eq:ode} becomes $p'(t) = -\gamma(p(t) - f({\eps\lambda(t)\beta(t)}))$,
where $\gamma^{-1}$ is the thermalization timescale arising from the coupling between system and bath, and $f(x) = (1+ e^x)^{-1}$ is the excited level population of the instantaneous Gibbs state.\\

We optimize $\< F\>$ of the QD heat engine using three different tools: RL, analytics in the fast-driving regime, and analytics in the slow-driving regime. 
The RL-based method allows us to numerically optimize $\< F\>$ without making any approximations on the dynamics, exploring all possible (time-discretized) time dependent controls $\beta(t)$ and $\lambda(t)$ subject to the constraints $\beta(t)\in[\beta_\text{H},\beta_\text{C}]$ and $\lambda(t)\in [\lambda_\text{min},\lambda_\text{max}]$, and identifying automatically also the optimal period $\tau$. The RL method, based on the soft actor-critic algorithm~\cite{haarnoja2018_pmlr} and generalized from \cite{Erdman2022,Erdman2023}, additionally includes the crucial impact of power fluctuations, and identifies Pareto-optimal cycles.
Machine learning methods have been employed for other quantum thermodynamic~\cite{Sgroi2021} and quantum control~\cite{bukov2018} tasks.

The fast-driving regime assumes that $\tau \ll \gamma^{-1}$.
Interestingly, without any assumption on the driving speed, we show that any trade-off between power and entropy production ($b=0$ in \eref{eq:F_def}) in the QD engine is maximized by Otto cycles in the fast-driving regime, i.e. switching between two values of $\beta(t)$ and $\lambda(t)$ ``as fast as possible''~\cite{cavina2021}. We thus expect such ``fast-Otto cycles'' to be nearly optimal in the high power regime. 

The slow-driving regime corresponds to the opposite limit,  i.e. $\tau \gg \gamma^{-1}$.  Since entropy production and power fluctuations can be minimized by considering quasi-static cycles, we expect this regime to be nearly optimal in the high efficiency and low fluctuations regime, i.e. for low values of $a$ in \eref{eq:F_def}. 
To make analytical progress in this regime, we maximize \eref{eq:F_def} assuming a finite-time Carnot cycle. The obtained results naturally generalize previous considerations for low-dissipation engines~\cite{esposito2010efficiency,Holubec2016,Abiuso2020} to account for the role of fluctuations. The main technical tool  is the geometric concept of ``thermodynamic length''\cite{Salamon1983,Abiuso2020Geometric} which yields the first order correction in $(\gamma\tau)^{-1}$ from the quasi-static limit.

\begin{figure}[!tb]
	\centering
	\includegraphics[width=\textwidth]{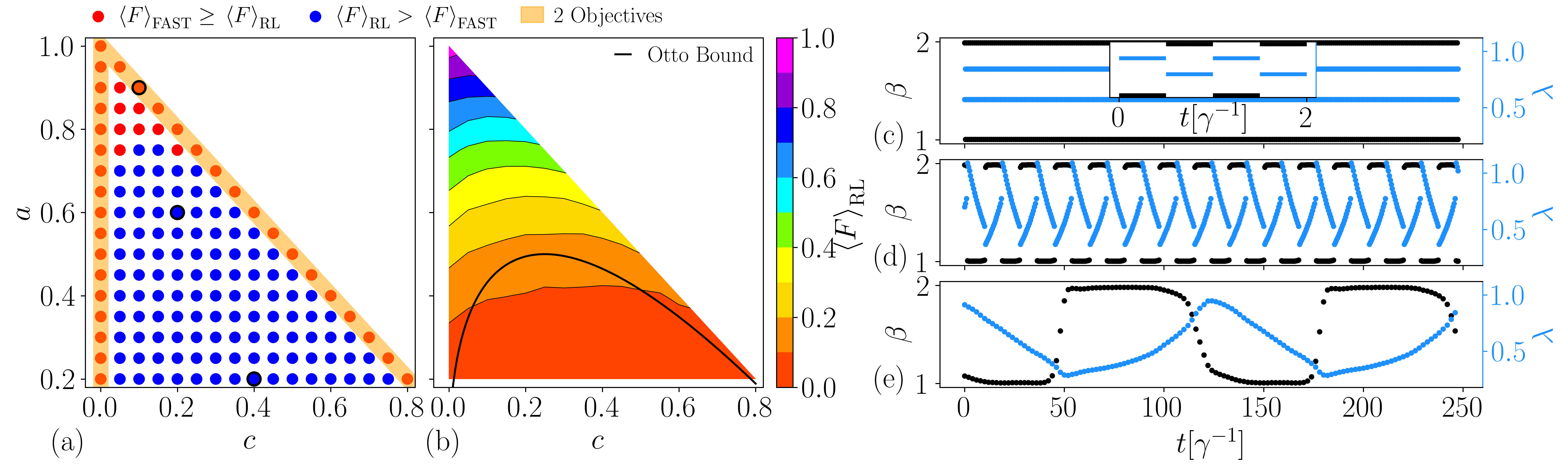}
	\caption{Optimization of $\ev{F}$ at different values of $a$ and $c$, with $b=1-a-c$, for a QD-based PDHE. Each dot in panel (a) displays, as a function of $c$ and $a$, whether $\ev{F}_\text{RL}>\ev{F}_\text{FAST}$ (blue dots) or not (red dots). Points with $a\sim 0$ are not displayed since, in such a regime, optimal cycles become infinitely long (to minimize entropy production and fluctuations) and the RL method does not converge reliably. (b): contour plot of $\ev{F}_\text{RL}$, as a function of $c$ and $a$, using the data-points of (a). The black line represents the curve below which $\ev{F}_\text{FAST} = 0$. (c,d,e): cycles, described by piece-wise constant values of $\beta$ (black dots) and $\lambda$ (blue dots) as a function of $t$, identified at the three values of $a$ and $c$ highlighted in black in panel (a) (respectively from top to bottom). The inset in panel (c) represents a zoom into the corresponding cycle, which is a fast-Otto cycle. Parameters: $\beta_\text{C} = 2$, $\beta_\text{H}=1$, $\lambda_\text{min}=0.2$,
		$\lambda_\text{max}=1.1$ and $\eps=2.5$.}
	\label{fig:fig_merit}
\end{figure}
We now present the results. Each point in \sfref{fig:fig_merit}{a} corresponds to a separate optimization of $\ev{F}$ with weights $c$ and $a$ displayed on the x-y axis. Since $b=1-a-c$, points lying on the sides of the triangle (highlighted in yellow) correspond to optimizing the trade-off between 2 objectives, whereas points inside the triangle take all 3 objectives into account. 
Denoting the figure of merit optimized with RL and with fast-Otto cycles with $\ev{F}_\text{RL}$ and $\ev{F}_\text{FAST}$,
in \sfref{fig:fig_merit}{a} we show blue (red) dots  when $\ev{F}_\text{RL} > \ev{F}_\text{FAST}$ ($\ev{F}_\text{RL} \leq \ev{F}_\text{FAST}$), while \sfref{fig:fig_merit}{b} is a contour plot of $\ev{F}_\text{RL}$. As expected, there are red dots when $b=0$ (along the hypotenuse), but it turns out that fast-Otto cycles are optimal also when~$c=0$. However, as soon as all 3 weights are finite, the optimal cycles identified with RL change abruptly and outperform fast-Otto cycles. 
Furthermore, we notice that while $\ev{F}_\text{RL}$ is positive for all values of the weights, $\ev{F}_\text{FAST}=0$ below the black curve shown in \sfref{fig:fig_merit}{b} for its analytic expression. 

To visualize the changes in protocol space, in \sfref{fig:fig_merit}{c,d,e} we show the cycles identified with RL at the three different values of the weights highlighted by a black circle in \sfref{fig:fig_merit}{a} (respectively from top to bottom). Since RL identifies piece-wise constant controls, the cycle is displayed as dots corresponding to the value of $\beta(t)$ (black dots) and $\lambda(t)$ (blue dots) at each small time-step. First, we notice that the inverse temperature abruptly switches between $\beta_\text{H}$ and $\beta_\text{C}$ for all values of the weights, so that in this engine no gain arises when smoothly varying the temperature.  
As expected, 
the cycle identified by RL in 
\sfref{fig:fig_merit}{c}, corresponding to the black point on the hypotenuse in \sfref{fig:fig_merit}{a},
is a fast-Otto cycle (a ``zoom'' in a short time interval is shown in the inset). However, moving down in weight space to the black dot at $a=0.6$ and $c=0.2$, we see that the corresponding cycle (\sfref{fig:fig_merit}{d}) now displays a finite period, with  linear modulations of $\lambda(t)$ at fixed temperatures, and a discontinuity of $\lambda(t)$ when switching between $\beta_\text{H}$ and $\beta_\text{C}$. The cycle in \sfref{fig:fig_merit}{e}, corresponding to the lowest black dot at $a=0.2$ and $c=0.4$, displays an extremely long period $\tau \approx 125 \gamma^{-1}$, which is far in the slow-driving regime. 
Optimal cycles, therefore, interpolate between the fast and the slow-driving regimes as we move in weight space (cf. \sfref{fig:fig_merit}{a}) from the sides to the lower and central region -- i.e. switching from 2 to 3 objectives. 

\begin{figure}[t]
	\centering
	\includegraphics[width=\textwidth]{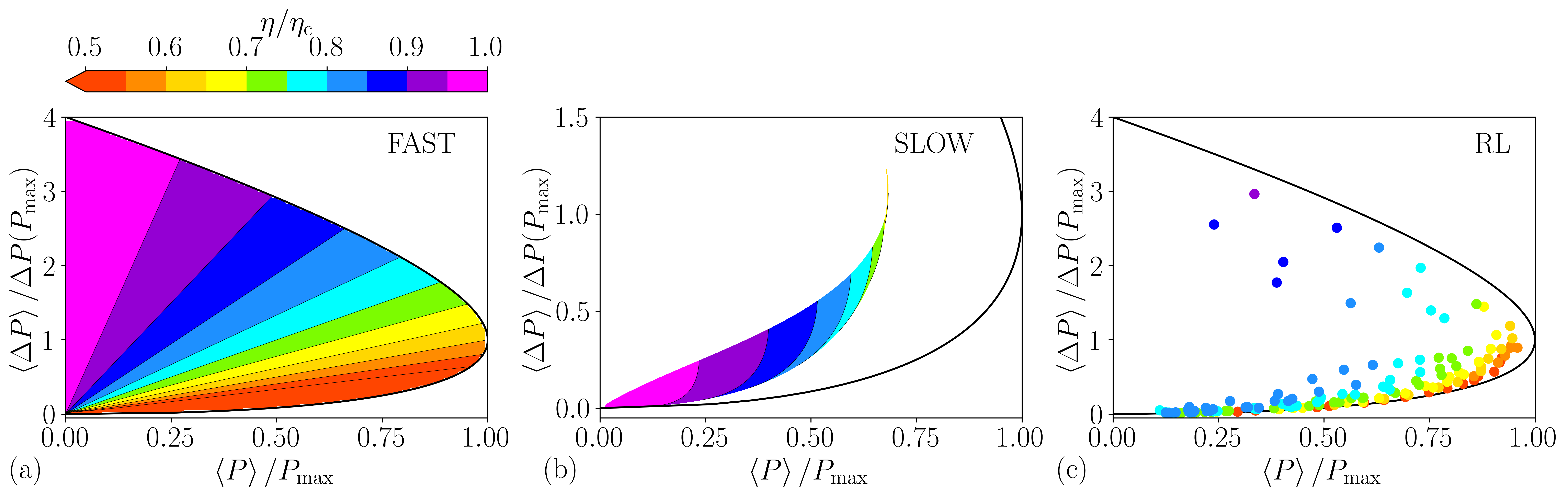}
	\caption{Pareto-front found optimizing $\ev{F}$ with fast-Otto cycles in the limit of small temperature differences (panel (a)), optimizing $\ev{F}$ in the slow-driving regime (panel (b)), and numerically using RL (panel (c)). The system parameters are as in \fref{fig:fig_merit}. All panel display $\ev{\Delta {P}}/\Delta P(P_\text{max})$ as a function of $\ev{P}/P_\text{max}$ (x-axis) and of $\eta/\eta_\text{c}$ (color). The black curve represents the outer border of the Pareto-front derived analytically.}
	\label{fig:full_pareto}
\end{figure}
In \fref{fig:full_pareto} we display the Pareto-front, i.e. we plot the value of ${P}/P_\text{max}$, $\eta/\eta_\text{c}$, and $\Delta {P}/\Delta P(P_\text{max})$ found maximizing $\ev{F}$ for various values of the weights. \sfref{fig:full_pareto}{a} is derived in the fast-driving regime assuming a small temperature difference, while \sfref{fig:full_pareto}{b} is derived in the slow-driving regime. The RL results, shown in \sfref{fig:full_pareto}{c}, correspond to the points in \sfref{fig:fig_merit}{a}.
First, we notice that, by definition of the Pareto front, the ``outer border'' corresponds to points where we only maximize the trade-off between the two objectives $\< P\>$ and $\ev{\Delta {P}}$. Since these points are optimized by fast-Otto cycles, the black border of \sfref{fig:full_pareto}{a}, also shown in \sfref{fig:full_pareto}{b,c}, is exact.
Moreover, in this setup, we can establish an exact mapping between the performance of a SSHE and of our PDHE operated with fast-Otto cycles. Since SSHE satisfy \eref{eq:tur_ineq}, also fast-Otto cycles have $\xi\leq 1$.
Furthermore, for small temperature differences,~$\xi=1$. This allows us to fully determine the internal part of the Pareto front in the fast-driving regime using the thermodynamic uncertainty relations, i.e. ${P}/P_\text{max} = (\Delta {P}/\Delta P (P_\text{max}))(\eta_\text{c}-\eta)/\eta$. Indeed, the linear contour lines in \sfref{fig:full_pareto}{a} stem from the linearity between ${P}$ and $\Delta {P}$, the angular coefficient being determined by the efficiency. 

Comparing \sfref{fig:full_pareto}{a,b}, we see where the fast and slow-driving regimes are optimal. 
As expected, the slow-driving Pareto front cannot reach the black border, especially in the high-power area, where fast-Otto cycles are optimal. However, in the low power and low fluctuation regime, cycles in the slow-driving substantially outperform fast-Otto cycles by delivering a higher efficiency (pink and purple regions in \sfref{fig:full_pareto}{b}).

Interestingly, the RL points in \sfref{fig:full_pareto}{c} capture the best features of both regimes. RL can describe the high-power and low fluctuation regime displaying both red and blue/green dots near the lower border. The red dots are fast-Otto cycles that are optimal exactly along the border but deliver a low efficiency. The blue/green dots instead are finite-time cycles that deliver a much higher efficiency by sacrificing a very small amount of power and fluctuations. This dramatic enhancement of the efficiency as we depart from the lower border is another signature of the abrupt change in optimal cycles.

\subsection{Violation of Thermodynamic Uncertainty Relation}
\begin{figure}[tb!]
	\centering
	\includegraphics[width=0.9\textwidth]{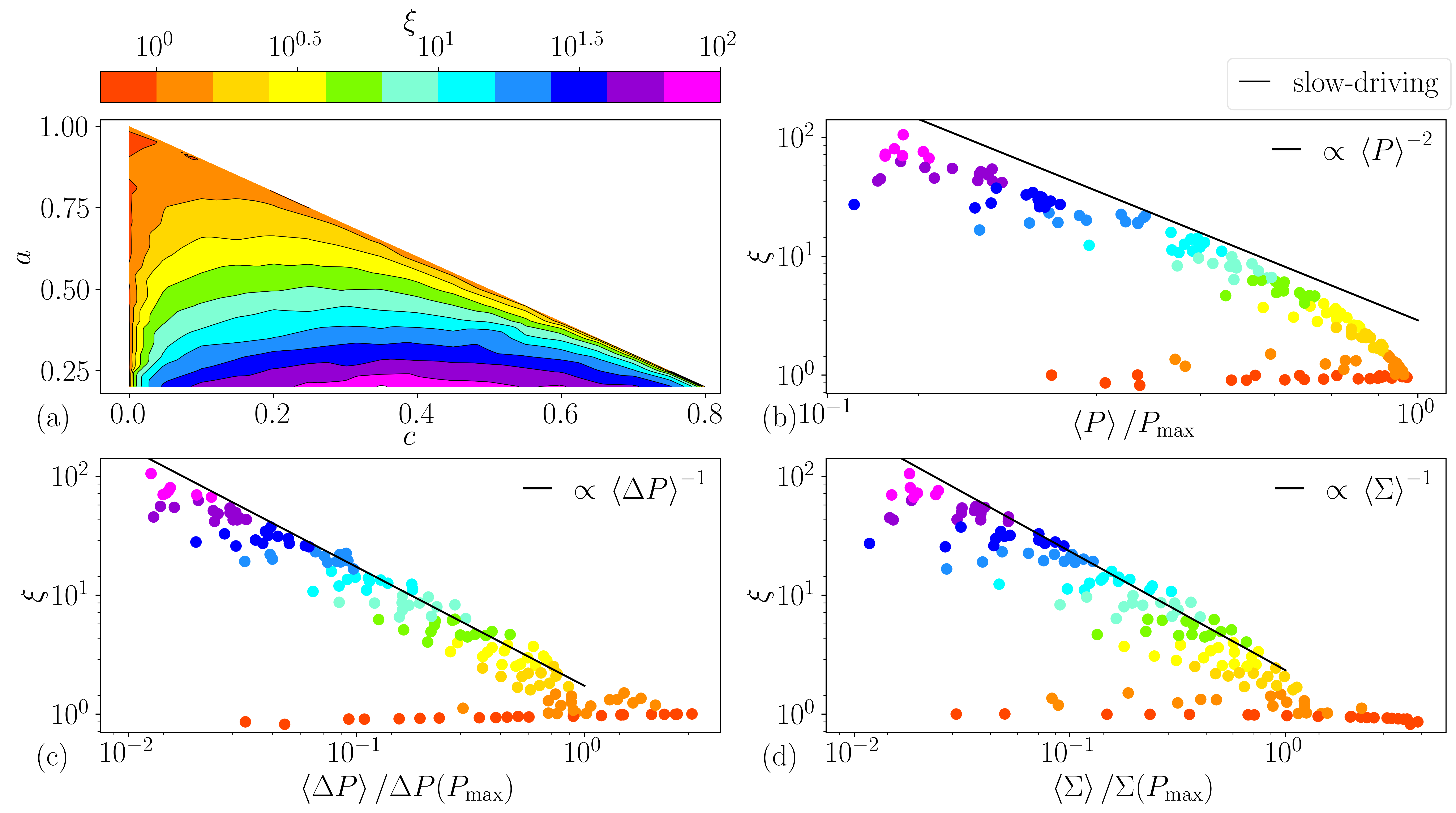}
	\caption{(a): contour plot of the SSHE thermodynamic uncertainty relationship ratio $\xi$ as a function of $c$ and $a$. (b,c,d): log-log plot of $\xi$, color mapped as in panel (a), as a function of $\ev{{P}}/P_\text{max}$, $\ev{\Delta {P}}/\Delta P( P_\text{max})$ and $\ev{\Sigma}/\Sigma(P_\text{max})$, respectively. 
		Every point corresponds to the same RL optimization performed in \fref{fig:fig_merit}. The black line is the behavior of $\xi$ derived analytically in the slow-driving regime for small values of $\ev{P}$, $\ev{\Delta P}$ and $\ev{\Sigma}$.}
	\label{fig:tur}
\end{figure}
At last, we analyze the behavior of the thermodynamic uncertainty relation ratio $\xi$ (cf. \eref{eq:tur_ineq}), which represents a relevant quantity combining the three objectives, computing it on Pareto-optimal cycles (recall that $\xi\leq 1$ for classical stochastic SSHE but  PDHE can violate this bound~\cite{barato2016}).
In \sfref{fig:tur}{a} we show a contour plot of $\xi$, computed with RL, as a function of~$a$ and~$c$. Because of the mapping between SSHE and fast-Otto cycles, we have $\xi = 1$ along the sides of the triangle, where only 2 objectives are optimized. However, this mapping breaks down for finite-time cycles, allowing us to observe a strong increase of $\xi$ 
in the green/purple region in \sfref{fig:tur}{a}. As shown in \fref{fig:fig_merit}, this region corresponds to long cycles operated in the slow-driving regime, where violations of thermodynamic uncertainty relations had already been reported. In \sfref{fig:tur}{b,c,d} we show a log-log plot of $\xi$ respectively as a function of $P/P_\text{max}$, $\Delta {P}/\Delta P(P_\text{max})$, and ${\Sigma}/\Sigma(P_\text{max})$ with the same color-map as in \sfref{fig:tur}{a}. We see that $\xi$ diverges in the limit of low power, low fluctuations, and low entropy production as a power law. 
Indeed, using the slow-driving approximation, we analytically prove that $\xi$ diverges as $\ev{P}^{-2}$, $\ev{\Delta {P}}^{-1}$, and $\ev{\Sigma}^{-1}$. Such relations, plotted as black lines, nicely agree with our RL results.


\section{Szilard Engine}
As opposed to the typical heat engine, \emph{information engines} function with a single thermal bath that is at equilibrium. Initially this might seem as if it would be violating a formulation of the second law of thermodynamics (cf. \sref{sec:second_law}), but no such violation is occurring. Information engines function by making use of a measurement instead of a cold bath~\cite{Bettmann2023}. The fact that this is possible becomes less surprising once one considers that there is a tight thermodynamic link between perfect measurements and zero-temperature baths~\cite{Taranto2023}.

The reason why there is no violation of the second law is that information is physical and, as a consequence, its manipulation has a thermodynamic cost~\cite{Landauer1991}. Likewise, information is a thermodynamic resource that can be exploited to generate  work, e.g., by means of information engines -- also known as \emph{Szilard engines}. These deep links between information and thermodynamics go back to seminal ideas by Maxwell, Szilard and Landauer~\cite{Landauer1961,Szilard1964}.  Their once thought-experiments can nowadays be realized in a variety of physical platforms driven by the development of the fields of stochastic and quantum thermodynamics~\cite{Ciliberto2017,Myers2022}. Pioneering  experimental works linking information and thermodynamics realized an information Szilard engine~\cite{Toyabe2010} and the erasure of information close to the Landauer limit~\cite{Brut2012} on single  colloidal particle, which were followed by several demonstrations in Brownian colloidal particles~\cite{Paneru2018} but also in quantum systems like single-electron quantum dots~\cite{Koski2014}, ultra-cold atoms~\cite{Kumar2018}, NMR~\cite{Camati2016} and superconducting circuits~\cite{Cottet2017}.

The steps in the cycle of a Szilard quantum dot engine can be broken down as follows: 0) The effective energy gap starts at $0$ and the system is always in contact with the bath. 1) Measure the occupation of the dot with the charge sensor. 2) If the outcome of the measurement at step 1 is that the dot in the state $\ket{0}$: quickly increase $\eps$ to a large value, if instead the outcome is $\ket{1}$ quickly decrease $\eps$ to a large negative value. 3) Decrease the energy gap in some finite time $\tau$ until it reaches $0$.\\
The protocol is designed to use the information of the measurement so that we are always in a position of gaining energy during step 3. However, by Landauer's principle, at the measurement in step 1 there is an implicit cost of $k_B T\ln 2$ that will be payed when the memory storing the result is erased.

Fully exploiting the thermodynamic content in information typically requires reversible, and hence infinitesimally slow, processes. This naturally raises the question of how information engines can be optimally driven in finite time and hence out-of-equilibrium conditions. This question has been extensively studied theoretically, including general optimization frameworks~\cite{Abiuso2020} as well as finite-time versions of the Landauer's principle~\cite{Rolandi2023Quantum}. At the experimental level, relevant progress has been achieved in the optimization of classical information engines based on optically trapped colloidal particles~\cite{Paneru2018PRE} whereas, for  quantum systems, a recent work exploited the concept of thermodynamic length to perform optimal Landauer erasure in the slow driving limit~\cite{Scandi2022}. 

The goal of this section is optimize -- similarly to the previous section -- and implement in a quantum dot experiment -- collaborating with the group of Natalia Ares -- a finite-time Szilard engine in the whole range of driving speeds: ranging from the slow (high efficiency, low power) to the fast (high power, low efficiency) driving regime. The optimal protocol is found from the theoretical results of~\cite{Esposito2010}, which interpolates between the two-jump protocols for fast driving~\cite{Rolandi2023Fast} and the geodesic protocols of~\cite{Scandi19} at slow driving. We also characterize work fluctuations, which play a dominant role at these scales~\cite{Jarzynski2011}. We observe that, whereas in the high efficiency regime work fluctuations disappear due to the fluctuation-dissipation regime, higher power comes inevitably with higher fluctuations.

\subsection{Experiment}
\begin{figure}[tb!]
	\centering
	\includegraphics[width=\textwidth]{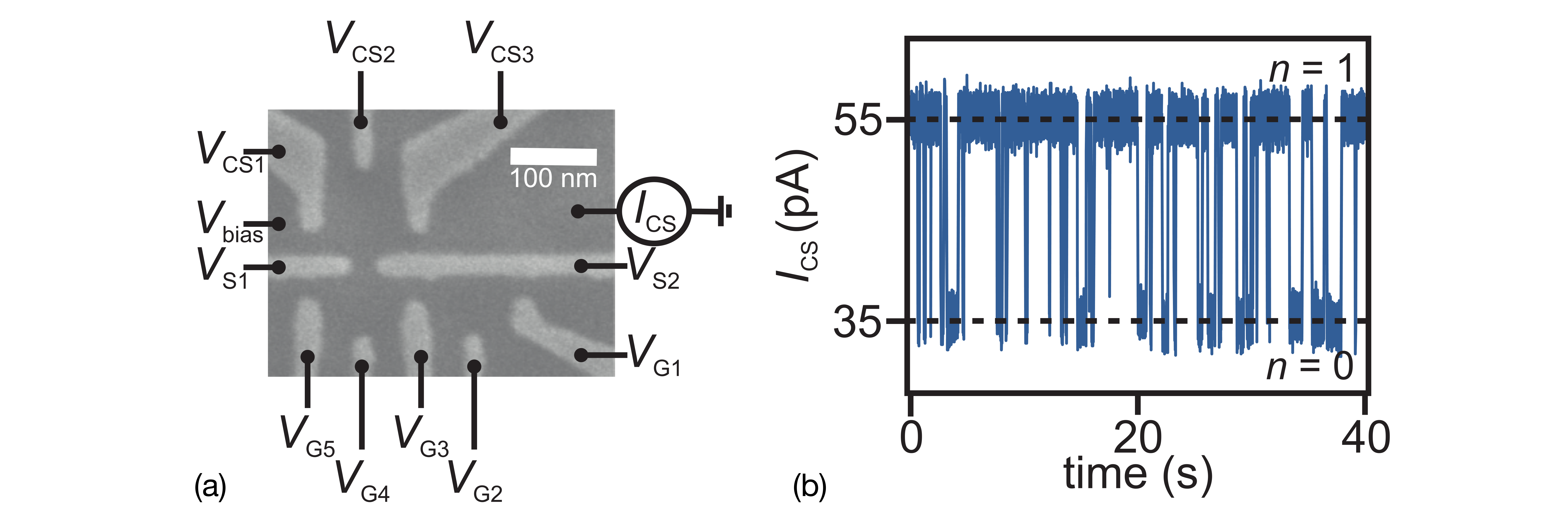}
	\caption{(a) The experimental device: a quantum dot system in a strained Ge quantum well. (b) Monitoring of the occupancy of the dot via the current $I_\text{CS}$.}
	\label{fig:info_experiment}
	\vspace{-15pt}
\end{figure}
Here we discuss the experimental device, which was realized by the group of Prof. N. Ares. The device is is shown in \sfref{fig:info_experiment}{a}. It consists of a quantum dot system in a strained Ge quantum well. An information bit is encoded in the occupancy of the right dot in the bottom array (QD1). This quantum dot QD1 is defined by applying voltages $V_\text{G1}$ and $V_\text{G3}$ that confine the electronic state. The discrete energy level $E$ is controlled using the plunger gate voltage $V_\text{G2}$. The left dot in the array is positioned in Coulomb blockage which limits the tunneling of QD1 to the right reservoir. Another quantum dot in the top array, defined using gates $V_\text{CS1-CS3}$, serves as a probe for the occupancy $n$ of QD1. The occupancy $n$ is monitored by measuring the current $I_\text{CS}$ through the charge sensor dot as shown in \sfref{fig:info_experiment}{b}.  

The experiment is performed in a regime where $n \in {0,1}$ i.e, when QD1 has an extra particle or not. The tunneling in and out rates are characterized by $\gamma_\text{in} = \Gamma_\text{in}f(\eps)$ and $\gamma_\text{out} = \Gamma_\text{out}(1-f(E))$, where $\Gamma_\text{in} = 3.5\ \text{Hz}$ and $\Gamma_\text{out} = 7\ \text{Hz}$. We note that $\Gamma_\text{out} \approx 2\Gamma_\text{in }$ indicating the spin degeneracy of the system.

In this section, we operate this device as a Szilard engine, the details of which are described in the next section. From an operational perspective this requires us to let the quantum dot system thermalize with the reservoir while keeping its energy at $E_0 = 2k_\text{B}T$ which corresponds to a $50\%-50\%$ occupation, where $T$ is the electron temperature. Then the charge sensor measures the instantaneous charge state, and the gate voltage $V_\text{G2}$ is modified to realize the optimal protocol.

\subsection{Optimization of a Szilard Engine}
In this section we will present how the setup can be used as a Szilard engine and then proceed to optimize it. The two-level system can be effectively described by the Hamiltonian
\begin{equation}
	\hat H(t) = \frac{1}{2}\eps(t)\hat \sigma_z~,
\end{equation}
where $\eps(t)$ is the effective gap between the energy levels, we define it so that it is $0$ when the occupation probability is $0.5$, therefore we have the relation $\eps = E + E_0$. This energy gap can be be externally controlled by the gate voltage $V_\text{G2}$. By denoting with $\ket{1}$ the excited state of $\hat H(t)$, $p(t) := \braket{1|\hat \rho(t)|1}$ is the probability of being in the excited state. The Markovian dynamical equation becomes
\begin{equation}\label{eq:markov_dot}
	\dot p(t) = \gamma(\omega(t) - p(t))~,
\end{equation}
where $\omega(t) = (1+e^{\beta \eps(t)})^{-1}$ is the excited level population of the instantaneous Gibbs state, $\gamma^{-1}$ is the thermalization timescale due to the interaction between system and bath, and $\beta = (k_B T)^{-1}$ is the inverse temperature of the bath. \Eref{eq:markov_dot} can be solved to find
\begin{equation}
	p(t) = p(0)e^{-\gamma t} + \gamma\int_0^t\!ds~\omega(s)e^{-\gamma(t-s)}~.
\end{equation}
Ref.~\cite{Esposito2010} shows that for the system at hand the exact dynamics are simple enough to obtain a general analytical solution for $\eps(t)$ and $p(t)$ that minimizes the work cost. Here we will focus on finding the optimal protocol $\eps_0(t)$ in the case we measure $\ket{0}$ in step 1. By spin-flip symmetry, the optimal protocol $\eps_1(t)$ in the case where we measure $\ket{1}$ is simply given by $\eps_1(t) = -\eps_0(t)$. Since we have to perform cycles, the symmetry of the problem imposes the boundary conditions $\eps(0) = \eps(\tau) = 0$.	From ref.~\cite{Esposito2010} optimal protocols satisfy	\begin{equation}\label{eq:p_eps}
	p(t) = \frac{1 - \sqrt{K}e^{\beta\eps(t)/2}}{1+e^{\beta\eps(t)}},\quad 0 < t < \tau~,
\end{equation}
where $K$ is an integration constant determined by the boundary conditions. For a given value of $K$, one can find $p(t) = F_K^{-1}(t)$, where $F_K(p)$ is a transcendental function.
By inserting \eref{eq:p_eps} into \eref{eq:work} we can express the work cost of optimal protocols as a function of their boundary conditions on $p$
\begin{equation}\label{eq:min_work}
	\min_{\eps(t)}\beta W[\eps(t)] = G(p(0))- G(p(\tau))~,
\end{equation}
where $G(p)$ is also a transcendental function that can be found in ref. \cite{Esposito2010}.

The measurement at step 1 sets the boundary condition $p(0) = 0$. Since at the start of the next cycle another measurement will be performed, we do not need to impose a boundary condition at $p(\tau)$. Therefore by replacing $p(\tau)$ with $F_K^{-1}(\tau)$ in \eref{eq:min_work} and minimizing with respect to $K$ we can find the optimal integration constant $\kappa_\tau$ (for a given protocol time $\tau$) which defines the optimal protocol:
\begin{equation}\label{eq:opt_eps}
	\beta\eps_0(t) = 2\ln\!\!\left[\frac{\sqrt{\kappa_\tau + 4F_{\kappa_\tau}^{-1}(t) (1-F_{\kappa_\tau}^{-1}(t))} -\! \sqrt{\kappa_\tau} }{2F_{\kappa_\tau}^{-1}(t)}\right]\!\!.
\end{equation}
Since the optimal integration constant is defined by
\begin{equation}
	\kappa_\tau := \argmax_{K} G(F_K^{-1}(\tau))~,
\end{equation}
we turned the functional minimization problem in \eref{eq:min_work} into a regular minimization problem, which is much simpler to handle numerically. It is worth noting that the optimal protocol features jumps at the start and the end of the protocol, as it can be noted from \eref{eq:opt_eps} that $\eps(t) > 0$ for $0<t<\tau$. In \sfref{fig:info_main}{a} we showcase these optimal protocols for a range of values of $\gamma\tau$ from the slow-driving regime to the fast-driving regime.

\subsection{Results and Discussion}
\begin{figure}[tb!]
	\centering
	\includegraphics[width=0.99\textwidth]{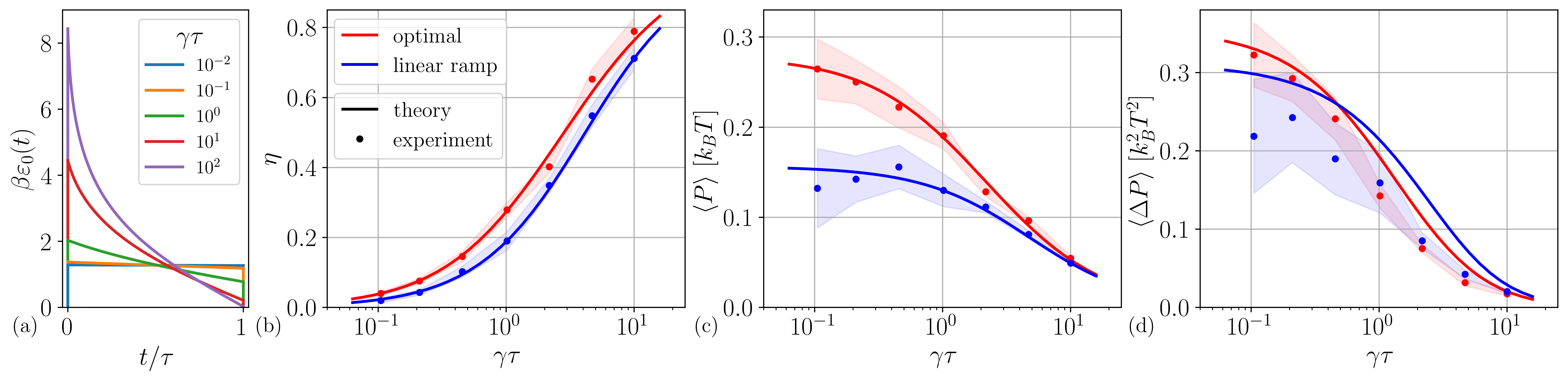}
	\vspace{-10pt}
	\caption{(a) Optimal protocol for different values of $\gamma\tau$. (b,c,d) Comparison of predicted and measured efficiency $\eta = -\beta W/\ln 2$ (b), power $\< P\> = - W/\tau$ (c), and power fluctuations $\<\Delta P\> = \text{Var}(P)$ (d) for implementations of the optimal protocol \eref{eq:opt_eps} (red) and a linear ramp $\eps(t) = 5k_B T(\tau-t)/\tau$ (blue) at different values of $\gamma\tau$ that range from the fast-driving regime to the slow-driving regime. The experimental data is represented with points and the shaded area corresponds to the statistical error of the measurement.}
	\vspace{-10pt}
	\label{fig:info_main}
\end{figure}
For the engine cycle to be truly closed, the information obtained from the measurement at step 1 will have to erased. Thus dissipating $k_BT\ln 2$ of heat into the environment because of Landauer's principle (cf. \cref{chap:landauer}). This gives us a simple formula for the efficiency of this engine
\begin{equation}
	\eta = \frac{-W}{k_BT\ln 2}~.
\end{equation}
The efficiency of the information engine reaches its maximum $\eta_C = 1$ in the static limit, which corresponds to the Carnot efficiency one obtains when setting the temperature of the cold bath to zero. This is simply one of many examples of the link between perfect measurements and zero temperature baths~\cite{Taranto2023}. 

It is interesting to note that this expression for efficiency implies that, for a given cycle length $\tau$, the optimization of power and efficiency coincide. Therefore here it is sufficient to maximize the work gained to optimize both the power and efficiency. In \sfref{fig:info_main}{b,c} we show as red solid lines the maximal efficiency and maximum power that can be achieved for a given value of $\gamma\tau$. We also compare the obtained results to the power and efficiency of a ``naive protocol'': a linear ramp from $\pm5k_B T$ to $0$ over the whole period $\tau$ -- the $+$ sign corresponds to measuring $\ket{0}$ and $-$ to measuring $\ket{1}$. We can se that the difference is not very significant in terms of efficiency. However, for power the optimal protocol has significant gains in the fast driving regime. In \sfref{fig:info_main}{d} we computed the power fluctuations $\<\Delta P\> = \text{Var}(P)$ (cf. \eref{eq:var_work}) for the optimal and naive protocols. Interestingly, we can see that in the fast driving regime the naive protocol has slightly better fluctuations than the optimal protocol.

For a series of values of $\gamma\tau$ between $10^{-1}$ and $10^{1}$ we implemented optimal protocols and linear ramps while measuring the dot's occupation. We can then use the dot occupation data to compute the work cost of a single round of the experiment, with multiple repetitions (in the thousands for the faster protocols and at least $200$ for the slower ones) we can compute the corresponding statistical cumulants. The resulting values are shown in \sfref{fig:info_main}{b-d} as dots, and the shaded area corresponds to the statistical error. We can see that the experiment is in total agreement with the predicted values in for the power and efficiency. However for the fluctuations it seems that the measured values are generally lower than the predicted values (more than $1\sigma$ away for most values). This bias could be explained with the fact that the calibration of the experiment drifts over time. But overall, the agreement between theory and experiment for the extracted work shows the feasibility of these optimal protocols in information engines.


\section{Fast Operations}
For the last section of this chapter we move away from heat engines and look at the optimization of operations in the fast driving regime. In particular, we will look at the problem of minimizing the dissipated work and minimizing the work fluctuations. However, typically it is not possible to minimize the dissipation and fluctuations simultaneously and a compromise must be chosen. Current research aims to understand the interplay and unavoidable trade-offs between dissipation and fluctuations in classical-stochastic and quantum thermodynamic systems~\cite{pietzonka2018,Miller2023}, and it remains an ongoing problem of how best to balance these two competing figures of merit in different scenarios.     

With regard to dissipation, or equivalently the average excess work done to drive a system out of equilibrium, optimal processes are well characterized in slowly driven or linear response regimes where methods from thermodynamic geometry can be employed \cite{Abiuso2020Geometric}. For classical stochastic systems operating in these close-to-equilibrium regimes the work fluctuation-dissipation relation holds \cite{Jarzynski1997}, implying that paths of minimal dissipation simultaneously minimize the resulting work fluctuations. While this is not always satisfied by non-classical systems due to the impact of quantum coherence \cite{Miller2019}, an alternative geometric approach has been recently developed that can determine minimum fluctuation protocols for slowly driven quantum systems \cite{Miller2020}. On the other hand, it is desirable to understand how to optimize systems beyond linear response driving and where shorter operation times are needed. The goal of this section is to establish a general optimization principle for minimizing both the average excess work and its fluctuations in rapidly driven small systems. 

The study of driven Brownian particles first hinted at a key feature of minimum-dissipation protocols for fast driving; such protocols contain discontinuous jumps in the system control parameters or degrees of freedom~\cite{Schmiedl2007opt}. This has been further evidenced in a range of other systems through either analytic or numerical treatments of finite-time thermodynamic optimization problems~\cite{Esposito2010,Aurell2011,Zulkowski2014,cavina2018}. More recently the optimality of these control parameter quenches has been proven \emph{in general} with regard to maximizing the power and efficiency of microscopic heat engines with fast operation cycles \cite{erdman2019_njp,cavina2021}, and furthermore proven optimal for minimizing the average excess work done on classical stochastic systems rapidly driven from equilibrium~\cite{Blaber2021}. In contrast to dissipation, little is known about how to minimize work fluctuations under rapid driving, nor is it known how these protocols compare to those with minimal dissipation. In this section we show that protocols with minimal fluctuations also consist of instantaneous jumps in the systems control parameters. Our result applies in full generality to any quantum or classical system whose generator is independent of the control parameter velocities. While sharing the same general design principle as minimal-dissipation protocols, these control variables typically need to jump to a distinct point in the parameter space, meaning that average excess work and work fluctuations cannot be simultaneously optimized. Furthermore, and as we will illustrate, one practical advantage of this approach is that it enables us to optimize driving protocols for complex many-body systems where exact results are lacking.  

In \sref{sec:expand} we start by deriving general expressions for the average excess work and its variance for rapidly driven quantum systems. Then in \sref{sec:optimal} we present the general Euler-Lagrange equations for finding optimal protocols in this fast driving regime and show that all solutions consist of discrete jumps in the control parameter space. We then explore different scenarios where this optimization scheme can be implemented; \sref{sec:closed} focuses on closed quantum systems, whereas \sref{sec:open_quantum_systems} concerns open quantum systems including erasure of a quantum dot and driving a classical and quantum Ising spin chain.

\subsection{Excess Work and Fluctuations}\label{sec:expand}
We will begin with a rather general treatment of a finite-dimensional quantum system subject to rapid time-dependent driving, which may be isolated or in contact with an environment. Following the notation presented in \sref{sec:geom} we decompose the Hamiltonian as follows
\begin{align}
	\hat H(\lambda)=\hat H_0+\lambda^j \hat X_j, \ \ \ \ \ \ t\in[0,\tau],
\end{align}
where $\hat H_0$ denotes a fixed Hamiltonian in the absence of driving, $\{\lambda^j\}_{j=1}^n$ are externally controllable parameters and $\{\hat X_j\}_{j=1}^n$ are the corresponding observables. For now we can assume the evolution is given by a Markovian generator of the form
\begin{align}\label{eq:gen}
	\frac{d\hat \rho(t)}{dt} = \mathcal{G}_{\vec{\lambda}_t}[\hat \rho(t)]; \ \ \ \ \hat\rho(0)=\hat\pi(\lambda_i).
\end{align}
with a thermal initial condition. The most notable part of this assumption is that the generator is \textit{independent of the velocity $d\lambda(t)/dt$}, and depends only on the local values of $\lambda(t)$. This is readily satisfied by isolated quantum systems evolving unitarily, adiabatically driven open quantum systems \cite{Albash2012} and Markovian dynamics for classical/quasi-classical systems driven by scalar potentials. On the other hand, open quantum systems driven non-adiabatically may not meet this requirement \cite{Dann2018}. We also stress that while our system is initially thermal, we place no restriction on the final state after the protocol has been applied.\\

As a quantifier for the degree of irreversibility associated with the process, the average \textit{excess} -- or dissipated, cf. \sref{sec:second_law} -- work done on the system is given by
\begin{align}
	W_{\text{ex}}&=W -\Delta F, 
\end{align}
where $\Delta F$ is the change in equilibrium free energy. The excess work disappears $W_{\text{ex}} \rightarrow 0$ in  quasi-static processes  where the system is always in thermal equilibrium, which also implies absence of work fluctuations  due to the work fluctuation-dissipation relation $\beta \sigma_W^2/2 = W_{\text{ex}}$ holding valid in this limit~\cite{Jarzynski1997}. For  non-equilibrium processes, both  $W_{\text{ex}}$ and $\sigma_W$ will become relevant, and we expect their magnitudes to increase with the speed of the process (i.e. as $\tau$ decreases).  
Our goal is then to investigate which protocols give the smallest values of average dissipation $W_{\text{ex}}$ and work fluctuations $\sigma_W$ respectively.  In general, computing and optimizing the work moments relies on knowing an exact solution to the dynamics \eref{eq:gen}. While this is not generally tractable,  we will demonstrate that this control problem becomes considerably simpler in fast driving regimes (i.e. when the overall time $\tau$ taken to go from $\lambda_i$ to $\lambda_f$ is small relative to the characteristic timescales of the system).  

We first quantify precisely what we mean by a fast protocol by defining a characteristic timescale $\tau_{\text{c}}$ for the generator given by \cite{cavina2021}
\begin{align}\label{eq:timescale}
	\tau_{\text{c}}^{-1}=\max_{0\leq t\leq \tau}||\mathcal{G}_{\lambda(t)}||,
\end{align}
where we introduce a norm
\begin{align}
	||\mathcal{G}_{\lambda}||=\max_{\tr{O}< \infty}\frac{||\mathcal{G}_{\lambda}[O]||_1}{||O||_1}
\end{align}
and $||A||_1=\tr{\sqrt{A^\dagger A}}$. For a finite-dimensional unitary generator, this parameter is bounded by the operator norm of the Hamiltonian, while for systems undergoing non-unitary dynamics with a unique fixed point then $\tau_{\text{c}}$ bounds the shortest relaxation timescale associated with the system. Overall, this gives us a definition of the fast driving regime which assumes that the total duration is short enough such that $\tau\ll \tau_c$. To see how this approximation impacts the work moments, let us start by seeing how it impacts the evolution. 
We can expand the solution to \eref{eq:gen} as a Dyson series:
\begin{align}\label{eq:dyson}
	\hat\rho(t)=\hat\pi(\lambda_i)+\sum^{\infty}_{n=1}\int^t_0\! dt_{n} \int^{t_{n}}_0\! dt_{n-1}...\int^{t_2}_{0}\! dt_1~  \mathcal{G}_{\lambda(t_{n})}\mathcal{G}_{\lambda(t_{n-1})}...\mathcal{G}_{\lambda(t_1)}[\hat\pi(\lambda_i)]~.
\end{align}
If one makes the integral parameters dimensionless by extracting $\tau$, with \eref{eq:timescale} one can see that the $n$-th term of the sum is of order $\mathcal O\!\left(\tau^n/\tau_c^n\right)$. Therefore we define
\begin{align}
	\hat\sigma(t)=\hat\pi(\lambda_i)+\int^t_0\! dt'~ \mathcal{G}_{\lambda (t')}[\hat\pi(\lambda_i)]~,
\end{align}
by construction $||\hat\rho(t)-\hat\sigma(t)||_1\leq\mathcal{O}\!\left(\tau^2/\tau_c^2\right)$. Therefore we can approximate the state with $\hat\sigma(t)$ so long as $\tau\ll \tau_c$. We can use this approximation to compute the excess work
\begin{equation}\label{eq:av_expand}
	W_{\text{ex}} = k_B T S(\hat\pi(\lambda_i)||\hat\pi(\lambda_f)) + \int^\tau_0\! dt~\frac{d\lambda^j(t)}{dt}\int^t_0\!dt'~ R_j(\lambda(t'))+\mathcal{O}\!\left(\tau^2/\tau_c^2\right)~.
\end{equation}
Where we use the relative entropy 
\begin{align}
	S(\hat\rho_1||\hat\rho_2)=\Tr[\hat\rho_1 \log \hat\rho_1]-\tr{\hat\rho_1 \log \hat\rho_2}~,
\end{align}
and defined the quantum \emph{initial force relaxation rate} (IFRR):
\begin{align}
	R_j(\lambda):=\big\<\mathcal{G}^\dagger_{\lambda}[\hat X_j]\big\>_{\lambda_i}~
\end{align}
where  $\< \hat X_j \>_{\lambda_i}=\Tr[\hat X_j\hat\pi(\lambda_i)]$ is the expectation value with respect to the initial equilibrium state. We can then do an integration by parts on \eref{eq:av_expand} to obtain
\begin{align}\label{eq:fastwork}
	 W_{\text{ex}}  = k_B T S(\hat\pi(\lambda_i)||\hat\pi(\lambda_f)) + \int^\tau_0\! dt~\big[\lambda_f - \lambda(t)\big]^j R_j(\lambda(t))+\mathcal{O}\!\left(\tau^2/\tau_c^2\right)~.
\end{align}
The first term represents the excess work from a perfect Hamiltonian quench \cite{Scandi2020}, while the second term gives the leading order correction for a protocol at finite speed. This expansion agrees with the results of \cite{Blaber2021} for classical Focker-Planck dynamics, now generalized to a fully quantum regime.\\ 
By applying the same procedure to work fluctuations \eref{eq:var_work} we obtain
\begin{multline}\label{eq:fast_var}
	\sigma_{W}^2 = k_B^2 T^2 V\big(\hat\pi(\lambda_i)||\hat\pi(\lambda_f)\big)
	+\int^\tau_0\! dt~\big[\lambda_f-\lambda(t)\big]^j \mathbf{G}_{jk}(\lambda(t))\big[\lambda_f-\lambda(t)\big]^k\\
	+\int^\tau_0\! dt~\big[\lambda_f-\lambda(t)\big]^j \mathbf{B}_{jk}(\lambda(t)) \big[\lambda(t)-\lambda_i\big]^k +\mathcal{O}\!\left(\tau^2/\tau_c^2\right)~.
\end{multline}
Where the first term is the \emph{relative entropy variance} \cite{Guarnieri2018}:
\begin{align}
	V(\hat\rho_1||\hat\rho_2)=\Tr[\hat\rho_1 (\log \hat\rho_1-\log\hat \rho_2)^2]-S^2(\hat\rho_1||\hat\rho_2)~,
\end{align}
$\mathbf{G}(\lambda)$ is the initial force correlation matrix, with elements
\begin{align}
	\mathbf{G}_{jk}(\lambda):=\frac{1}{2} \big\<\mathcal{G}^\dagger_{\lambda}\big[\{\Delta\hat X_j,\Delta\hat X_k\}\big]\big\>_{\lambda_i}~,
\end{align}
where $\{\hat X,\hat Y\}=\hat X \hat Y + \hat Y \hat X$ is the anti-commutator and we define shifted force observables as $\Delta \hat X_j:=\hat X_j-\big<\hat X_j\big\>_{\lambda_i}$. Finally, $\mathbf{B}(\lambda)$ is another correlation function given by
\begin{align}
	\mathbf{B}_{jk}(\lambda):=\big<\{\mathcal{G}^\dagger_{\lambda}[\Delta\hat X_j],\Delta\hat X_k\}\big\>_{\lambda_i}~.
\end{align}

As we saw with the average excess work, the first term here is what one would expect for work fluctuations via an instantaneous quench \cite{Scandi2020}, while the two integral terms are the leading order correction for a finite speed protocol. The expressions \eref{eq:fastwork} and \eref{eq:fast_var} are the first main result of this section, and will now form the basis for finding optimal protocols in the fast driving regime.

\subsection{Optimality of Instantaneous Jump Protocols}
\label{sec:optimal}

Our aim is now to determine control protocols that minimize the dissipated work and the work fluctuations. Since the zero-th order terms in the expansions of the previous section depend only on the boundary conditions of the protocol, it is useful to define the \emph{short-term power savings} \cite{Blaber2021}
\begin{align}\label{eq:power_save}
	P_{\text{save}}:=\tau^{-1}\left[k_B T S\big(\hat\pi(\lambda_i)||\hat\pi(\lambda_f)\big)- W_{\text{ex}}\right]~,
\end{align}
which quantifies any additional reduction to the rate of work done provided by the finite-time protocol beyond that of an instantaneous quench. In a similar fashion we also introduce the \emph{short-term constancy savings},
\begin{align}\label{eq:const}
	C_{\text{save}}:=\tau^{-1}\left[k_B^2 T^2 V\big(\hat\pi(\lambda_i)||\hat\pi(\lambda_f)\big)-\sigma_{W}^2\right]~,
\end{align}
This measures the reductions to the rate of work fluctuations in a short-time protocol. These are now the two objectives to maximize in our control problem. Using our short-time approximations to both the average excess work \eref{eq:fastwork} and work fluctuations \eref{eq:fast_var}, a general optimization principle becomes immediately apparent for this regime. Since the integrands appearing in \eref{eq:fastwork} and \eref{eq:fast_var} are each independent of the control velocity $d\lambda/dt$, we can infer that optimal protocols will consist of an instantaneous jump from $\lambda_i$ to a point in the parameter space, remaining there for the total duration $\tau$ and concluding with another instantaneous jump to the final boundary point $\lambda_f$. We will denote the control values that maximize $P_{\text{save}}$ and $C_{\text{save}}$ respectively by $\xi$ and $\Lambda$, which are determined by the solutions to the following \textit{distinct} Euler-Lagrange equations:
\begin{align}\label{eq:EL_av}
	R_j(\xi) = \frac{\partial}{\partial\lambda^j} \bigg(\big[\lambda_f - \vec{\xi}\big]^k R_k(\lambda)\bigg)\bigg|_{\lambda=\xi} 
\end{align}
and
\begin{align}\label{eq:EL_var}
	\frac{\partial}{\partial\lambda^j}\bigg(\big[\lambda_f-\lambda\big]^k \mathbf{G}_{kl}(\lambda)\big[\lambda_f-\lambda\big]^l\bigg)\bigg|_{\lambda=\Lambda}= \frac{\partial}{\partial\lambda^j}\bigg(\big[\lambda-\lambda_f\big]^k \mathbf{B}_{kl}(\lambda) \big[\lambda-\lambda_i\big]^l\bigg)\bigg|_{\lambda=\Lambda}.
\end{align}
The maximal short-term power savings are then given by
\begin{align}\label{eq:max_p}
	P_{\text{save}}\leq P^{*}_{\text{save}}:=\big[\xi-\lambda_f\big]^j R_j(\xi)~,
\end{align}
which is saturated via the jump protocol $\lambda(t)=\lambda_i+[\xi-\lambda_i]\theta(t)+[\lambda_f-\xi]\theta(t-\tau)$, where $\theta(t)$ denotes the Heaviside step function. The optimality of such processes was proven in \cite{Blaber2021} for classical systems. We have here shown that the same result applies to quantum mechanical systems, provided that the dynamical generator \eref{eq:gen} remains independent of $d\lambda/dt$. As a more significant result, we can now see that the same is true for fluctuations, albeit with a different choice of point in the parameter space. The maximum short-term constancy savings are given by
\begin{align}\label{eq:max_c}
	C_{\text{save}}\leq C^{*}_{\text{save}}:= \big[\Lambda-\lambda_f\big]^j\bigg( \mathbf{G}_{jk}(\Lambda)\big[\lambda_f-\Lambda\big]^l+  \mathbf{B}_{jk}(\Lambda) \big[\Lambda-\lambda_i\big]^k\bigg)~.
\end{align}
which is saturated by jumping to $\Lambda$ instead: $\lambda(t)=\lambda_i+[\Lambda-\lambda_i]\theta(t)+[\lambda_f-\Lambda]\theta(t-\tau)$.\\
In general the values of $\xi$ and $\Lambda$ will not typically coincide, implying a trade-off between minimized excess work versus minimal fluctuations. This can remain the case even in quasi-classical regimes where only changes to the energy levels of the system are allowed. This should be contrasted with slow driving or linear response regimes, which allow for simultaneous optimization of the average and variance due to the validity of the fluctuation dissipation relation in the absence of quantum friction \cite{Speck2004,Miller2019}. However, depending on the particular Hamiltonian parameters and dynamics it is still possible to find situations where $\xi=\Lambda$ and simultaneous optimization is possible, as we will highlight in subsequent sections. 

Before we proceed it is important to highlight some consistency requirements needed to implement a jump protocol. As we are restricted to operating in fast driving regimes, this places restrictions on the set of points one can jump to in order to ensure that the Taylor expansions remain valid. We can compute the exact excess work and fluctuations induced by the jump protocols and compare that to the truncated expressions \eref{eq:fastwork} and \eref{eq:fast_var}. We find that the error of the dissipation expansion is bounded by $\Delta h(\xi)\mathcal{O}\!\left(\tau^2/\tau_c^2\right)$, while for the fluctuations expansions it is bounded by $\Delta h(\Lambda)^2\mathcal{O} \!\left(\tau^2/\tau_c^2\right)$. Where we defined
\begin{align}
	\Delta h(\lambda):=2\max\big\{||\hat H(\lambda)-\hat H(\lambda_i)||_1,||\hat H(\lambda_f)-\hat H(\vec{\xi})||_1\big\}~.
\end{align}
This tells us that one cannot jump arbitrarily far from the boundary points $\lambda_i$, $\lambda_f$ as this would lead to a large $\Delta h$ and hence invalidate the fast driving approximation. Therefore any freedom in setting the magnitude of $\xi$ and $\Lambda$ must take these bounds into account, discounting arbitrarily large values of both $P^{*}_{\text{save}}$ and $C^{*}_{\text{save}}$. In the remainder of this chapter we will demonstrate the utility of these jump protocols in a range of different types of system.     

\subsection{Closed Quantum Systems}
\label{sec:closed}
As a starting point we consider an isolated quantum system whose dynamics are given by the time-dependent Liouville-von Neumann equation:
\begin{align}
	\label{eq:evclosed}
	\mathcal{G}_{\lambda}[\hat\rho] = -\frac{i}{\hbar}\big[\hat H(\lambda),\hat\rho\big]~.
\end{align}
The work statistics of quenched isolated systems are well studied, particularly in the context of many-body quantum systems~\cite{Fusco2014}. Our formalism can now be used to calculate the leading short-time corrections to the excess work and fluctuations arising when the (instantaneous) quenches are replaced by fast Hamiltonian ramps, and then subsequently minimize them using the appropriate jump protocols outlined in the previous section. For closed, finite dimensional systems it is clear that the characteristic time scale is $\tau_c \sim\hbar/E_{\max}(\lambda)$, where $E_{\max}(\lambda)$ denotes the maximum energy eigenvalue of $\hat H(\lambda)$. The relevant initial force relaxation rate and correlation functions are found to be
\begin{align}
	R_k(\lambda) &= -\frac{i}{\hbar}\big\<\big[ \hat X_k,\hat H(\lambda)\big]\big\>_{\lambda_i}~,\\
	\big[\mathbf{G}(\lambda)\big]_{jk} &= -\frac{i}{2\hbar}\big\<\big[\big\{\Delta \hat X_j,\Delta \hat X_k\big\},\hat H(\lambda)\big]\big\>_{\lambda_i}~,\\
	\big[\mathbf{B}(\lambda)\big]_{jk} &= -\frac{i}{\hbar}\big\<\big\{\big[\Delta \hat X_j,\hat H(\lambda)\big],\Delta \hat X_k\big\}\big\>_{\lambda_i}~.
\end{align}
The  short-time power savings are then
\begin{align}
	P_{\text{save}}:=\frac{i}{\tau\hbar}\int^\tau_0 dt \  \big\<\big[\hat H(\lambda_f),\hat H(\lambda(t))\big]\big\>_{\lambda_i} \label{eq:P_unitary}
\end{align}
while the constancy savings are
\begin{multline}\label{eq:C_unitary}
	C_{\text{save}}:=\frac{i}{\tau\hbar}\int^\tau_0 dt \ \Big(
	\big\<\big[\hat H(\lambda_f)^2,\hat H(\lambda(t))\big]\big\>_{\lambda_i} - \big\<\big\{\hat H(\lambda_i), \big[\hat H(\lambda_f),\hat H(\lambda(t))\big]\big\}\big\>_{\lambda_i} \\ 
	- 2\big\< \hat H(\lambda_f)-\hat H(\lambda_i)\big\>_{\lambda_i}\big\<\big[\hat H(\lambda_f),\hat H(\lambda(t))\big]\big\>_{\lambda_i}
	\Big).
\end{multline}
We can already see from \eref{eq:P_unitary} and \eref{eq:C_unitary} that if $\hat H(\lambda_f)$ and $\hat H(\lambda_i)$ commute, or $\lambda(t)$ is chosen such that $\hat H(\lambda(t))$ commutes with either $\hat H(\lambda_f)$ or $\hat  H(\lambda_i)$, then the integrand is exactly $0$ - which directly follows using the cyclic property of the trace. Therefore, if $\lambda(t)$ is a linear combination of $\lambda_i$ and $\lambda_f$, the first order correction vanishes. An immediate consequence is that a naive protocol that linearly interpolates between the initial and final Hamiltonian in a closed system is equivalent to a quench up to first order in driving speed.

We now choose a jump protocol to maximize either variable. We can notice that both $P_{\text{save}}$ and $C_{\text{save}}$ are linear in the control variables. Therefore the respective gradients are independent of $\lambda$. Which implies that the optimal points $\xi$ and $\Lambda$ are vectors pointing in the direction of the respective gradients, with the norm chosen as large as possible. However, as argued in the previous section, the larger this norm is chosen, the larger the error of the approximation is. In particular, setting $\|\xi\|\gg \|\lambda_i\|, \|\lambda_f\|$ gives $\|\xi\|\propto \Delta h(\xi)$ while $\tau_c \propto 1/\|\xi\|$. This implies that the error on the expansion of $P_\text{save}$ scales as $\mathcal{O}(\|\xi\|^3\tau^2)$, which clearly limits how large the norm can be chosen relative to the duration of the protocol. A similar argument applies to the constancy savings and norm of the optimal point $\Lambda$. 

We can make some further inferences about the relation between the different jumps $\xi$ and $\Lambda$. We can show that
\begin{align}\label{eq:grads_unitary}
	(\lambda_i)^j \frac{\partial}{\partial\lambda^j} P_{\text{save}}=(\lambda_i)^j \frac{\partial}{\partial\lambda^j} C_{\text{save}}=(\lambda_f)^j \frac{\partial}{\partial\lambda^j} P_{\text{save}}=(\lambda_f)^j \frac{\partial}{\partial\lambda^j} C_{\text{save}}=0~,
\end{align}
which means that both gradients are orthogonal to $\lambda_i$ and $\lambda_f$. As was said before, this implies that if the protocol consists of a linear combination of $\hat H_i$ and $\hat H_f$ then the correction will be zero. But if the Hamiltonian has $d\leq 2$ controllable parameters, it is impossible for $\lambda(t)$ to be linearly independent from $\lambda_i$ and $\lambda_f$. Therefore, regardless of the type of driving, with $d\leq 2$ controllable parameters the correction is always zero.

It is interesting to consider what happens when we can control exactly three parameters, $d=3$. \Eref{eq:grads_unitary} constrains the gradients of $P_\text{save}$ and $C_\text{save}$ to be parallel, which implies
\begin{align}
	dP_{\text{save}} =\pm \frac{||\nabla_{\lambda} P_{\text{save}}  ||}{||\nabla_{\lambda} C_{\text{save}}  ||} \ dC_{\text{save}}~,
\end{align}
where the sign is positive if the gradients are oriented in the same direction and negative otherwise. If the sign is positive, we can optimize fluctuations and excess work simultaneously with $\xi=\Lambda$; if the sign is negative, we have a direct trade-off between savings in power and constancy.

\subsubsection*{Driven Qubit}
\begin{figure}[tb!]
	\centering
	\includegraphics[width=\textwidth]{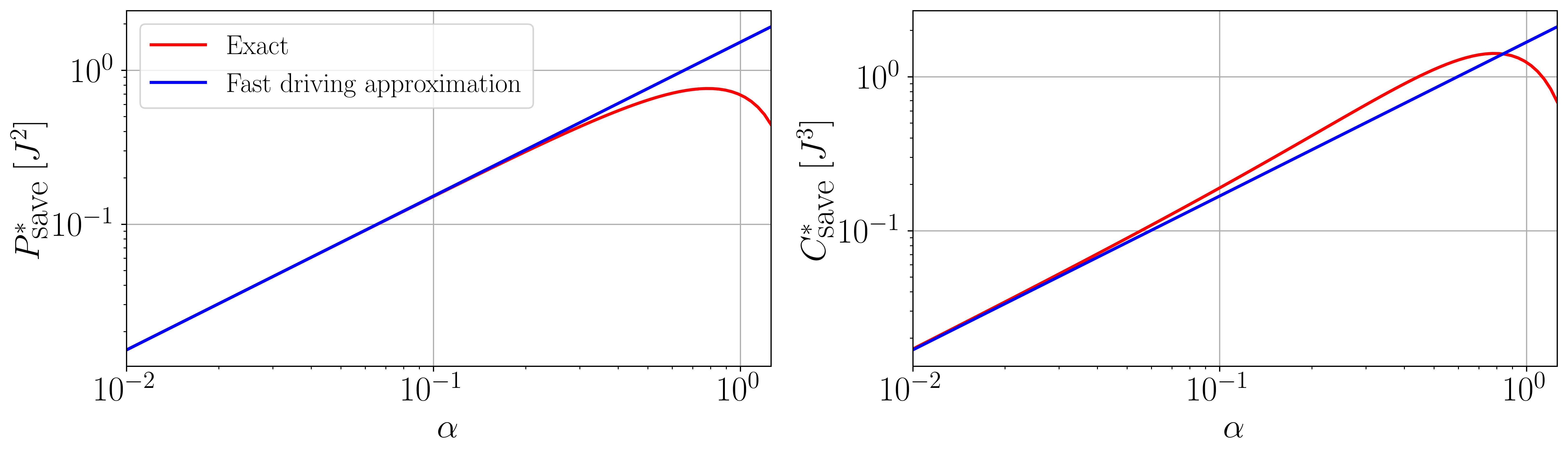}
	\vspace{-5pt}
	\caption{Comparison of $P_{\text{save}}^*$ and $C_{\text{save}}^*$ (in units of $J^2$ and $J^3$ respectively) in the exact case and fast driving approximation as a function of $\alpha$ for a jump protocol with $H_i = J\sigma^x$, $H_f = J\sigma^z$ and $J\tau = \beta J= 1$.}\label{fig:qubit}
	\vspace{-10pt}
\end{figure}

As an illustrative example, we can compute and optimize the excess work and work fluctuations of a qubit that is undergoing controlled unitary evolution. The most general Hamiltonian for a qubit is
\begin{equation}
	\hat H(\lambda) = J \vec\lambda \cdot \vec\sigma~,
\end{equation}
where $\vec\sigma = (\hat\sigma_x,\hat\sigma_y,\hat\sigma_z)$ is the Pauli vector, $J$ is an energy scale and $\vec\lambda = (\lambda^x,\lambda^y,\lambda^z)$ are dimensionless parameters that characterize the Hamiltonian. We find that the power and constancy savings can be optimized simultaneously with $\vec\xi =\vec\Lambda= \alpha\,\vec\lambda_i \wedge \vec\lambda_f$, where $\alpha>0$ controls the norm of the resulting Hamiltonian. We find the optimal values
\begin{align}\label{eq:p_opt_q}
	P_{\text{save}}^*&=2\alpha J^2\sin^2\phi \|\lambda_i\| \|\lambda_f\|^2 \tanh(\beta J \|\lambda_i\|)~,\\
	\label{eq:c_opt_q}
	C^*_{\text{save}} &=4\alpha J^3\sin^2\phi \|\lambda_i\|^2 \|\lambda_f\|^2 \left[1 - \tanh(\beta J \|\lambda_i\|)^2  \left(1- \frac{\|\lambda_f\|}{\|\lambda_i\|}\cos\phi\right) \right]~,
\end{align}
where $\phi$ is the angle between $\lambda_i$ and $\lambda_f$. The magnitude of $\alpha$ has to be chosen in such a way that that the error of the approximation remains small. A sufficient condition is then given by choosing $\alpha \ll (J\tau |\sin\phi| \|\lambda_i\| \|\lambda_f\|)^{-1}$. This is illustrated in \fref{fig:qubit}, in which we compare the results of \eref{eq:p_opt_q} and \eref{eq:c_opt_q} to the exact calculation of $P_{\text{save}}$ and $C_{\text{save}}$ for jump protocols in a qubit. The boundary conditions were set to $\hat H_i = J\hat\sigma_x$, $\hat H_f = J\hat \sigma_z$ and the relevant constants are set to $\tau J = \beta J = 1$. Then the condition for the validity of the approximation becomes $\alpha \ll 1$, indeed we can see from the figure that as $\alpha$ approaches $\mathcal{O}(1)$ the approximation breaks down. 

It is important to stress that for higher dimensional closed systems, simultaneous optimization of $P_{\text{save}}$ and $C_{\text{save}}$ cannot always be guaranteed despite what we observe in the case of a qubit. 

\subsection{Open Quantum Systems}
\label{sec:open_quantum_systems}

We now move to open quantum systems. Our framework can be applied to any  Markovian Lindblad equation of the form \eref{eq:gen}, in which the generator $\mathcal{G}_{\lambda(t)}$ is independent of the velocity $d\lambda(t)/dt$. Here we illustrate it for the simple evolution: 
\begin{align} \label{eq:markov_relaxation}
	\mathcal{G}_{\lambda}[~\cdot~  ]=\frac{\hat\pi(\lambda)\tr{~\cdot~} -(~\cdot~)}{\tau^{eq}}~,
\end{align}
which describes a decay of the state $\hat\rho$ into the instantaneous Gibbs state $\hat\pi(\lambda)$ with a timescale $\tau^{eq}$. This dissipative evolution naturally arises in collisional models~\cite{Bumer2019} and also describes some systems weakly interacting with a reservoir with a sufficiently flat spectral density~\cite{Esposito2010}. For this type of dynamics we find some more illuminating expressions for the various terms appearing in the leading corrections to the excess work and variance. Firstly, the initial force relaxation rate becomes
\begin{align}
	R_j(\lambda):=\frac{\<\hat X_j\>_{\lambda}-\<\hat X_j\>_{\lambda_i}}{\tau^{eq}}~.
\end{align}
This demonstrates that $R_j(\lambda)$ quantifies the average rate at which each expectation $\< \hat X_j\>$ changes from its initial value relative to the characteristic timescale $\tau^{eq}$. Furthermore, the correlation functions become
\begin{align}
	\big[\mathbf{G}(\lambda)\big]_{jk}=\frac{\mathcal{F}_{jk}(\lambda)-\mathcal{F}_{jk}(\lambda_i)}{\tau^{eq}}+\bigg(\< \hat X_j \>_{\lambda} \ R_k(\lambda)+\< \hat X_k \>_{\lambda} \ R_j(\lambda)\bigg)~.
\end{align}
and 
\begin{align}
	\big[\mathbf{B}(\lambda)\big]_{jk}=-\frac{2}{\tau^{eq}}\mathcal{F}_{jk}(\lambda)~,
\end{align}
where $\mathcal{F}_{jk}(\lambda)$ is the symmetric covariance defined as
\begin{align}\label{eq:metric_fast}
	\mathcal{F}_{jk}(\lambda):=\frac{1}{2}\<\{\hat X_j, \hat X_k\}\>_\lambda-\< \hat X_j\>_{\lambda}\< \hat X_k\>_{\lambda}~.
\end{align}
This function defines a metric tensor on the manifold of control parameters, and was first introduced in \cite{Janyszek1986} as a means of quantifying the geometric structure of thermal states. In quasi-classical regimes where $[\hat X_j,\hat X_k]=0$, this metric becomes proportional to the well-known thermodynamic metric, also known as the Fisher information matrix of the thermal state. \\
We will now illustrate these results with the optimization of three different systems: a driven quantum dot, and two Ising spin chains.

\subsubsection*{Fast Erasure of a Single Bit}
A driven quantum dot interacting weakly with an environment is a paradigmatic example of a system that can be described by the simple dynamics \eref{eq:markov_relaxation}~\cite{Esposito2010}. In that case the Hamiltonian is given by $\hat H(\eps)=\frac{1}{2}\eps\hat\sigma_z$ with a single control variable $\lambda(t)=\eps(t)$ given by the energy gap of the two-level system. The optimal finite-time thermodynamics of such  systems has been well studied with regard to minimizing average dissipation in Landauer erasure, including a recent experimental implementation in a driven single dot~\cite{Scandi2022}. In the present context, we apply our results to a rapid bit-erasure process.  The boundary conditions for erasure are then $\eps_i = 0$ and $\beta \eps_f \gg 1$, which leads to the following expressions for the power and constancy savings
\begin{align}
	P_{\text{save}}&= \frac{k_B T}{\tau^{eq}}\int_0^1 ds~ \big(\beta\eps_f-\beta\eps(s)\big)\bigg(\frac{1}{2} - \frac{1}{1+e^{\beta\eps(s)}} \bigg)~,\\
	C_{\text{save}}&= \frac{k_B^2 T^2}{2\tau^{eq}} \int_0^1 ds~ \beta\eps(s)\big(\beta\eps_f-\beta\eps(s)\big)~.
\end{align}
We now seek to find the optimal energy gaps to jump to in order to maximize either $P_{\text{save}}$ or $C_{\text{save}}$. It will become clear in this case that the power and constancy savings cannot be simultaneously maximized, and so the distinct gaps are denoted by $\xi$ and $\Lambda$ respectively. Maximizing $P_{\text{save}}$ amounts to solving the following transcendental equation $\frac{1}{2}-\frac{1}{1+e^{\beta\xi}} = \frac{(\beta \eps_f - \beta\xi) e^{\beta \xi}}{(1+e^{\beta\xi})^2}$.
In the limit of $\beta \eps_f \gg 1$ we can solve it analytically up to terms $\mathcal{O}(\beta^{-1}\eps_f^{-1}\ln{\beta\eps_f})$ and find the optimal jump $\eps\rightarrow\xi = \beta^{-1}\ln{2\beta\eps_f}$. Maximum power savings are thus
\begin{align}
	P^*_{\text{save}} \simeq\frac{\eps_f}{2\tau^{eq}}~.
\end{align}
For this power-optimized jump let us denote the resulting sub-optimal constancy savings by $C_{\text{save}}^\xi$:
\begin{align}
	C_{\text{save}}^\xi=\frac{\eps^2_f}{\tau^{eq}}  \frac{\ln{2\beta\eps_f}}{2\beta \eps_f}~.
\end{align}
On the other hand, to maximize the constancy savings we need to choose a jump to $\eps \rightarrow\Lambda = \eps_f/2$ instead. This yields
\begin{align}
	C_{\text{save}}^*=\frac{\eps_f^2}{8\tau^{eq}}~, \ \ \ \ \ \ \ \ \ \ \ P^\Lambda_{\text{save}} = \frac{\eps_f}{4\tau^{eq}}~.
\end{align}
where the sub-optimal savings in power are denoted $P^\Lambda_{\text{save}}$. Clearly there exists a significant trade-off between these two choices of optimal protocol, with power-optimized jumps causing no improvement to the constancy while constancy-optimized jumps reducing the potential power savings by a factor of $1/2$. Further comparison can be made with that of a naive linear driving $\eps(t)= \eps_f t/\tau$, which results in savings given by 
\begin{align}
	P_{\text{naive}} \simeq\frac{\eps_f}{4\tau^{eq}}~, \ \ \ \ \ \ \ \ \ \ \ C_{\text{naive}} =\frac{\eps_f^2}{12\tau^{eq}}~,
\end{align}
where we again drop terms of order $\mathcal{O}(\beta^{-1}\eps_f^{-1}\ln{\beta\eps_f})$. Therefore we can see that choosing an optimal jump for the excess work leads to an improvement factor of $1/2$, and choosing the optimal jump for the fluctuations gives an improvement factor of~$3/2$, each indicating significant improvements over a naive protocol. However, two unexpected observations here are that naive protocols are able to achieve larger savings in constancy than that of the power optimized protocol, and also achieve the same level of power savings to the constancy-optimized protocol. This emphasizes that improvements to one objective do not necessarily translate into improvements of the other.

\subsubsection*{Dissipative Classical Ising Chain}
The strength of our approach is that it enables to deal with more complex systems, where exact solutions for minimizing dissipation and/or fluctuations are lacking -- in contrast to the previous example.  This is illustrated now for an Ising chain weakly coupled to a bath with dynamics \eref{eq:markov_relaxation}. We note that optimal driving protocols for classical spin chains have been devised in the slow driving regime, and now we complement such results by studying the opposite fast-driving regime. We first consider a classical spin chain,
\begin{equation}
	\hat H(\eps) = J\sum_{i=1}^n \left( \varepsilon\hat\sigma_z^{(i)} - \hat\sigma_z^{(i)} \hat\sigma_z^{(i+1)} \right)~,
\end{equation}
where $J$ is the energy scale and $\varepsilon$ is a dimensionless parameter which can be interpreted as the strength of an external magnetic field. By assuming periodic boundary conditions, we can compute the partition function in the thermodynamic limit $n\to \infty$:
\begin{equation}
	\lim_{n\rightarrow\infty}\frac{1}{n}\text{ln} \ Z = \beta J + \text{ln}\!\left[\cosh(\beta J\varepsilon) + \sqrt{\sinh(\beta J\varepsilon)^2 + e^{-4\beta J}}\right]~.
\end{equation}
The relevant force here is then the total $\hat X= J\sum_i \hat\sigma_z^{(i)} $. We now identify the following the relations
\begin{align}
	-\frac{1}{\beta}\frac{\partial}{\partial \varepsilon}\log Z &= \tr{\hat X \hat \pi(\varepsilon)}~,\\
	\frac{1}{\beta^2}\frac{\partial^2}{\partial \varepsilon^2}\log Z &= \tr{\hat X^2 \hat\pi(\varepsilon)} - \tr{\hat X \hat\pi(\varepsilon)}^2 ~.
\end{align}
These allow us to compute the first order corrections to the excess work and the fluctuations per site of the protocol in the thermodynamic limit from \eref{eq:fastwork} and \eref{eq:fast_var}. We find
\begin{align}
	\tau^{eq} R(\varepsilon) &=  \frac{\sinh(\beta J \varepsilon_i)}{\sqrt{\sinh(\beta J\varepsilon_i)^2 + e^{-4\beta J}}} - \frac{\sinh(\beta J \varepsilon)}{\sqrt{\sinh(\beta J\varepsilon)^2 + e^{-4\beta J}}}~,\\
	\tau^{eq} G(\varepsilon) &= e^{-4\beta J}\left.\frac{\cosh(\beta Jx)}{(\sinh(\beta Jx)^2 + e^{-4\beta J})^{3/2}}\right|_{x=\eps_i}^{x=\eps}  + (\tau^{eq} R(\varepsilon))^2 ,\\
	\tau^{eq} B(\varepsilon) &= -2 e^{-4\beta J} \frac{\cosh(\beta J \varepsilon_i)}{(\sinh(\beta J\varepsilon_i)^2 + e^{-4\beta J})^{3/2}}~.
\end{align}
\begin{figure}[tb!]
	\centering
	\includegraphics[width=\textwidth]{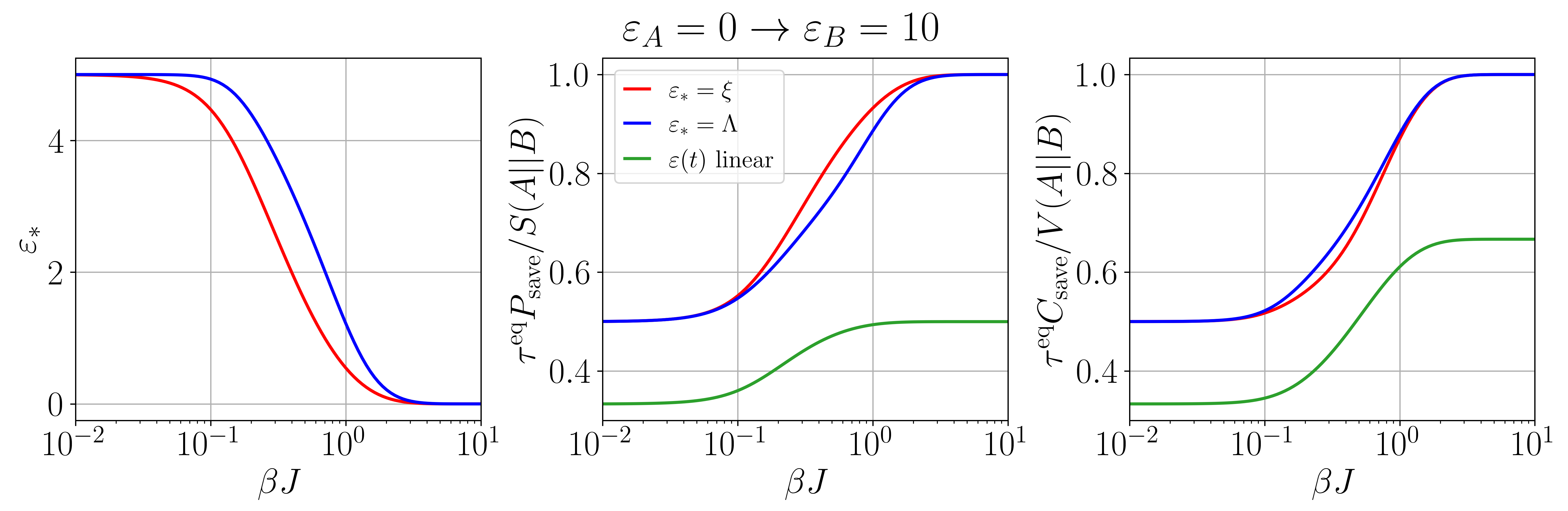}
	\vspace{-5pt}
	\caption{(left) Optimal value of $\varepsilon_*$ for a protocol going from $\varepsilon_i = 0$ to $\varepsilon_f=10$. (center and right) Relative power and constancy savings in three protocols going from $\varepsilon_i = 0$ to $\varepsilon_f = 10$. We compare protocols that optimize excess work and fluctuations to a protocol that varies linearly the value of $\varepsilon$.}
	\label{fig:opt_Cising}
	\vspace{-10pt}
\end{figure}  
It is now a case of substituting these into the two different Euler-Lagrange equations \eref{eq:EL_av} and \eref{eq:EL_var} to determine the optimal points $\xi$ and $\Lambda$ needed in each jump protocol, with solutions found numerically for a process that brings $\varepsilon$ from $\varepsilon_i=0$ to $\varepsilon_f=10$ -- i.e. turning on the magnetic field. In \fref{fig:opt_Cising} (left) we display the optimal field strength $\varepsilon_*=\{\xi,\Lambda\}$ that maximizes either the power or constancy savings. We can notice that in the limits of high and low temperatures they coincide, while we cannot maximize them simultaneously in between these regimes. In \fref{fig:opt_Cising} (center) we plot the power savings $P^*_{\text{save}}$ relative to the zero-th order contribution $k_B T S\big(\pi(\lambda_i)||\pi(\lambda_f)\big)/\tau_c$, while \fref{fig:opt_Cising} (right) displays the  constancy savings $C^*_{\text{save}}$ in units of $k_B^2 T^2 V\big(\pi(\lambda_i)||\pi(\lambda_f)\big)/\tau_c$. Both plots show the relative savings depending on whether we choose to optimize the power or constancy, and this is also compared to the savings achieved by taking a naive linear driving $\eps(t)=\eps_i(1-t/\tau)+\eps_f t/\tau$. In this case we can see that there is only a modest difference between the $\eps_i\rightarrow\xi\rightarrow \eps_f$ and $\eps_i\rightarrow\Lambda\rightarrow \eps_f$ jump protocols, and they each perform considerably better than the naive approach, contrasting with what we observed for the driven quantum dot. This highlights the importance of optimal control in many-body open quantum systems. It is also interesting to note that in the low temperature regime the best protocol becomes to simply do a quench directly to $\eps_f$ for both power and fluctuations.

\subsubsection*{Ising Chain in Transverse Field}
We will conclude with a final example covering the remaining scenario of an open quantum system where the control is such that the Hamiltonian may not commute at different times, so that $[\hat H(\lambda(t)),\hat H(\lambda(t'))]\neq 0$. This non-commutativity implies the presence of quantum friction, which is a distinctly non-classical contribution to the work done to drive the system that arises from allowing transitions between energy eigenstates. 
For this purpose we will consider a  dissipative Ising chain with simple dynamics \eref{eq:markov_relaxation}, though this time we apply a transverse field along the x-axis that can be controlled in time. We note that optimal driving protocols for this model have been considered in the slow driving regime~\cite{Scandi2020}, and the results presented here in the fast driving regime  are hence complementary. In particular, we will  focus on performing drivings close to a quantum phase transition, which has also been considered in previous works~\cite{Fusco2014}. 

The Hamiltonian of the system is
\begin{equation}
	\hat H(g) = -J\sum_{i=1}^n \left(\hat\sigma_z^{(i)} \hat\sigma_z^{(i+1)} + g\hat\sigma_x^{(i)}\right)~,
\end{equation}
where $J$ is the energy scale and $g$ is a dimensionless parameter which can be interpreted as an external (transverse) magnetic field. Clearly such a model will generate quantum friction as we vary the strength $g$ in time. Assuming again periodic boundary conditions, we can compute the spectrum of the system with a Jordan-Wigner transformation \cite{Sachdev2011}. Then by taking the thermodynamic limit the partition function is given by
\begin{equation}
	\lim_{n\rightarrow\infty}\frac{1}{n}\log Z = \int_0^{2\pi}dk~ \log\!\left[2\cosh\frac{\beta\epsilon_k}{2}\right]~,
\end{equation}
where $\epsilon_k$ is the eigen-energy corresponding to the momentum $k$
\begin{equation}
	\epsilon_k = 2J\sqrt{1+g^2-2g\cos k}~.
\end{equation}
At zero temperature and $g=1$ this system presents a phase transition from an ordered ferromagnetic phase to a quantum paramagnetic phase. We will focus on studying protocols that take the system across this point by changing $g$ at finite temperature. The relevant force this time is $\hat X = -J\sum_i \hat\sigma_x^{(i)}$, and we can use the relations
\begin{align}
	-\frac{1}{\beta}\frac{\partial}{\partial g}\log Z &= \tr{\hat X \hat\pi(g)}~,\\
	\frac{1}{\beta^2}\frac{\partial^2}{\partial g^2}\log Z &= \tr{ \hat X^2 \hat\pi(g)} - \tr{\hat X \hat\pi(g)}^2. 
\end{align}
The first order corrections to the excess work and the fluctuations per site of the protocol are now computed within the thermodynamic limit, giving us
\begin{align}
	\tau^{eq} R(g_*) &= -\frac{1}{2}\left.\int_0^{2\pi}dk~ \dot\epsilon_k \tanh \frac{\beta\epsilon_k}{2}\right|^{g=g_*}_{g=g_i} ~,\\
	\tau^{eq} G(g_*) &= \frac{J}{2}\left.\int_0^{2\pi}dk~ \ddot\epsilon_k \tanh\frac{\beta\epsilon_k}{2} + \frac{\dot\epsilon_k^2}{2J} \cosh^{-2}\frac{\beta\epsilon_k}{2}\right|^{g=g_*}_{g=g_i} + (\tau^{eq} R(g_*))^2 ,\\
	\tau^{eq} B(g_*) &= -J\left.\int_0^{2\pi}dk~ \ddot\epsilon_k \tanh\frac{\beta\epsilon_k}{2} + \frac{\dot\epsilon_k^2}{2J} \cosh^{-2}\frac{\beta\epsilon_k}{2}\right|_{g=g_i} ~,
\end{align}
where $\dot \epsilon_k = \frac{d \epsilon_k}{dg}$. Substituting into the Euler-Lagrange equations \eref{eq:EL_av} and \eref{eq:EL_var} and solving numerically, we can determine the instantaneous jumps $g_i\rightarrow\xi\rightarrow g_f$ and $g_i\rightarrow\Lambda\rightarrow g_f$ for maximizing the respective power and constancy savings. We set our boundary conditions to be $g_i=0$ and $g_f=3$ so that we turn on the transverse magnetic field and cross the phase transition point at $g=1$. Similarly to the classical case, we also compare these optimal protocols to a ``naive'' protocol in which the parameter is varied linearly in time, $g(t)=g_i(1-t/\tau)+g_f t/\tau$.

\begin{figure}[tb!]
	\centering
	\includegraphics[width=\textwidth]{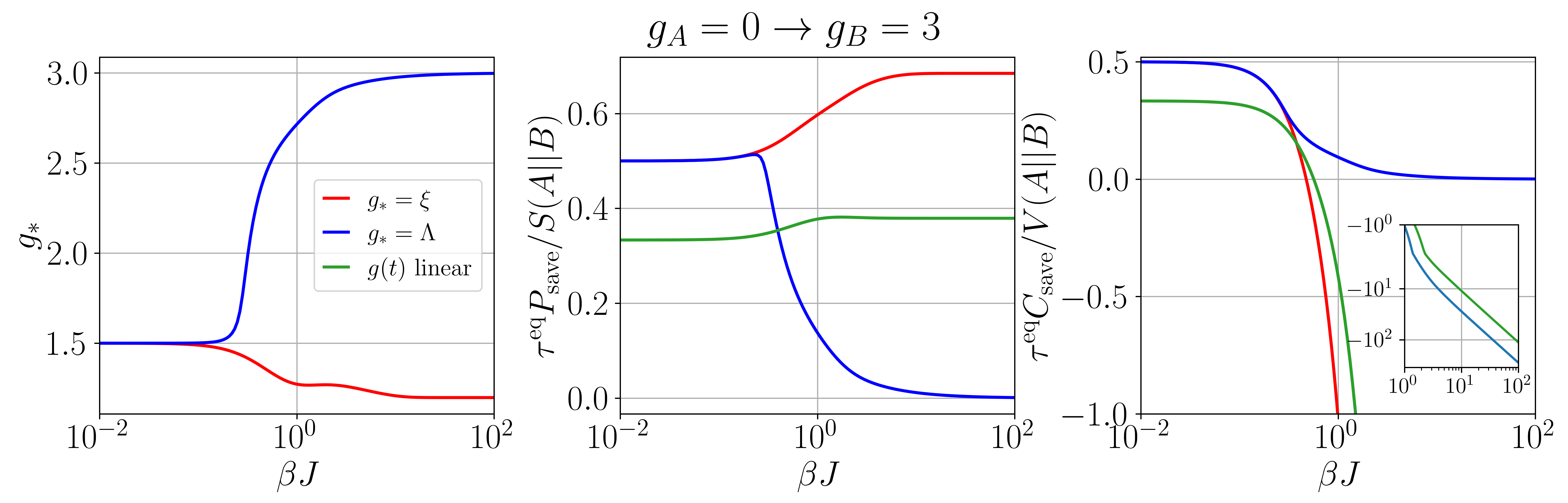}
	\vspace{-5pt}
	\caption{(left) Optimal value of $g_*=\{\xi,\Lambda\}$ for a protocol going from $g_i = 0$ to $g_f=3$. (center) Relative power savings in three protocols going from $g_i = 0$ to $g_f = 3$ and (right) relative constancy savings for the same three protocols. In each figure we compare these optimal protocols to one that varies linearly the value of $g$.}
	\label{fig:opt_Qising}
	\vspace{-10pt}
\end{figure}

In \fref{fig:opt_Qising} (left) we display the optimal fields strength of $g_*=\{\xi,\Lambda\}$, which noticeably coincide in the limit of high temperatures like we saw with the classical Ising chain. On the other hand at low temperatures they no longer coincide, indicating a distinctly non-classical feature of this example and demonstrating that simultaneous optimization is no longer possible. In \fref{fig:opt_Qising} we compare these two choices of protocol to linear driving and plot the resulting power savings (center) and constancy savings (right). Since temperature is finite, the phase transition is washed out but we can still observe a signature in the power and constancy savings occurring at lower temperatures, where we see that the two quantities move significantly further apart. One dramatic feature is the fact that optimizing the power savings results in the constancy savings becoming significantly negative at lower temperatures beyond the phase transition, indicating a large growth in overall work fluctuations above that of an infinitely fast quench. On the other hand, if we choose to maximize the constancy savings we see that this drops to zero alongside the power savings at low temperatures. This indicates that the system is highly sensitive to the choice of protocol when driven close to a quantum phase transition. 

\section{Conclusions}
In this chapter we studied three examples of thermodynamic optimization.
First, we introduced a general framework, described by  \eqsref{eq:p_dp}{eq:ode}, to identify Pareto-optimal cycles between power, efficiency, and power fluctuations in quantum or classical stochastic heat engines. As opposed to previous works, we account for the crucial impact of power fluctuations, which modify non-trivially the optimal driving solution.
We then employed RL to optimize a quantum dot-based heat engine solving its exact finite-time and out-of-equilibrium dynamics.
We observed an abrupt change in Pareto-optimal cycles when switching from the optimization of $2$ objectives, where Otto cycles in the fast-driving regime are optimal, to $3$ objectives, where the optimal cycles have a finite period.
This feature, which shares analogies with the phase transition in protocol space observed in Ref.~\cite{solon2018}, corresponds to a large enhancement of one of the objectives while only slightly decreasing the other ones.
Furthermore, we find an exact mapping between Otto cycles in the fast-driving regime and SSHEs, implying that a violation of the thermodynamic uncertainty relation ratio $\xi$ in \eref{eq:tur_ineq} requires the optimization of all $3$ objectives.
We then find that $\xi$ becomes arbitrarily large in the slow-driving regime. Cycles found with RL display the best features analytically identified in the fast and slow driving regimes.\\

Secondly, we optimized and implemented an information engine, which represents a special class on its own among thermal engines that explicitly demonstrates the intricate connection between information and thermodynamics. We successfully implemented an optimized finite-time Szilard engine within a quantum dot system. The experimental setup consisted of a quantum dot system strained in a Germanium quantum well, where the occupancy of the dot can be manipulated and monitored. The optimization is based on maximizing the extracted energy for a given, arbitrary, cycle length. For the given system, it allows for the simultaneous optimization of power and efficiency. These optimal protocols showed significant improvements over a naive linear ramp protocol at all driving speeds, in particular in the fast driving regime. Additionally, we examined the work fluctuations generated by these optimal protocols, and observed that higher power inevitably comes with higher fluctuations.
The experimental results corroborated the theoretical predictions, showing a high degree in precision of the (indirect) measurement of extracted work. The study described in this section showcased the successful optimization and implementation of a paradigmatic example of a quantum heat engine.

Finally, we have derived approximations for the average excess work done \eref{eq:fastwork} to rapidly drive a small system out of equilibrium along with the resulting work fluctuations \eref{eq:fast_var}. This has been derived under the assumption that $(i)$~the dynamics can be described by a Markovian generator that is independent of the velocities in the time-dependent control parameters, and $(ii)$~the duration of the process is short relative to the characteristic timescale of the dynamics, $\tau_c\gg \tau$. Under these approximations we were able to prove that rapid processes that minimize either the average excess work, or work fluctuations, under fixed boundary conditions consist of two instantaneous jumps in the system control parameters, generalizing the results obtained for the fast driving regime in the first section of the chapter. Also bringing a contrast with the smooth geodesic paths that are optimal on slow driving and linear response regimes \cite{Sivak2012,Miller2019}. Protocols that minimize the excess work cost, jump from the initial configuration to a point determined from the Euler-Lagrange equation \eref{eq:EL_av}, stay there for the duration $\tau$ then jump to the final boundary value. Protocols that minimize the work fluctuations follow the same pattern, but jump to an alternative point satisfying a different Euler-Lagrange equation \eref{eq:EL_var}. We have seen that in general, these points do not coincide which indicates a trade-off between the optimal values of the average and variance. These results extend the work of \cite{Blaber2021} to show that jump protocols continue to be optimal when one also cares about keeping fluctuations minimal, while also extending this approach to the full quantum regime. In particular, it is worth emphasizing that, beyond the standard scenario of a driven system in contact with a Markovian environment, our approach also applies to closed quantum driven systems where the form of minimally dissipative driving processes remain less explored~\cite{Deffner2020}.   
Due to its generality, our optimization scheme can be used to improve the control of complicated chemical, biological and quantum many-body systems whenever short operation times are desired. This has been illustrated by minimizing both excess work and fluctuations for a classical and quantum spin chain where an external magnetic field is rapidly changed in time. When driving the system close to a quantum phase transition, we found  that optimizing over driving protocols leads to substantial gains (see \fref{fig:opt_Qising}). \\
A number of improvements and generalizations to our approach are warranted. For open quantum systems, it is important to note that protocols with non-commuting Hamiltonians may not adequately be described by adiabatic Lindblad equations such as \eref{eq:markov_relaxation} when operating in the fast driving regime \cite{erdman2019_njp}. Interestingly, since non-adiabatic corrections can potentially lead to a dependence on the control velocities \cite{Dann2018}, this would imply that instantaneous jumps are not necessarily optimal in these cases.
	\cleardoublepage
	\chapter{Conclusion and Outlook}\label{chap:conclusion}
\epigraph{We always have, as we imagine, ideas that a certain thing is true, but we don't know what it means for something to be true. You may say, ``I know what it means for it to be true that the atom is in the box; it's in the box.'' But we have found that it isn't in the box. And what you may say is, ``I know what you mean by `is in the box'. It means that if you put a detector there it'll click.'' No, that isn't what it means either.}{--- Richard Feynman, \emph{The Character of Physical Law}}

In this thesis we have covered a set of topics that shed light on the fundamental limits of the irreversibility of thermodynamic protocols. The general theme that links these results is the minimization of dissipation for finite-time operations. This is a fundamentally difficult question in thermodynamics -- whether quantum or classical -- because it requires to perform functional optimization over the solution of the dynamics of the system and environment -- which is often not solvable. Through this question we were able to probe one of the most fundamental aspects of thermodynamics: how close one can get to reversible protocols with finite resources? Furthermore, thanks to the universality of the question of minimizing work, we were also able to address some more ``applied'' optimization problems for mesoscopic systems.

We started by presenting a self-contained review on the laws of thermodynamics in \cref{chap:thermo_laws}. We then presented the mathematical formalism that is used in the rest of the thesis. In particular, in \sref{sec:geom} we generalized the results of \cite{Scandi19} to obtain a derivation of the thermodynamic geometry technique that applies to all types of dynamics -- i.e. for both open and closed quantum systems; thus giving a formal setting for the work minimization of thermodynamic protocols in the slow-driving regime, which has been used multiple times in the rest of the thesis.

In \cref{chap:landauer} we applied this extended formalism to obtain a finite-time correction to Landauer's bound on the dissipated energy required to erase information. While previous works obtaining a finite-time correction to the bound already existed, these were limited to the Markovian regime~\cite{Scandi19,VanVu2022,Zhen2021}. The main result of the chapter  \eref{MainResult} is valid beyond such approximations. Indeed, it unveils the need for strong coupling for the optimal energy management of erasure processes. Furthermore, despite the result being derived in a specific setting, it reveals the general form of a universally valid finite-time correction to Landauer's bound with the emergence of \emph{Planckian time}.

Remarkably, the results of \cref{chap:collective} show how one can use the framework of geometric thermodynamics to unveil a novel type of collective effects that allow to drastically reduce the amount of energy that is dissipated when performing a task on a collection of systems simultaneously. Collective effects are a well known and important phenomenon in physics, ranging from phase transition to quantum entanglement~\cite{Campaioli2018,Vroylandt2017,Souza2022Collective,Niedenzu_2018,Chiaracane2020}. It is often the case that one can exploit these effects to improve the result or cost of a given task, leading to a notion of collective advantages: the outcome of a task is improved when performed globally on a collection of systems instead of each system individually. Indeed, our results show that if one wishes to perform multiple tasks on multiples systems, then it is possible to exploit classical correlations between these systems to drastically suppress the amount of energy that is dissipated into the environment. To put it simply, one manages to ``share the losses'' between the individual systems. With sufficient amount of control, this suppression is strong enough to obtain a \emph{vanishing dissipation per system} in the thermodynamic limit. This is an astounding result, as it seems to go counter to the third law of thermodynamics, but on a technical level there seems to be no contradiction as the remaining finite-time dissipation is always strictly positive.

Finally, in \cref{chap:opt} we focused on the thermodynamic optimization of mesoscopic systems. We showed how to fully optimize two types of nano-scale thermal engines and fast operations on a system. The optimization of thermal engines and, more generally, operations at the nano-scale are questions that allows us to probe the fundamental limits of the trade-offs between relevant physical costs and desiderata~\cite{Holubec2014,Mehboudi2022,pietzonka2018,Abiuso2020,Miller2023}. In steady state heat engines, these trade-offs are captured by thermodynamic uncertainty relations~\cite{barato2015,pietzonka2018}.
First, we studied how one can fully optimize all the relevant thermodynamic aspects of the simplest possible thermal engine: a periodically driven quantum dot engine. Indeed, with the use of analytical techniques and reinforcement learning we were able to characterize the optimization trade-off of power, efficiency and power fluctuations of a quantum dot engine in all driving regimes. Second, we characterized and optimized the performance of an information engine at all driving speeds. Information engines stand in a class of their own, as they make use of measurements on the system instead of a cold reservoir. Furthermore, we confirmed the feasibility of the optimal protocols we computed, with an experimental implementation thanks to the collaboration of the group of prof. N. Ares. Last, but not least, we developed a general framework for the optimization of work and work fluctuations in rapidly driven systems. Our results showed that optimal fast protocols, in both open and closed system dynamics, consist of two instantaneous jumps in the control parameters -- one at start and one at the end of the protocol. Interestingly, this is true whether we are aiming to optimize work or work fluctuations -- or any trade-off of the two quantities, but it is seldom the case that the optimal ``jump point'' is the same for both objectives. We showcased this phenomenon by applying the framework to compute the optimal fast protocols across phase transitions of many-body quantum systems.\\

A substantial portion of the presented results relies on the framework of geometric thermodynamics. Therefore, it is very natural to ask how could one could generalize these results beyond the slow-driving regime -- particularly in the case of the finite-time correction to Landauer's principle. However, beyond purely technical extensions of the results, these works shed light on further questions regarding the fundamental limit of irreversible operations. First, whether one can characterize the dependence of the term $a$ -- in the main result of the chapter on Landauer's principle \eref{MainResult} -- on the type of interaction with the environment, or even a general non-trivial lower bound to it. Furthermore, the appearance of Planckian time -- which is conjectured to be the fastest relaxation time scale possible in many-body systems -- in the bound seems to suggest that it also has a role in limiting the speed of information processing and metrology.

Second, the results of \cref{chap:collective} show that it is possible to exploit correlations between multiple systems to exploit collective effects to drastically reduce energy dissipated in a thermodynamic operation. However, this is heavily dependent on the relaxation dynamics of the system. Therefore, it is left as a future challenge to characterize how these collective advantages could arise from a microscopic model of interaction with the thermal bath. This is a crucial step to eventually prove experimentally the presence of such collective advantages, which could be used one day to limit the overheating of chips in super-computers~\cite{McMahon23}. Furthermore, it is still unclear if these collective advantages are fully consistent with the third law of thermodynamics. Therefore it possible that these results help shed light on the thermodynamic consistency of master equations~\cite{Potts2021,Soret2022}. For example, if these collective advantages cannot be derived from a microscopic model, they could be taken as a witness of thermodynamic inconsistency of some master equations. Thus opening another avenue on the question of which constraints one needs to impose on a master equation to ensure its thermodynamic consistency. Conversely,
it is also possible that inconsistencies do not arise, which could be because of a divergence in the \emph{complexity} of the protocols that realize these collective advantages -- whose role in the third law of thermodynamics has recently been receiving increased attention~\cite{Taranto2023}.

	\cleardoublepage
	\phantomsection

	\renewcommand*{\bibfont}{\footnotesize}
	\addcontentsline{toc}{chapter}{Bibliography}
	\printbibliography

@article{Einstein1905,
  author =        {Einstein, A.},
  journal =       {Annalen der Physik},
  month =         jan,
  number =        {6},
  pages =         {132–148},
  publisher =     {Wiley},
  title =         {\"{U}ber einen die Erzeugung und Verwandlung des
                   Lichtes betreffenden heuristischen Gesichtspunkt},
  volume =        {322},
  year =          {1905},
  doi =           {10.1002/andp.19053220607},
  issn =          {1521-3889},
  url =           {http://dx.doi.org/10.1002/andp.19053220607},
}

@book{vonNeuman1932,
  address =       {Princeton, NJ},
  author =        {von Neumann, John},
  month =         feb,
  publisher =     {Princeton University Press},
  title =         {Mathematical foundations of quantum mechanics},
  year =          {1932},
  language =      {en},
}

@article{vonNeumann2010,
  author =        {von Neumann, J.},
  journal =       {The European Physical Journal H},
  number =        {2},
  pages =         {201–237},
  publisher =     {Springer Science and Business Media LLC},
  title =         {Proof of the ergodic theorem and the H-theorem in
                   quantum mechanics: Translation of: Beweis des
                   Ergodensatzes und des H-Theorems in der neuen
                   Mechanik},
  volume =        {35},
  year =          {2010},
  doi =           {10.1140/epjh/e2010-00008-5},
  issn =          {2102-6467},
  url =           {http://dx.doi.org/10.1140/epjh/e2010-00008-5},
}

@article{Landauer1961,
  author =        {Landauer, R.},
  journal =       {IBM Journal of Research and Development},
  number =        {3},
  pages =         {183--191},
  publisher =     {Ibm},
  title =         {Irreversibility and heat generation in the computing
                   process},
  volume =        {5},
  year =          {1961},
  doi =           {10.1147/rd.53.0183},
}

@article{Jarzynski1997,
  author =        {Jarzynski, C},
  journal =       {Phys. Rev. Lett.},
  number =        {14},
  pages =         {2690},
  title =         {{Nonequilibrium equality for free energy
                   differences}},
  volume =        {78},
  year =          {1997},
  doi =           {10.1103/PhysRevLett.78.2690},
  url =           {https://doi.org/10.1103/PhysRevLett.78.2690},
}

@article{Crooks1999,
  author =        {Crooks, Gavin E.},
  journal =       {Phys. Rev. E},
  month =         {Sep},
  pages =         {2721--2726},
  publisher =     {American Physical Society},
  title =         {Entropy production fluctuation theorem and the
                   nonequilibrium work relation for free energy
                   differences},
  volume =        {60},
  year =          {1999},
  doi =           {10.1103/PhysRevE.60.2721},
  url =           {https://link.aps.org/doi/10.1103/PhysRevE.60.2721},
}

@article{Ronagel2016,
  author =        {Roßnagel, Johannes and Dawkins, Samuel T. and
                   Tolazzi, Karl N. and Abah, Obinna and Lutz, Eric and
                   Schmidt-Kaler, Ferdinand and Singer, Kilian},
  journal =       {Science},
  month =         apr,
  number =        {6283},
  pages =         {325–329},
  publisher =     {American Association for the Advancement of Science
                   (AAAS)},
  title =         {A single-atom heat engine},
  volume =        {352},
  year =          {2016},
  doi =           {10.1126/science.aad6320},
  issn =          {1095-9203},
  url =           {http://dx.doi.org/10.1126/science.aad6320},
}

@article{Rademacher2022,
  author =        {Rademacher, Markus and Konopik, Michael and
                   Debiossac, Maxime and Grass, David and Lutz, Eric and
                   Kiesel, Nikolai},
  journal =       {Phys. Rev. Lett.},
  month =         feb,
  number =        {7},
  publisher =     {American Physical Society (APS)},
  title =         {Nonequilibrium Control of Thermal and Mechanical
                   Changes in a Levitated System},
  volume =        {128},
  year =          {2022},
  doi =           {10.1103/physrevlett.128.070601},
  issn =          {1079-7114},
  url =           {http://dx.doi.org/10.1103/PhysRevLett.128.070601},
}

@article{Scandi19,
  author =        {Scandi, Matteo and {Perarnau-Llobet}, Mart{\'i}},
  journal =       {Quantum},
  month =         oct,
  pages =         {197},
  title =         {Thermodynamic Length in Open Quantum Systems},
  volume =        {3},
  year =          {2019},
  doi =           {10.22331/q-2019-10-24-197},
}

@article{VanVu2022,
  author =        {Van Vu, T. and Saito, K.},
  journal =       {Phys. Rev. Lett.},
  month =         {Jan},
  pages =         {010602},
  publisher =     {American Physical Society},
  title =         {Finite-Time Quantum {L}andauer Principle and Quantum
                   Coherence},
  volume =        {128},
  year =          {2022},
  doi =           {10.1103/PhysRevLett.128.010602},
  url =           {https://link.aps.org/doi/10.1103/PhysRevLett.128.010602},
}

@article{Zhen2021,
  author =        {Zhen, Yi-Zheng and Egloff, Dario and Modi, Kavan and
                   Dahlsten, Oscar},
  journal =       {Phys. Rev. Lett.},
  month =         nov,
  number =        {19},
  publisher =     {American Physical Society (APS)},
  title =         {Universal Bound on Energy Cost of Bit Reset in Finite
                   Time},
  volume =        {127},
  year =          {2021},
  doi =           {10.1103/physrevlett.127.190602},
  issn =          {1079-7114},
  url =           {http://dx.doi.org/10.1103/PhysRevLett.127.190602},
}

@article{Rolandi2023Quantum,
  author =        {Rolandi, Alberto and Perarnau-Llobet, Mart{\'{i}}},
  journal =       {{Quantum}},
  month =         nov,
  pages =         {1161},
  publisher =     {{Verein zur F{\"{o}}rderung des Open Access
                   Publizierens in den Quantenwissenschaften}},
  title =         {Finite-time {L}andauer principle beyond weak
                   coupling},
  volume =        {7},
  year =          {2023},
  doi =           {10.22331/q-2023-11-03-1161},
  issn =          {2521-327X},
  url =           {https://doi.org/10.22331/q-2023-11-03-1161},
}

@article{Rolandi2023Collective,
  author =        {Rolandi, Alberto and Abiuso, Paolo and
                   Perarnau-Llobet, Martí},
  journal =       {Phys. Rev. Lett.},
  month =         nov,
  number =        {21},
  publisher =     {American Physical Society (APS)},
  title =         {Collective Advantages in Finite-Time Thermodynamics},
  volume =        {131},
  year =          {2023},
  doi =           {10.1103/physrevlett.131.210401},
  issn =          {1079-7114},
  url =           {http://dx.doi.org/10.1103/PhysRevLett.131.210401},
}

@incollection{Campaioli2018,
  author =        {Francesco Campaioli and Felix A. Pollock and
                   Sai Vinjanampathy},
  booktitle =     {Fundamental Theories of Physics},
  pages =         {207--225},
  publisher =     {Springer International Publishing},
  title =         {Quantum Batteries},
  year =          {2018},
  doi =           {10.1007/978-3-319-99046-0_8},
  url =           {https://doi.org/10.1007/978-3-319-99046-0_8},
}

@article{Vroylandt2017,
  author =        {Hadrien Vroylandt and Massimiliano Esposito and
                   Gatien Verley},
  journal =       {{EPL} (Europhysics Letters)},
  month =         nov,
  number =        {3},
  pages =         {30009},
  publisher =     {{IOP} Publishing},
  title =         {Collective effects enhancing power and efficiency},
  volume =        {120},
  year =          {2017},
  doi =           {10.1209/0295-5075/120/30009},
  url =           {https://doi.org/10.1209/0295-5075/120/30009},
}

@article{Souza2022Collective,
  author =        {Souza, Leonardo da Silva and Manzano, Gonzalo and
                   Fazio, Rosario and Iemini, Fernando},
  journal =       {Phys. Rev. E},
  month =         {Jul},
  pages =         {014143},
  publisher =     {American Physical Society},
  title =         {Collective effects on the performance and stability
                   of quantum heat engines},
  volume =        {106},
  year =          {2022},
  doi =           {10.1103/PhysRevE.106.014143},
  url =           {https://link.aps.org/doi/10.1103/PhysRevE.106.014143},
}

@article{Niedenzu_2018,
  author =        {Wolfgang Niedenzu and Gershon Kurizki},
  journal =       {New Journal of Physics},
  month =         {nov},
  number =        {11},
  pages =         {113038},
  publisher =     {IOP Publishing},
  title =         {Cooperative many-body enhancement of quantum thermal
                   machine power},
  volume =        {20},
  year =          {2018},
  doi =           {10.1088/1367-2630/aaed55},
  url =           {https://dx.doi.org/10.1088/1367-2630/aaed55},
}

@article{Chiaracane2020,
  author =        {Cecilia Chiaracane and Mark T. Mitchison and
                   Archak Purkayastha and G{\'{e}}raldine Haack and
                   John Goold},
  journal =       {Phys. Rev. Res.},
  month =         jan,
  number =        {1},
  publisher =     {American Physical Society ({APS})},
  title =         {Quasiperiodic quantum heat engines with a mobility
                   edge},
  volume =        {2},
  year =          {2020},
  doi =           {10.1103/physrevresearch.2.013093},
  url =           {https://doi.org/10.1103/physrevresearch.2.013093},
}

@article{Holubec2014,
  author =        {Holubec, Viktor},
  journal =       {Journal of Statistical Mechanics: Theory and
                   Experiment},
  month =         may,
  number =        {5},
  pages =         {P05022},
  publisher =     {IOP Publishing},
  title =         {An exactly solvable model of a stochastic heat
                   engine: optimization of power, power fluctuations and
                   efficiency},
  volume =        {2014},
  year =          {2014},
  doi =           {10.1088/1742-5468/2014/05/p05022},
  issn =          {1742-5468},
  url =           {http://dx.doi.org/10.1088/1742-5468/2014/05/p05022},
}

@article{Mehboudi2022,
  author =        {Mehboudi, Mohammad and Miller, Harry J. D.},
  journal =       {Phys. Rev. A},
  month =         {Jun},
  pages =         {062434},
  publisher =     {American Physical Society},
  title =         {Thermodynamic length and work optimization for
                   Gaussian quantum states},
  volume =        {105},
  year =          {2022},
  doi =           {10.1103/PhysRevA.105.062434},
  url =           {https://link.aps.org/doi/10.1103/PhysRevA.105.062434},
}

@article{pietzonka2018,
  author =        {Pietzonka, Patrick and Seifert, Udo},
  journal =       {Phys. Rev. Lett.},
  month =         may,
  number =        {19},
  publisher =     {American Physical Society (APS)},
  title =         {Universal Trade-Off between Power, Efficiency, and
                   Constancy in Steady-State Heat Engines},
  volume =        {120},
  year =          {2018},
  doi =           {10.1103/physrevlett.120.190602},
  issn =          {1079-7114},
  url =           {http://dx.doi.org/10.1103/PhysRevLett.120.190602},
}

@article{Abiuso2020,
  author =        {Abiuso, Paolo and Perarnau-Llobet, Mart\'{\i}},
  journal =       {Phys. Rev. Lett.},
  month =         {Mar},
  pages =         {110606},
  publisher =     {American Physical Society},
  title =         {Optimal Cycles for Low-Dissipation Heat Engines},
  volume =        {124},
  year =          {2020},
  doi =           {10.1103/PhysRevLett.124.110606},
  url =           {https://link.aps.org/doi/10.1103/PhysRevLett.124.110606},
}

@article{Miller2023,
  author =        {Miller, Harry J. D. and Perarnau Llobet, Martí},
  journal =       {SciPost Physics},
  month =         apr,
  number =        {4},
  publisher =     {Stichting SciPost},
  title =         {Finite-time bounds on the probabilistic violation of
                   the second law of thermodynamics},
  volume =        {14},
  year =          {2023},
  doi =           {10.21468/scipostphys.14.4.072},
  issn =          {2542-4653},
  url =           {http://dx.doi.org/10.21468/SciPostPhys.14.4.072},
}

@article{barato2015,
  author =        {Barato, Andre C. and Seifert, Udo},
  journal =       {Phys. Rev. Lett.},
  month =         apr,
  number =        {15},
  publisher =     {American Physical Society (APS)},
  title =         {Thermodynamic Uncertainty Relation for Biomolecular
                   Processes},
  volume =        {114},
  year =          {2015},
  doi =           {10.1103/physrevlett.114.158101},
  issn =          {1079-7114},
  url =           {http://dx.doi.org/10.1103/PhysRevLett.114.158101},
}

@article{Erdman2023Pareto,
  author =        {Erdman, Paolo A. and Rolandi, Alberto and
                   Abiuso, Paolo and Perarnau-Llobet, Mart\'{\i} and
                   No\'e, Frank},
  journal =       {Phys. Rev. Res.},
  month =         {Apr},
  pages =         {L022017},
  publisher =     {American Physical Society},
  title =         {Pareto-optimal cycles for power, efficiency and
                   fluctuations of quantum heat engines using
                   reinforcement learning},
  volume =        {5},
  year =          {2023},
  doi =           {10.1103/PhysRevResearch.5.L022017},
  url =           {https://link.aps.org/doi/10.1103/PhysRevResearch.5.L022017},
}

@article{Rolandi2023Fast,
  author =        {Rolandi, Alberto and Perarnau Llobet, Martí and
                   Miller, Harry J. D.},
  journal =       {New Journal of Physics},
  title =         {Optimal control of dissipation and work fluctuations
                   for rapidly driven systems},
  year =          {2023},
  doi =           {10.1088/1367-2630/ace2e3},
  url =           {http://iopscience.iop.org/article/10.1088/1367-2630/ace2e3},
}

@book{Fermi56,
  author =        {Fermi, Enrico},
  publisher =     {Dover Publications},
  title =         {Thermodynamics},
  year =          {1956},
  isbn =          {9780486603612},
}

@article{Dann2023,
  author =        {Dann, Roie and Kosloff, Ronnie},
  journal =       {New Journal of Physics},
  month =         apr,
  number =        {4},
  pages =         {043019},
  publisher =     {IOP Publishing},
  title =         {Unification of the first law of quantum
                   thermodynamics},
  volume =        {25},
  year =          {2023},
  doi =           {10.1088/1367-2630/acc967},
  issn =          {1367-2630},
  url =           {http://dx.doi.org/10.1088/1367-2630/acc967},
}

@article{Jarzynski2011,
  author =        {Christopher Jarzynski},
  journal =       {Annual Review of Condensed Matter Physics},
  month =         mar,
  number =        {1},
  pages =         {329--351},
  publisher =     {Annual Reviews},
  title =         {Equalities and Inequalities: Irreversibility and the
                   Second Law of Thermodynamics at the Nanoscale},
  volume =        {2},
  year =          {2011},
  doi =           {10.1146/annurev-conmatphys-062910-140506},
  url =           {https://doi.org/10.1146/annurev-conmatphys-062910-140506},
}

@article{Clausius1865,
  author =        {Clausius, R.},
  journal =       {Annalen der Physik},
  number =        {7},
  pages =         {353-400},
  title =         {Ueber verschiedene für die Anwendung bequeme Formen
                   der Hauptgleichungen der mechanischen Wärmetheorie},
  volume =        {201},
  year =          {1865},
  doi =           {10.1002/andp.18652010702},
  url =           {https://onlinelibrary.wiley.com/doi/abs/10.1002/
                  andp.18652010702},
}

@article{Strasberg2021,
  author =        {Strasberg, Philipp and Winter, Andreas},
  journal =       {PRX Quantum},
  month =         aug,
  number =        {3},
  publisher =     {American Physical Society (APS)},
  title =         {First and Second Law of Quantum Thermodynamics: A
                   Consistent Derivation Based on a Microscopic
                   Definition of Entropy},
  volume =        {2},
  year =          {2021},
  doi =           {10.1103/prxquantum.2.030202},
  issn =          {2691-3399},
  url =           {http://dx.doi.org/10.1103/PRXQuantum.2.030202},
}

@article{PerarnauLlobet2015,
  author =        {Perarnau-Llobet, Martí and Hovhannisyan, Karen V. and
                   Huber, Marcus and Skrzypczyk, Paul and Tura, Jordi and
                   Acín, Antonio},
  journal =       {Phys. Rev. E},
  month =         oct,
  number =        {4},
  publisher =     {American Physical Society (APS)},
  title =         {Most energetic passive states},
  volume =        {92},
  year =          {2015},
  doi =           {10.1103/physreve.92.042147},
  issn =          {1550-2376},
  url =           {http://dx.doi.org/10.1103/PhysRevE.92.042147},
}

@article{Brunner2012,
  author =        {Brunner, Nicolas and Linden, Noah and Popescu, Sandu and
                   Skrzypczyk, Paul},
  journal =       {Phys. Rev. E},
  month =         {May},
  pages =         {051117},
  publisher =     {American Physical Society},
  title =         {Virtual qubits, virtual temperatures, and the
                   foundations of thermodynamics},
  volume =        {85},
  year =          {2012},
  doi =           {10.1103/PhysRevE.85.051117},
  url =           {https://link.aps.org/doi/10.1103/PhysRevE.85.051117},
}

@article{zulkowski2013optimal,
  author =        {P. R. Zulkowski and D. A. Sivak and M. R. DeWeese},
  journal =       {PloS one},
  number =        {12},
  pages =         {e82754},
  publisher =     {Public Library of Science},
  title =         {Optimal control of transitions between nonequilibrium
                   steady states},
  volume =        {8},
  year =          {2013},
  doi =           {10.1371/journal.pone.0082754},
}

@article{Landi2021rev,
  author =        {Landi, Gabriel T. and Paternostro, Mauro},
  journal =       {Rev. Mod. Phys.},
  month =         {Sep},
  pages =         {035008},
  publisher =     {American Physical Society},
  title =         {Irreversible entropy production: From classical to
                   quantum},
  volume =        {93},
  year =          {2021},
  doi =           {10.1103/RevModPhys.93.035008},
  url =           {https://link.aps.org/doi/10.1103/RevModPhys.93.035008},
}

@article{Kox2006,
  author =        {Kox, A.J.},
  journal =       {Studies in History and Philosophy of Science Part B:
                   Studies in History and Philosophy of Modern Physics},
  month =         mar,
  number =        {1},
  pages =         {101–114},
  publisher =     {Elsevier BV},
  title =         {Confusion and clarification: Albert Einstein and
                   Walther Nernst’s Heat Theorem, 1911–1916},
  volume =        {37},
  year =          {2006},
  doi =           {10.1016/j.shpsb.2005.10.001},
  issn =          {1355-2198},
  url =           {http://dx.doi.org/10.1016/j.shpsb.2005.10.001},
}

@article{Masanes2017,
  author =        {Llu{\'{\i}}s Masanes and Jonathan Oppenheim},
  journal =       {Nature Communications},
  month =         mar,
  number =        {1},
  publisher =     {Springer Science and Business Media {LLC}},
  title =         {A general derivation and quantification of the third
                   law of thermodynamics},
  volume =        {8},
  year =          {2017},
  doi =           {10.1038/ncomms14538},
  url =           {https://doi.org/10.1038/ncomms14538},
}

@article{Wilming2017,
  author =        {Wilming, Henrik and Gallego, Rodrigo},
  journal =       {Phys. Rev. X},
  month =         nov,
  number =        {4},
  publisher =     {American Physical Society (APS)},
  title =         {Third Law of Thermodynamics as a Single Inequality},
  volume =        {7},
  year =          {2017},
  doi =           {10.1103/physrevx.7.041033},
  issn =          {2160-3308},
  url =           {http://dx.doi.org/10.1103/PhysRevX.7.041033},
}

@article{Nernst1906,
  author =        {Nernst, W.},
  journal =       {Nachrichten von der Gesellschaft der Wissenschaften
                   zu Göttingen, Mathematisch-Physikalische Klasse},
  pages =         {1-40},
  title =         {Ueber die Berechnung chemischer Gleichgewichte aus
                   thermischen Messungen},
  volume =        {1906},
  year =          {1906},
  url =           {http://eudml.org/doc/58630},
}

@article{Taranto2023,
  author =        {Philip Taranto and Faraj Bakhshinezhad and
                   Andreas Bluhm and Ralph Silva and Nicolai Friis and
                   Maximilian P.E. Lock and Giuseppe Vitagliano and
                   Felix C. Binder and Tiago Debarba and
                   Emanuel Schwarzhans and Fabien Clivaz and
                   Marcus Huber},
  journal =       {{PRX} Quantum},
  month =         mar,
  number =        {1},
  publisher =     {American Physical Society ({APS})},
  title =         {Landauer Versus Nernst: What is the True Cost of
                   Cooling a Quantum System?},
  volume =        {4},
  year =          {2023},
  doi =           {10.1103/prxquantum.4.010332},
  url =           {https://doi.org/10.1103/prxquantum.4.010332},
}

@article{Manzano2020,
  author =        {Manzano, Daniel},
  journal =       {AIP Advances},
  month =         feb,
  number =        {2},
  publisher =     {AIP Publishing},
  title =         {A short introduction to the Lindblad master equation},
  volume =        {10},
  year =          {2020},
  doi =           {10.1063/1.5115323},
  issn =          {2158-3226},
  url =           {http://dx.doi.org/10.1063/1.5115323},
}

@article{Abiuso2020Geometric,
  author =        {Paolo Abiuso and Harry J. D. Miller and
                   Mart{\'{\i}} Perarnau-Llobet and Matteo Scandi},
  journal =       {Entropy},
  month =         sep,
  number =        {10},
  pages =         {1076},
  publisher =     {{MDPI} {AG}},
  title =         {Geometric Optimisation of Quantum Thermodynamic
                   Processes},
  volume =        {22},
  year =          {2020},
  doi =           {10.3390/e22101076},
  url =           {https://doi.org/10.3390/e22101076},
}

@article{Cresser2021,
  author =        {Cresser, J. D. and Anders, J.},
  journal =       {Phys. Rev. Lett.},
  month =         {Dec},
  pages =         {250601},
  publisher =     {American Physical Society},
  title =         {Weak and Ultrastrong Coupling Limits of the Quantum
                   Mean Force Gibbs State},
  volume =        {127},
  year =          {2021},
  doi =           {10.1103/PhysRevLett.127.250601},
  url =           {https://link.aps.org/doi/10.1103/PhysRevLett.127.250601},
}

@article{Eisert2015,
  author =        {J. Eisert and M. Friesdorf and C. Gogolin},
  journal =       {Nature Physics},
  month =         feb,
  number =        {2},
  pages =         {124--130},
  publisher =     {Springer Science and Business Media {LLC}},
  title =         {Quantum many-body systems out of equilibrium},
  volume =        {11},
  year =          {2015},
  doi =           {10.1038/nphys3215},
  url =           {https://doi.org/10.1038/nphys3215},
}

@misc{jake_drawing,
  author =        {Xuereb, Jake},
  month =         aug,
  title =         {Illustration of the geometric thermodynamics
                   manifold},
  year =          {2024},
}

@book{Boullion1971,
  author =        {Boullion, T L and Odell, P L},
  month =         jul,
  publisher =     {Wiley-Interscience},
  title =         {Generalized inverse matrices},
  year =          {1971},
}

@article{Schmiedl2007opt,
  author =        {Schmiedl, Tim and Seifert, Udo},
  journal =       {Phys. Rev. Lett.},
  month =         {Mar},
  pages =         {108301},
  publisher =     {American Physical Society},
  title =         {Optimal Finite-Time Processes In Stochastic
                   Thermodynamics},
  volume =        {98},
  year =          {2007},
  doi =           {10.1103/PhysRevLett.98.108301},
  url =           {https://link.aps.org/doi/10.1103/PhysRevLett.98.108301},
}

@article{Esposito2010,
  author =        {M. Esposito and R. Kawai and K. Lindenberg and
                   C. Van den Broeck},
  journal =       {{EPL} (Europhysics Letters)},
  month =         jan,
  number =        {2},
  pages =         {20003},
  publisher =     {{IOP} Publishing},
  title =         {Finite-time thermodynamics for a single-level quantum
                   dot},
  volume =        {89},
  year =          {2010},
  doi =           {10.1209/0295-5075/89/20003},
  url =           {https://doi.org/10.1209/0295-5075/89/20003},
}

@article{Sagawa2009,
  author =        {Sagawa, Takahiro and Ueda, Masahito},
  journal =       {Phys. Rev. Lett.},
  month =         {Jun},
  pages =         {250602},
  publisher =     {American Physical Society},
  title =         {Minimal Energy Cost for Thermodynamic Information
                   Processing: Measurement and Information Erasure},
  volume =        {102},
  year =          {2009},
  doi =           {10.1103/PhysRevLett.102.250602},
  url =           {https://link.aps.org/doi/10.1103/PhysRevLett.102.250602},
}

@article{Esposito2011,
  author =        {M. Esposito and C. Van den Broeck},
  journal =       {{EPL} (Europhysics Letters)},
  month =         aug,
  number =        {4},
  pages =         {40004},
  publisher =     {{IOP} Publishing},
  title =         {Second law and Landauer principle far from
                   equilibrium},
  volume =        {95},
  year =          {2011},
  doi =           {10.1209/0295-5075/95/40004},
  url =           {https://doi.org/10.1209/0295-5075/95/40004},
}

@misc{ciampini2021experimental,
  author =        {Ciampini, Mario A. and Wenzl, Tobias and
                   Konopik, Michael and Thalhammer, Gregor and
                   Aspelmeyer, Markus and Lutz, Eric and
                   Kiesel, Nikolai},
  publisher =     {arXiv},
  title =         {Experimental nonequilibrium memory erasure beyond
                   Landauer's bound},
  year =          {2021},
  doi =           {10.48550/ARXIV.2107.04429},
  url =           {https://arxiv.org/abs/2107.04429},
}

@article{Scandi2022,
  author =        {Scandi, Matteo and Barker, David and
                   Lehmann, Sebastian and Dick, Kimberly A. and
                   Maisi, Ville F. and Perarnau-Llobet, Mart\'{\i}},
  journal =       {Phys. Rev. Lett.},
  month =         {Dec},
  pages =         {270601},
  publisher =     {American Physical Society},
  title =         {Minimally Dissipative Information Erasure in a
                   Quantum Dot via Thermodynamic Length},
  volume =        {129},
  year =          {2022},
  doi =           {10.1103/PhysRevLett.129.270601},
  url =           {https://link.aps.org/doi/10.1103/PhysRevLett.129.270601},
}

@article{vanvu2023,
  author =        {Van Vu, Tan and Saito, Keiji},
  journal =       {Phys. Rev. X},
  month =         {Feb},
  pages =         {011013},
  publisher =     {American Physical Society},
  title =         {Thermodynamic Unification of Optimal Transport:
                   Thermodynamic Uncertainty Relation, Minimum
                   Dissipation, and Thermodynamic Speed Limits},
  volume =        {13},
  year =          {2023},
  doi =           {10.1103/PhysRevX.13.011013},
  url =           {https://link.aps.org/doi/10.1103/PhysRevX.13.011013},
}

@article{VanVu2021,
  author =        {Van Vu, Tan and Hasegawa, Yoshihiko},
  journal =       {Phys. Rev. Lett.},
  month =         {Jan},
  pages =         {010601},
  publisher =     {American Physical Society},
  title =         {Geometrical Bounds of the Irreversibility in
                   Markovian Systems},
  volume =        {126},
  year =          {2021},
  doi =           {10.1103/PhysRevLett.126.010601},
  url =           {https://link.aps.org/doi/10.1103/PhysRevLett.126.010601},
}

@article{Zulkowski2014,
  author =        {Zulkowski, P. R. and DeWeese, M. R.},
  journal =       {Phys. Rev. E},
  month =         {May},
  pages =         {052140},
  publisher =     {American Physical Society},
  title =         {Optimal finite-time erasure of a classical bit},
  volume =        {89},
  year =          {2014},
  doi =           {10.1103/PhysRevE.89.052140},
  url =           {https://link.aps.org/doi/10.1103/PhysRevE.89.052140},
}

@article{Zhen2022,
  author =        {Zhen, Y.-Z. and Egloff, D. and Modi, K. and
                   Dahlsten, O.},
  journal =       {Phys. Rev. E},
  month =         {Apr},
  pages =         {044147},
  publisher =     {American Physical Society},
  title =         {Inverse linear versus exponential scaling of work
                   penalty in finite-time bit reset},
  volume =        {105},
  year =          {2022},
  doi =           {10.1103/PhysRevE.105.044147},
  url =           {https://link.aps.org/doi/10.1103/PhysRevE.105.044147},
}

@article{Proesmans2020,
  author =        {Proesmans, K. and Ehrich, J. and Bechhoefer, J.},
  journal =       {Phys. Rev. Lett.},
  month =         {Sep},
  pages =         {100602},
  publisher =     {American Physical Society},
  title =         {Finite-Time {L}andauer Principle},
  volume =        {125},
  year =          {2020},
  doi =           {10.1103/PhysRevLett.125.100602},
  url =           {https://link.aps.org/doi/10.1103/PhysRevLett.125.100602},
}

@article{Rivas2020,
  author =        {Rivas, \'Angel},
  journal =       {Phys. Rev. Lett.},
  month =         {Apr},
  pages =         {160601},
  publisher =     {American Physical Society},
  title =         {Strong Coupling Thermodynamics of Open Quantum
                   Systems},
  volume =        {124},
  year =          {2020},
  doi =           {10.1103/PhysRevLett.124.160601},
  url =           {https://link.aps.org/doi/10.1103/PhysRevLett.124.160601},
}

@article{PerarnauLlobet2018,
  author =        {M. Perarnau-Llobet and H. Wilming and A. Riera and
                   R. Gallego and J. Eisert},
  journal =       {Phys. Rev. Lett.},
  month =         mar,
  number =        {12},
  publisher =     {American Physical Society ({APS})},
  title =         {Strong Coupling Corrections in Quantum
                   Thermodynamics},
  volume =        {120},
  year =          {2018},
  doi =           {10.1103/physrevlett.120.120602},
  url =           {https://doi.org/10.1103/physrevlett.120.120602},
}

@book{schaller2014open,
  author =        {Schaller, Gernot},
  publisher =     {Springer},
  title =         {Open quantum systems far from equilibrium},
  volume =        {881},
  year =          {2014},
  doi =           {10.1007/978-3-319-03877-3},
}

@article{Hartnoll22,
  author =        {Hartnoll, Sean A. and Mackenzie, Andrew P.},
  journal =       {Rev. Mod. Phys.},
  month =         {Nov},
  pages =         {041002},
  publisher =     {American Physical Society},
  title =         {Colloquium: Planckian dissipation in metals},
  volume =        {94},
  year =          {2022},
  doi =           {10.1103/RevModPhys.94.041002},
  url =           {https://link.aps.org/doi/10.1103/RevModPhys.94.041002},
}

@article{Rochette2019,
  author =        {S. Rochette and M. Rudolph and A.-M. Roy and
                   M. J. Curry and G. A. Ten Eyck and R. P. Manginell and
                   J. R. Wendt and T. Pluym and S. M. Carr and
                   D. R. Ward and M. P. Lilly and M. S. Carroll and
                   M. Pioro-Ladri{\`{e}}re},
  journal =       {Applied Physics Letters},
  month =         feb,
  number =        {8},
  pages =         {083101},
  publisher =     {{AIP} Publishing},
  title =         {Quantum dots with split enhancement gate tunnel
                   barrier control},
  volume =        {114},
  year =          {2019},
  doi =           {10.1063/1.5091111},
  url =           {https://doi.org/10.1063/1.5091111},
}

@article{Covito2018,
  author =        {F. Covito and F. G. Eich and R. Tuovinen and
                   M. A. Sentef and A. Rubio},
  journal =       {Journal of Chemical Theory and Computation},
  month =         apr,
  number =        {5},
  pages =         {2495--2504},
  publisher =     {American Chemical Society ({ACS})},
  title =         {Transient Charge and Energy Flow in the Wide-Band
                   Limit},
  volume =        {14},
  year =          {2018},
  doi =           {10.1021/acs.jctc.8b00077},
  url =           {https://doi.org/10.1021/acs.jctc.8b00077},
}

@misc{Bettmann2024Meso,
  author =        {Bettmann, Laetitia P. and Kewming, Michael J. and
                   Landi, Gabriel T. and Goold, John and
                   Mitchison, Mark T.},
  publisher =     {arXiv},
  title =         {Quantum stochastic thermodynamics in the
                   mesoscopic-leads formulation},
  year =          {2024},
  doi =           {10.48550/ARXIV.2404.06426},
  url =           {https://arxiv.org/abs/2404.06426},
}

@article{Timpanaro2020,
  author =        {Timpanaro, Andr\'e M. and Santos, Jader P. and
                   Landi, Gabriel T.},
  journal =       {Phys. Rev. Lett.},
  month =         {Jun},
  pages =         {240601},
  publisher =     {American Physical Society},
  title =         {Landauer's Principle at Zero Temperature},
  volume =        {124},
  year =          {2020},
  doi =           {10.1103/PhysRevLett.124.240601},
  url =           {https://link.aps.org/doi/10.1103/PhysRevLett.124.240601},
}

@book{Fox87,
  author =        {L. Fox and D. F. Mayers},
  publisher =     {Springer Netherlands},
  title =         {Numerical Solution of Ordinary Differential
                   Equations},
  year =          {1987},
  doi =           {10.1007/978-94-009-3129-9},
  url =           {https://doi.org/10.1007/978-94-009-3129-9},
}

@article{Giovannetti2006,
  author =        {Giovannetti, Vittorio and Lloyd, Seth and
                   Maccone, Lorenzo},
  journal =       {Phys. Rev. Lett.},
  month =         {Jan},
  pages =         {010401},
  publisher =     {American Physical Society},
  title =         {Quantum Metrology},
  volume =        {96},
  year =          {2006},
  doi =           {10.1103/PhysRevLett.96.010401},
  url =           {https://link.aps.org/doi/10.1103/PhysRevLett.96.010401},
}

@misc{maria_drawing,
  author =        {Pereira, Maria Ana},
  month =         jan,
  title =         {Illustration individual and steamlined production},
  year =          {2024},
}

@article{Salamon1985,
  author =        {Peter Salamon and James D. Nulton and
                   R. Stephen Berry},
  journal =       {The Journal of Chemical Physics},
  month =         mar,
  number =        {5},
  pages =         {2433--2436},
  publisher =     {{AIP} Publishing},
  title =         {Length in statistical thermodynamics},
  volume =        {82},
  year =          {1985},
  doi =           {10.1063/1.448337},
  url =           {https://doi.org/10.1063/1.448337},
}

@article{Crooks2007,
  author =        {Crooks, Gavin E.},
  journal =       {Phys. Rev. Lett.},
  month =         sep,
  number =        {10},
  publisher =     {American Physical Society (APS)},
  title =         {Measuring Thermodynamic Length},
  volume =        {99},
  year =          {2007},
  doi =           {10.1103/physrevlett.99.100602},
  issn =          {1079-7114},
  url =           {http://dx.doi.org/10.1103/PhysRevLett.99.100602},
}

@article{Scandi2020,
  author =        {Scandi, Matteo and Miller, Harry J. D. and
                   Anders, Janet and Perarnau-Llobet, Mart\'{\i}},
  journal =       {Phys. Rev. Research},
  month =         {Jun},
  pages =         {023377},
  publisher =     {American Physical Society},
  title =         {Quantum work statistics close to equilibrium},
  volume =        {2},
  year =          {2020},
  doi =           {10.1103/PhysRevResearch.2.023377},
  url =           {https://link.aps.org/doi/10.1103/PhysRevResearch.2.023377},
}

@article{Abiuso2024,
  author =        {Abiuso, Paolo and Andrea Erdman, Paolo and
                   Ronen, Michael and Noé, Frank and Haack, Géraldine and
                   Perarnau-Llobet, Martí},
  journal =       {Quantum Science and Technology},
  month =         apr,
  number =        {3},
  pages =         {035008},
  publisher =     {IOP Publishing},
  title =         {Optimal thermometers with spin networks},
  volume =        {9},
  year =          {2024},
  doi =           {10.1088/2058-9565/ad37d3},
  issn =          {2058-9565},
  url =           {http://dx.doi.org/10.1088/2058-9565/ad37d3},
}

@article{Seifert2012,
  author =        {Udo Seifert},
  journal =       {Reports on Progress in Physics},
  month =         nov,
  number =        {12},
  pages =         {126001},
  publisher =     {{IOP} Publishing},
  title =         {Stochastic thermodynamics, fluctuation theorems and
                   molecular machines},
  volume =        {75},
  year =          {2012},
  doi =           {10.1088/0034-4885/75/12/126001},
  url =           {https://doi.org/10.1088/0034-4885/75/12/126001},
}

@article{Blaber2020Skewed,
  author =        {Steven Blaber and David A. Sivak},
  journal =       {The Journal of Chemical Physics},
  month =         dec,
  number =        {24},
  pages =         {244119},
  publisher =     {{AIP} Publishing},
  title =         {Skewed thermodynamic geometry and optimal free energy
                   estimation},
  volume =        {153},
  year =          {2020},
  doi =           {10.1063/5.0033405},
  url =           {https://doi.org/10.1063/5.0033405},
}

@misc{dechant2019thermodynamic,
  author =        {Dechant, Andreas and Sakurai, Yohei},
  publisher =     {arXiv},
  title =         {Thermodynamic interpretation of Wasserstein distance},
  year =          {2019},
  doi =           {10.48550/ARXIV.1912.08405},
  url =           {https://arxiv.org/abs/1912.08405},
}

@article{Carnot1824,
  author =        {Carnot, S.},
  journal =       {Annales scientifiques de l’École normale
                   supérieure},
  pages =         {393–457},
  publisher =     {Societe Mathematique de France},
  title =         {Réflexions sur la puissance motrice du feu et sur
                   les machines propres à développer cette puissance},
  volume =        {1},
  year =          {1824},
  doi =           {10.24033/asens.88},
  issn =          {1873-2151},
  url =           {http://dx.doi.org/10.24033/asens.88},
}

@article{McMahon23,
  author =        {McMahon, Peter L.},
  journal =       {Nature Reviews Physics},
  month =         oct,
  number =        {12},
  pages =         {717–734},
  publisher =     {Springer Science and Business Media LLC},
  title =         {The physics of optical computing},
  volume =        {5},
  year =          {2023},
  doi =           {10.1038/s42254-023-00645-5},
  issn =          {2522-5820},
  url =           {http://dx.doi.org/10.1038/s42254-023-00645-5},
}

@article{Das21,
  author =        {Das, Arpan and Bera, Anindita and Chakraborty, Sagnik and
                   Chruściński, Dariusz},
  journal =       {Phys. Rev. A},
  month =         oct,
  number =        {4},
  publisher =     {American Physical Society (APS)},
  title =         {Thermodynamics and the quantum speed limit in the
                   non-Markovian regime},
  volume =        {104},
  year =          {2021},
  doi =           {10.1103/physreva.104.042202},
  issn =          {2469-9934},
  url =           {http://dx.doi.org/10.1103/PhysRevA.104.042202},
}

@article{Aurell2011,
  author =        {Aurell, Erik and Mej\'{\i}a-Monasterio, Carlos and
                   Muratore-Ginanneschi, Paolo},
  journal =       {Phys. Rev. Lett.},
  month =         {Jun},
  pages =         {250601},
  publisher =     {American Physical Society},
  title =         {Optimal Protocols and Optimal Transport in Stochastic
                   Thermodynamics},
  volume =        {106},
  year =          {2011},
  doi =           {10.1103/PhysRevLett.106.250601},
  url =           {https://link.aps.org/doi/10.1103/PhysRevLett.106.250601},
}

@article{cavina2018,
  author =        {Cavina, Vasco and Mari, Andrea and Carlini, Alberto and
                   Giovannetti, Vittorio},
  journal =       {Phys. Rev. A},
  month =         jul,
  number =        {1},
  publisher =     {American Physical Society (APS)},
  title =         {Optimal thermodynamic control in open quantum
                   systems},
  volume =        {98},
  year =          {2018},
  doi =           {10.1103/physreva.98.012139},
  issn =          {2469-9934},
  url =           {http://dx.doi.org/10.1103/PhysRevA.98.012139},
}

@article{erdman2019_njp,
  author =        {Erdman, P A and Cavina, V and Fazio, R and Taddei, F and
                   Giovannetti, V},
  journal =       {New Journal of Physics},
  month =         oct,
  number =        {10},
  pages =         {103049},
  publisher =     {IOP Publishing},
  title =         {Maximum power and corresponding efficiency for
                   two-level heat engines and refrigerators: optimality
                   of fast cycles},
  volume =        {21},
  year =          {2019},
  doi =           {10.1088/1367-2630/ab4dca},
  issn =          {1367-2630},
  url =           {http://dx.doi.org/10.1088/1367-2630/ab4dca},
}

@article{geva1992,
  author =        {Geva, Eitan and Kosloff, Ronnie},
  journal =       {The Journal of Chemical Physics},
  month =         feb,
  number =        {4},
  pages =         {3054–3067},
  publisher =     {AIP Publishing},
  title =         {A quantum-mechanical heat engine operating in finite
                   time. A model consisting of spin-1/2 systems as the
                   working fluid},
  volume =        {96},
  year =          {1992},
  doi =           {10.1063/1.461951},
  issn =          {1089-7690},
  url =           {http://dx.doi.org/10.1063/1.461951},
}

@article{feldmann2000,
  author =        {Feldmann, Tova and Kosloff, Ronnie},
  journal =       {Phys. Rev. E},
  month =         may,
  number =        {5},
  pages =         {4774–4790},
  publisher =     {American Physical Society (APS)},
  title =         {Performance of discrete heat engines and heat pumps
                   in finite time},
  volume =        {61},
  year =          {2000},
  doi =           {10.1103/physreve.61.4774},
  issn =          {1095-3787},
  url =           {http://dx.doi.org/10.1103/PhysRevE.61.4774},
}

@article{Ye2022,
  author =        {Ye, Zhuolin and Cerisola, Federico and Abiuso, Paolo and
                   Anders, Janet and Perarnau-Llobet, Martí and
                   Holubec, Viktor},
  journal =       {Phys. Rev. Res.},
  month =         nov,
  number =        {4},
  publisher =     {American Physical Society (APS)},
  title =         {Optimal finite-time heat engines under constrained
                   control},
  volume =        {4},
  year =          {2022},
  doi =           {10.1103/physrevresearch.4.043130},
  issn =          {2643-1564},
  url =           {http://dx.doi.org/10.1103/PhysRevResearch.4.043130},
}

@article{Miller2019,
  author =        {Miller, Harry J. D. and Scandi, Matteo and
                   Anders, Janet and Perarnau-Llobet, Mart\'{\i}},
  journal =       {Phys. Rev. Lett.},
  month =         {Dec},
  pages =         {230603},
  publisher =     {American Physical Society},
  title =         {Work Fluctuations in Slow Processes: Quantum
                   Signatures and Optimal Control},
  volume =        {123},
  year =          {2019},
  doi =           {10.1103/PhysRevLett.123.230603},
  url =           {https://link.aps.org/doi/10.1103/PhysRevLett.123.230603},
}

@article{giazotto2006,
  author =        {Giazotto, Francesco and Heikkil\"{a}, Tero T. and
                   Luukanen, Arttu and Savin, Alexander M. and
                   Pekola, Jukka P.},
  journal =       {Reviews of Modern Physics},
  month =         mar,
  number =        {1},
  pages =         {217–274},
  publisher =     {American Physical Society (APS)},
  title =         {Opportunities for mesoscopics in thermometry and
                   refrigeration: Physics and applications},
  volume =        {78},
  year =          {2006},
  doi =           {10.1103/revmodphys.78.217},
  issn =          {1539-0756},
  url =           {http://dx.doi.org/10.1103/RevModPhys.78.217},
}

@article{Benenti2017,
  author =        {Benenti, Giuliano and Casati, Giulio and Saito, Keiji and
                   Whitney, Robert S.},
  journal =       {Physics Reports},
  month =         jun,
  pages =         {1–124},
  publisher =     {Elsevier BV},
  title =         {Fundamental aspects of steady-state conversion of
                   heat to work at the nanoscale},
  volume =        {694},
  year =          {2017},
  doi =           {10.1016/j.physrep.2017.05.008},
  issn =          {0370-1573},
  url =           {http://dx.doi.org/10.1016/j.physrep.2017.05.008},
}

@article{Agarwalla2018,
  author =        {Agarwalla, Bijay Kumar and Segal, Dvira},
  journal =       {Phys. Rev. B},
  month =         {Oct},
  pages =         {155438},
  publisher =     {American Physical Society},
  title =         {Assessing the validity of the thermodynamic
                   uncertainty relation in quantum systems},
  volume =        {98},
  year =          {2018},
  doi =           {10.1103/PhysRevB.98.155438},
  url =           {https://link.aps.org/doi/10.1103/PhysRevB.98.155438},
}

@article{barato2016,
  author =        {Barato, Andre C. and Seifert, Udo},
  journal =       {Phys. Rev. X},
  month =         dec,
  number =        {4},
  publisher =     {American Physical Society (APS)},
  title =         {Cost and Precision of Brownian Clocks},
  volume =        {6},
  year =          {2016},
  doi =           {10.1103/physrevx.6.041053},
  issn =          {2160-3308},
  url =           {http://dx.doi.org/10.1103/PhysRevX.6.041053},
}

@article{Koyuk2018,
  author =        {Koyuk, Timur and Seifert, Udo and Pietzonka, Patrick},
  journal =       {Journal of Physics A: Mathematical and Theoretical},
  month =         dec,
  number =        {2},
  pages =         {02LT02},
  publisher =     {IOP Publishing},
  title =         {A generalization of the thermodynamic uncertainty
                   relation to periodically driven systems},
  volume =        {52},
  year =          {2018},
  doi =           {10.1088/1751-8121/aaeec4},
  issn =          {1751-8121},
  url =           {http://dx.doi.org/10.1088/1751-8121/aaeec4},
}

@article{proesmans2017,
  author =        {Proesmans, Karel and Van den Broeck, Christian},
  journal =       {EPL (Europhysics Letters)},
  month =         jul,
  number =        {2},
  pages =         {20001},
  publisher =     {IOP Publishing},
  title =         {Discrete-time thermodynamic uncertainty relation},
  volume =        {119},
  year =          {2017},
  doi =           {10.1209/0295-5075/119/20001},
  issn =          {1286-4854},
  url =           {http://dx.doi.org/10.1209/0295-5075/119/20001},
}

@article{miller2021c,
  author =        {Miller, Harry J. D. and Mohammady, M. Hamed and
                   Perarnau-Llobet, Martí and Guarnieri, Giacomo},
  journal =       {Phys. Rev. E},
  month =         may,
  number =        {5},
  publisher =     {American Physical Society (APS)},
  title =         {Joint statistics of work and entropy production along
                   quantum trajectories},
  volume =        {103},
  year =          {2021},
  doi =           {10.1103/physreve.103.052138},
  issn =          {2470-0053},
  url =           {http://dx.doi.org/10.1103/PhysRevE.103.052138},
}

@article{lindblad1976,
  author =        {Lindblad, G.},
  journal =       {Communications in Mathematical Physics},
  month =         jun,
  number =        {2},
  pages =         {119–130},
  publisher =     {Springer Science and Business Media LLC},
  title =         {On the generators of quantum dynamical semigroups},
  volume =        {48},
  year =          {1976},
  doi =           {10.1007/bf01608499},
  issn =          {1432-0916},
  url =           {http://dx.doi.org/10.1007/BF01608499},
}

@book{breuer2002,
  author =        {Breuer, Heinz-Peter and Petruccione, Francesco},
  month =         jan,
  publisher =     {Oxford University PressOxford},
  title =         {The Theory of Open Quantum Systems},
  year =          {2007},
  doi =           {10.1093/acprof:oso/9780199213900.001.0001},
  isbn =          {9780191706349},
  url =           {http://dx.doi.org/10.1093/acprof:oso/
                  9780199213900.001.0001},
}

@article{sothmann2015,
  author =        {Sothmann, Bj\"{o}rn and Sánchez, Rafael and
                   Jordan, Andrew N},
  journal =       {Nanotechnology},
  month =         dec,
  number =        {3},
  pages =         {032001},
  publisher =     {IOP Publishing},
  title =         {Thermoelectric energy harvesting with quantum dots},
  volume =        {26},
  year =          {2014},
  doi =           {10.1088/0957-4484/26/3/032001},
  issn =          {1361-6528},
  url =           {http://dx.doi.org/10.1088/0957-4484/26/3/032001},
}

@article{Alicki1979,
  author =        {Alicki, R},
  journal =       {Journal of Physics A: Mathematical and General},
  month =         may,
  number =        {5},
  pages =         {L103–L107},
  publisher =     {IOP Publishing},
  title =         {The quantum open system as a model of the heat
                   engine},
  volume =        {12},
  year =          {1979},
  doi =           {10.1088/0305-4470/12/5/007},
  issn =          {1361-6447},
  url =           {http://dx.doi.org/10.1088/0305-4470/12/5/007},
}

@inproceedings{haarnoja2018_pmlr,
  author =        {Haarnoja, Tuomas and Zhou, Aurick and Abbeel, Pieter and
                   Levine, Sergey},
  booktitle =     {Proceedings of the 35th International Conference on
                   Machine Learning},
  editor =        {Dy, Jennifer and Krause, Andreas},
  month =         {Jul},
  pages =         {1861--1870},
  publisher =     {PMLR},
  series =        {Proceedings of Machine Learning Research},
  title =         {Soft Actor-Critic: Off-Policy Maximum Entropy Deep
                   Reinforcement Learning with a Stochastic Actor},
  volume =        {80},
  year =          {2018},
  url =           {https://proceedings.mlr.press/v80/haarnoja18b.html},
}

@article{Erdman2022,
  author =        {Paolo A. Erdman and Frank No{\'{e}}},
  journal =       {npj Quantum Information},
  month =         jan,
  number =        {1},
  publisher =     {Springer Science and Business Media {LLC}},
  title =         {Identifying optimal cycles in quantum thermal
                   machines with reinforcement-learning},
  volume =        {8},
  year =          {2022},
  doi =           {10.1038/s41534-021-00512-0},
  url =           {https://doi.org/10.1038/s41534-021-00512-0},
}

@article{Erdman2023,
  author =        {Erdman, Paolo A and Noé, Frank},
  editor =        {Editor, A},
  journal =       {PNAS Nexus},
  month =         aug,
  number =        {8},
  publisher =     {Oxford University Press (OUP)},
  title =         {Model-free optimization of power/efficiency tradeoffs
                   in quantum thermal machines using reinforcement
                   learning},
  volume =        {2},
  year =          {2023},
  doi =           {10.1093/pnasnexus/pgad248},
  issn =          {2752-6542},
  url =           {http://dx.doi.org/10.1093/pnasnexus/pgad248},
}

@article{Sgroi2021,
  author =        {Sgroi, Sofia and Palma, G. Massimo and
                   Paternostro, Mauro},
  journal =       {Phys. Rev. Lett.},
  month =         jan,
  number =        {2},
  publisher =     {American Physical Society (APS)},
  title =         {Reinforcement Learning Approach to Nonequilibrium
                   Quantum Thermodynamics},
  volume =        {126},
  year =          {2021},
  doi =           {10.1103/physrevlett.126.020601},
  issn =          {1079-7114},
  url =           {http://dx.doi.org/10.1103/physrevlett.126.020601},
}

@article{bukov2018,
  author =        {Bukov, Marin and Day, Alexandre G. R. and
                   Sels, Dries and Weinberg, Phillip and
                   Polkovnikov, Anatoli and Mehta, Pankaj},
  journal =       {Phys. Rev. X},
  month =         sep,
  number =        {3},
  publisher =     {American Physical Society (APS)},
  title =         {Reinforcement Learning in Different Phases of Quantum
                   Control},
  volume =        {8},
  year =          {2018},
  doi =           {10.1103/physrevx.8.031086},
  issn =          {2160-3308},
  url =           {http://dx.doi.org/10.1103/PhysRevX.8.031086},
}

@article{cavina2021,
  author =        {Cavina, Vasco and Erdman, Paolo A. and Abiuso, Paolo and
                   Tolomeo, Leonardo and Giovannetti, Vittorio},
  journal =       {Phys. Rev. A},
  month =         sep,
  number =        {3},
  publisher =     {American Physical Society (APS)},
  title =         {Maximum-power heat engines and refrigerators in the
                   fast-driving regime},
  volume =        {104},
  year =          {2021},
  doi =           {10.1103/physreva.104.032226},
  issn =          {2469-9934},
  url =           {http://dx.doi.org/10.1103/PhysRevA.104.032226},
}

@article{esposito2010efficiency,
  author =        {Esposito, Massimiliano and Kawai, Ryoichi and
                   Lindenberg, Katja and Van den Broeck, Christian},
  journal =       {Phys. Rev. Lett.},
  month =         oct,
  number =        {15},
  publisher =     {American Physical Society (APS)},
  title =         {Efficiency at Maximum Power of Low-Dissipation Carnot
                   Engines},
  volume =        {105},
  year =          {2010},
  doi =           {10.1103/physrevlett.105.150603},
  issn =          {1079-7114},
  url =           {http://dx.doi.org/10.1103/PhysRevLett.105.150603},
}

@article{Holubec2016,
  author =        {Viktor Holubec and Artem Ryabov},
  journal =       {Journal of Statistical Mechanics: Theory and
                   Experiment},
  month =         jul,
  number =        {7},
  pages =         {073204},
  publisher =     {{IOP} Publishing},
  title =         {Maximum efficiency of low-dissipation heat engines at
                   arbitrary power},
  volume =        {2016},
  year =          {2016},
  doi =           {10.1088/1742-5468/2016/07/073204},
  url =           {https://doi.org/10.1088/1742-5468/2016/07/073204},
}

@article{Salamon1983,
  author =        {Salamon, Peter and Berry, R. Stephen},
  journal =       {Phys. Rev. Lett.},
  month =         sep,
  number =        {13},
  pages =         {1127–1130},
  publisher =     {American Physical Society (APS)},
  title =         {Thermodynamic Length and Dissipated Availability},
  volume =        {51},
  year =          {1983},
  doi =           {10.1103/physrevlett.51.1127},
  issn =          {0031-9007},
  url =           {http://dx.doi.org/10.1103/PhysRevLett.51.1127},
}

@article{Bettmann2023,
  author =        {Bettmann, Laetitia P. and Kewming, Michael J. and
                   Goold, John},
  journal =       {Phys. Rev. E},
  month =         apr,
  number =        {4},
  publisher =     {American Physical Society (APS)},
  title =         {Thermodynamics of a continuously monitored
                   double-quantum-dot heat engine in the repeated
                   interactions framework},
  volume =        {107},
  year =          {2023},
  doi =           {10.1103/physreve.107.044102},
  issn =          {2470-0053},
  url =           {http://dx.doi.org/10.1103/PhysRevE.107.044102},
}

@article{Landauer1991,
  author =        {Landauer, Rolf},
  journal =       {Physics Today},
  month =         may,
  number =        {5},
  pages =         {23–29},
  publisher =     {AIP Publishing},
  title =         {Information is Physical},
  volume =        {44},
  year =          {1991},
  doi =           {10.1063/1.881299},
  issn =          {1945-0699},
  url =           {http://dx.doi.org/10.1063/1.881299},
}

@article{Szilard1964,
  author =        {Szilard, Leo},
  journal =       {Behavioral Science},
  number =        {4},
  pages =         {301–310},
  publisher =     {Wiley},
  title =         {On the decrease of entropy in a thermodynamic system
                   by the intervention of intelligent beings},
  volume =        {9},
  year =          {1964},
  doi =           {10.1002/bs.3830090402},
  issn =          {1099-1743},
  url =           {http://dx.doi.org/10.1002/bs.3830090402},
}

@article{Ciliberto2017,
  author =        {Ciliberto, S.},
  journal =       {Phys. Rev. X},
  month =         {Jun},
  pages =         {021051},
  publisher =     {American Physical Society},
  title =         {Experiments in Stochastic Thermodynamics: Short
                   History and Perspectives},
  volume =        {7},
  year =          {2017},
  doi =           {10.1103/PhysRevX.7.021051},
  url =           {https://link.aps.org/doi/10.1103/PhysRevX.7.021051},
}

@article{Myers2022,
  author =        {Myers, Nathan M. and Abah, Obinna and
                   Deffner, Sebastian},
  journal =       {AVS Quantum Science},
  month =         apr,
  number =        {2},
  publisher =     {American Vacuum Society},
  title =         {Quantum thermodynamic devices: From theoretical
                   proposals to experimental reality},
  volume =        {4},
  year =          {2022},
  doi =           {10.1116/5.0083192},
  issn =          {2639-0213},
  url =           {http://dx.doi.org/10.1116/5.0083192},
}

@article{Toyabe2010,
  author =        {Toyabe, Shoichi and Sagawa, Takahiro and
                   Ueda, Masahito and Muneyuki, Eiro and Sano, Masaki},
  journal =       {Nature Physics},
  month =         nov,
  number =        {12},
  pages =         {988–992},
  publisher =     {Springer Science and Business Media LLC},
  title =         {Experimental demonstration of information-to-energy
                   conversion and validation of the generalized
                   Jarzynski equality},
  volume =        {6},
  year =          {2010},
  doi =           {10.1038/nphys1821},
  issn =          {1745-2481},
  url =           {http://dx.doi.org/10.1038/nphys1821},
}

@article{Brut2012,
  author =        {Bérut, Antoine and Arakelyan, Artak and
                   Petrosyan, Artyom and Ciliberto, Sergio and
                   Dillenschneider, Raoul and Lutz, Eric},
  journal =       {Nature},
  month =         mar,
  number =        {7388},
  pages =         {187–189},
  publisher =     {Springer Science and Business Media LLC},
  title =         {Experimental verification of Landauer’s principle
                   linking information and thermodynamics},
  volume =        {483},
  year =          {2012},
  doi =           {10.1038/nature10872},
  issn =          {1476-4687},
  url =           {http://dx.doi.org/10.1038/nature10872},
}

@article{Paneru2018,
  author =        {Paneru, Govind and Lee, Dong Yun and Tlusty, Tsvi and
                   Pak, Hyuk Kyu},
  journal =       {Phys. Rev. Lett.},
  month =         {Jan},
  pages =         {020601},
  publisher =     {American Physical Society},
  title =         {Lossless Brownian Information Engine},
  volume =        {120},
  year =          {2018},
  doi =           {10.1103/PhysRevLett.120.020601},
  url =           {https://link.aps.org/doi/10.1103/PhysRevLett.120.020601},
}

@article{Koski2014,
  author =        {Koski, Jonne V. and Maisi, Ville F. and
                   Pekola, Jukka P. and Averin, Dmitri V.},
  journal =       {PNAS},
  month =         sep,
  number =        {38},
  pages =         {13786--13789},
  title =         {Experimental realization of a {Szilard} engine with a
                   single electron},
  volume =        {111},
  year =          {2014},
  doi =           {10.1073/pnas.1406966111},
  issn =          {0027-8424, 1091-6490},
}

@article{Kumar2018,
  author =        {Kumar, Aishwarya and Wu, Tsung-Yao and
                   Giraldo, Felipe and Weiss, David S.},
  journal =       {Nature},
  month =         sep,
  number =        {7721},
  pages =         {83–87},
  publisher =     {Springer Science and Business Media LLC},
  title =         {Sorting ultracold atoms in a three-dimensional
                   optical lattice in a realization of Maxwell’s
                   demon},
  volume =        {561},
  year =          {2018},
  doi =           {10.1038/s41586-018-0458-7},
  issn =          {1476-4687},
  url =           {http://dx.doi.org/10.1038/s41586-018-0458-7},
}

@article{Camati2016,
  author =        {Camati, Patrice A. and Peterson, John P. S. and
                   Batalhão, Tiago B. and Micadei, Kaonan and
                   Souza, Alexandre M. and Sarthour, Roberto S. and
                   Oliveira, Ivan S. and Serra, Roberto M.},
  journal =       {Phys. Rev. Lett.},
  month =         dec,
  number =        {24},
  publisher =     {American Physical Society (APS)},
  title =         {Experimental Rectification of Entropy Production by
                   Maxwell’s Demon in a Quantum System},
  volume =        {117},
  year =          {2016},
  doi =           {10.1103/physrevlett.117.240502},
  issn =          {1079-7114},
  url =           {http://dx.doi.org/10.1103/PhysRevLett.117.240502},
}

@article{Cottet2017,
  author =        {Cottet, Nathanaël and Jezouin, Sébastien and
                   Bretheau, Landry and Campagne-Ibarcq, Philippe and
                   Ficheux, Quentin and Anders, Janet and
                   Auffèves, Alexia and Azouit, Rémi and
                   Rouchon, Pierre and Huard, Benjamin},
  journal =       {Proceedings of the National Academy of Sciences},
  month =         jul,
  number =        {29},
  pages =         {7561–7564},
  publisher =     {Proceedings of the National Academy of Sciences},
  title =         {Observing a quantum Maxwell demon at work},
  volume =        {114},
  year =          {2017},
  doi =           {10.1073/pnas.1704827114},
  issn =          {1091-6490},
  url =           {http://dx.doi.org/10.1073/pnas.1704827114},
}

@article{Paneru2018PRE,
  author =        {Paneru, Govind and Lee, Dong Yun and Park, Jong-Min and
                   Park, Jin Tae and Noh, Jae Dong and Pak, Hyuk Kyu},
  journal =       {Phys. Rev. E},
  month =         {Nov},
  pages =         {052119},
  publisher =     {American Physical Society},
  title =         {Optimal tuning of a Brownian information engine
                   operating in a nonequilibrium steady state},
  volume =        {98},
  year =          {2018},
  doi =           {10.1103/PhysRevE.98.052119},
  url =           {https://link.aps.org/doi/10.1103/PhysRevE.98.052119},
}

@article{Miller2020,
  author =        {Harry J.{\hspace{0.167em}}D. Miller and
                   Giacomo Guarnieri and Mark T. Mitchison and
                   John Goold},
  journal =       {Phys. Rev. Lett.},
  month =         oct,
  number =        {16},
  publisher =     {American Physical Society ({APS})},
  title =         {Quantum Fluctuations Hinder Finite-Time Information
                   Erasure near the Landauer Limit},
  volume =        {125},
  year =          {2020},
  doi =           {10.1103/physrevlett.125.160602},
  url =           {https://doi.org/10.1103/physrevlett.125.160602},
}

@article{Blaber2021,
  author =        {Blaber, Steven and Louwerse, Miranda D. and
                   Sivak, David A.},
  journal =       {Phys. Rev. E},
  month =         {Aug},
  pages =         {L022101},
  publisher =     {American Physical Society},
  title =         {Steps minimize dissipation in rapidly driven
                   stochastic systems},
  volume =        {104},
  year =          {2021},
  doi =           {10.1103/PhysRevE.104.L022101},
  url =           {https://link.aps.org/doi/10.1103/PhysRevE.104.L022101},
}

@article{Albash2012,
  author =        {Albash, Tameem and Boixo, Sergio and Lidar, Daniel A and
                   Zanardi, Paolo},
  journal =       {New Journal of Physics},
  month =         dec,
  number =        {12},
  pages =         {123016},
  publisher =     {IOP Publishing},
  title =         {Quantum adiabatic Markovian master equations},
  volume =        {14},
  year =          {2012},
  doi =           {10.1088/1367-2630/14/12/123016},
  issn =          {1367-2630},
  url =           {http://dx.doi.org/10.1088/1367-2630/14/12/123016},
}

@article{Dann2018,
  author =        {Dann, Roie and Levy, Amikam and Kosloff, Ronnie},
  journal =       {Phys. Rev. A},
  month =         nov,
  number =        {5},
  publisher =     {American Physical Society (APS)},
  title =         {Time-dependent Markovian quantum master equation},
  volume =        {98},
  year =          {2018},
  doi =           {10.1103/physreva.98.052129},
  issn =          {2469-9934},
  url =           {http://dx.doi.org/10.1103/PhysRevA.98.052129},
}

@article{Guarnieri2018,
  author =        {Guarnieri, G and Ng, N. H.Y. and Modi, K and
                   Eisert, J and Paternostro, M and Goold, J},
  journal =       {Phys. Rev. E},
  number =        {5},
  pages =         {050101},
  title =         {{Quantum work statistics and resource theories:
                   Bridging the gap through R{\'{e}}nyi divergences}},
  volume =        {99},
  year =          {2019},
  doi =           {10.1103/PhysRevE.99.050101},
  issn =          {24700053},
}

@article{Speck2004,
  author =        {Speck, Thomas and Seifert, Udo},
  journal =       {Phys. Rev. E},
  pages =         {066112},
  title =         {{Distribution of work in isothermal non-equilibrium
                   processes}},
  volume =        {70},
  year =          {2004},
  url =           {https://journals.aps.org/pre/abstract/10.1103/
                  PhysRevE.70.066112},
}

@article{Fusco2014,
  author =        {Fusco, L and Pigeon, S and Apollaro, T J G and
                   Xuereb, A and Mazzola, L and Campisi, M and
                   Ferraro, A and Paternostro, M and Chiara, G De},
  journal =       {Phys. Rev. X},
  pages =         {1--18},
  title =         {{Assessing the Nonequilibrium Thermodynamics in a
                   Quenched Quantum Many-Body System via Single
                   Projective Measurements}},
  volume =        {031029},
  year =          {2014},
  doi =           {10.1103/PhysRevX.4.031029},
}

@article{Bumer2019,
  author =        {Elisa B\"{a}umer and Mart{\'{\i}} Perarnau-Llobet and
                   Philipp Kammerlander and Henrik Wilming and
                   Renato Renner},
  journal =       {Quantum},
  month =         jun,
  pages =         {153},
  publisher =     {Verein zur Forderung des Open Access Publizierens in
                   den Quantenwissenschaften},
  title =         {Imperfect Thermalizations Allow for Optimal
                   Thermodynamic Processes},
  volume =        {3},
  year =          {2019},
  doi =           {10.22331/q-2019-06-24-153},
  url =           {https://doi.org/10.22331/q-2019-06-24-153},
}

@article{Janyszek1986,
  author =        {Janyszek, Henryk},
  journal =       {Reports on Mathematical Physics},
  month =         aug,
  number =        {1},
  pages =         {11–19},
  publisher =     {Elsevier BV},
  title =         {On the geometrical structure of the generalized
                   quantum Gibbs states},
  volume =        {24},
  year =          {1986},
  doi =           {10.1016/0034-4877(86)90037-6},
  issn =          {0034-4877},
  url =           {http://dx.doi.org/10.1016/0034-4877(86)90037-6},
}

@book{Sachdev2011,
  author =        {Sachdev, Subir},
  month =         apr,
  publisher =     {Cambridge University Press},
  title =         {Quantum Phase Transitions},
  year =          {2011},
  doi =           {10.1017/cbo9780511973765},
  isbn =          {9780511973765},
  url =           {http://dx.doi.org/10.1017/CBO9780511973765},
}

@article{solon2018,
  author =        {Solon, Alexandre P. and Horowitz, Jordan M.},
  journal =       {Phys. Rev. Lett.},
  month =         may,
  number =        {18},
  publisher =     {American Physical Society (APS)},
  title =         {Phase Transition in Protocols Minimizing Work
                   Fluctuations},
  volume =        {120},
  year =          {2018},
  doi =           {10.1103/physrevlett.120.180605},
  issn =          {1079-7114},
  url =           {http://dx.doi.org/10.1103/PhysRevLett.120.180605},
}

@article{Sivak2012,
  author =        {Sivak, David A. and Crooks, Gavin E.},
  journal =       {Phys. Rev. L},
  number =        {19},
  pages =         {190602 (2012)},
  title =         {{Thermodynamic metrics and optimal paths}},
  volume =        {108},
  year =          {2012},
  doi =           {10.1103/PhysRevLett.108.190602},
  isbn =          {0031-9007},
  issn =          {00319007},
}

@article{Deffner2020,
  author =        {Sebastian Deffner and Marcus V. S. Bonan{\c{c}}a},
  journal =       {{EPL} (Europhysics Letters)},
  month =         aug,
  number =        {2},
  pages =         {20001},
  publisher =     {{IOP} Publishing},
  title =         {Thermodynamic control {\textemdash}An old paradigm
                   with new applications},
  volume =        {131},
  year =          {2020},
  doi =           {10.1209/0295-5075/131/20001},
  url =           {https://doi.org/10.1209/0295-5075/131/20001},
}

@article{Potts2021,
  author =        {Patrick P Potts and Alex Arash Sand Kalaee and
                   Andreas Wacker},
  journal =       {New Journal of Physics},
  month =         {dec},
  number =        {12},
  pages =         {123013},
  publisher =     {IOP Publishing},
  title =         {A thermodynamically consistent Markovian master
                   equation beyond the secular approximation},
  volume =        {23},
  year =          {2021},
  doi =           {10.1088/1367-2630/ac3b2f},
  url =           {https://dx.doi.org/10.1088/1367-2630/ac3b2f},
}

@article{Soret2022,
  author =        {Soret, Ariane and Cavina, Vasco and
                   Esposito, Massimiliano},
  journal =       {Phys. Rev. A},
  month =         {Dec},
  pages =         {062209},
  publisher =     {American Physical Society},
  title =         {Thermodynamic consistency of quantum master
                   equations},
  volume =        {106},
  year =          {2022},
  doi =           {10.1103/PhysRevA.106.062209},
  url =           {https://link.aps.org/doi/10.1103/PhysRevA.106.062209},
}

%

\end{document}